\documentclass[superscriptaddress, nofootinbib, amsmath, amssymb]{revtex4}
\usepackage{graphicx}
\usepackage{wasysym}
\usepackage[colorlinks=true,citecolor=blue,linkcolor=blue,urlcolor=blue]{hyperref}
\usepackage{color}
\begin{document}
\title{Degravitation, Orbital Dynamics and the Effective Barycentre}
\author{Alain Dirkes}
\email{dirkes@fias.uni-frankfurt.de}
\affiliation{Frankfurt Institute for Advanced Studies (FIAS), Goethe University Frankfurt,\\ Ruth-Moufang Str.1, Frankfurt am Main, D-60438, Germany}
\begin{abstract}
In this article we present a particular theory of gravity in which Einstein's field equations are modified by promoting Newton's constant $G$ to a covariant differential operator $G_\Lambda(\Box_g)$. The general idea was obviously outlined for the first time in \cite{Dvali1, Dvali2, Barvinsky1, Barvinsky2} and originates from the quest of finding a mechanism that is able to {\it degravitate} the vacuum energy on cosmological scales. We suggest in this manuscript a precise covariant coupling model which acts like a high-pass filter with a macroscopic distance filter scale $\sqrt{\Lambda}$. In the context of this specific theory of gravity we review some cosmological aspects before we briefly recall the effective relaxed Einstein equations outlined for the first time in \cite{Alain1}. We present a general procedure to determine the gravitational potentials for a far away wave zone field point. Moreover we work out the modified orbital dynamics of a binary-system as well as the effective 1.5 post-Newtonian barycentre for a generic $n$-body system. We notice that it is always possible to recover the corresponding general relativistic results in the limit of vanishing nonlocal modification parameters. 
\end{abstract}
\maketitle
\section{Introduction:}
In this section we will briefly recall the nonlocally modified Einstein field equations, introduced for the first time in \cite{Alain1}, before we outline how the vacuum energy is effectively {\it degravitated} on cosmological scales. We will close this chapter by presenting a succinct cosmological model in which we worked out the effective Friedmann-Lema\^{i}tre equation. In the second chapter we will quickly review the standard relaxed Einstein equations and their solutions in terms of a post-Newtonian expansion. In the third chapter we will work out the effective wave equation and provide a formal solution for a far away wave zone field point. Chapter four is devoted to the study of the nonlocally modified effective energy-momentum pseudotensor. In chapter five we determined the effective orbital dynamics of a binary-system and we determined an upper bound for one of the two, {\it \`{a} priori}, free parameters. In the penultimate chapter we combine the results worked out previously in this article in order to compute the effective 1.5 post-Newtonian barycentre for a generic $n$-body system and in particular we compute the position vectors of a two-body system at the same order of accuracy. It should be noticed that most of the chapters presented in this article have a separate appendix-section in which we outline additional computational details.
\subsection{The nonlocally modified Einstein field equations:}
Albert Einstein's elegant gravitational field equations, $G_{\alpha\beta}\,=\,\frac{8\pi}{c^4} \ G \ T_{\alpha\beta}$, \cite{Einstein1} can be concisely summarized by John A. Wheeler's eminent words {\it matter tells spacetime how to curve and spacetime tells matter how to move}. The gravitational field $g(x)$ is contained within the famous Einstein curvature tensor $G^{\alpha\beta}$ and $T^{\alpha\beta}$ is the energy-momentum tensor of the materiel source. The correlation between the gravitational field produced by a matter source term was already discovered long before Einstein published his theory of general relativity (GR) and can be elegantly summarized by the well known Poisson equation. The latter however is purely phenomenological, whereas the theory of general relativity provides, via the concept of spacetime curvature, a deeper understanding of the true nature of gravity. Only one year after the final formulation of the theory of general relativity, Einstein predicted the existence of gravitational waves, generated by time variations of the mass quadrupole moment of the source. Although the direct experimental detection is extremely challenging because of the waves' remarkably small amplitude \cite{Einstein2, Einstein3}, gravitational radiation has been measured indirectly since the mid seventies of the past century in the context of binary-systems \cite{Taylor1, Burgay1, Stairs1, Stairs2, Taylor2}. Precisely one century after Einstein's theoretical prediction, an international collaboration of scientists (LIGO Scientific Collaboration and Virgo Collaboration) reported the first direct observation of gravitational waves \cite{LIGO1,LIGO2,LIGO3}. The wave signal GW150914 was detected independently by the two LIGO detectors and its basic features point to the coalescence of two stellar black holes. Despite the considerable success of Einstein's theory in describing the gravitational field in the context of astrophysics and cosmology, some challenges remain yet unsolved. The most prominent questions that need to be addressed are the missing mass problem, the dark energy problem, the physical interpretation of black hole curvature singularities or the question of how to unify quantum mechanics and general relativity. In order to circumvent some of these issues many potentially viable alternative theories of gravity have been developed over the past decades \cite{Will1, Esposito1, Clifton1, Tsujikawa1, Woodard1, BertiBuonannoWill}. In this sense we will in the remaining part of this chapter briefly resume the nonlocally modified theory of gravity outlined for the first time in \cite{Alain1}. The main difference between our modified theory of gravity and the standard field equations is that we promote the Newton's gravitational constant to a covariant differential operator,
\begin{equation}
\label{NonlocalEinstein}
G_{\alpha\beta}\,=\, \frac{8\pi}{c^4} \ G_\Lambda(\Box_g) \ T_{\alpha\beta},
\end{equation}
where $\Box_g=\nabla^{\alpha}\nabla_\alpha$ is the covariant d'Alembert operator and $\sqrt{\Lambda}$ is the scale at which infrared (IR) modifications become important. The generic idea to modify Einstein's field equations in this way was apparently formulated for the first time in \cite{Dvali1,Barvinsky1,Dvali2,Barvinsky2} in framework of the cosmological constant problem \cite{Weinberg1}. The concept of a varying gravitational coupling parameter dates back to early works of Dirac \cite{Dirac1} and Jordan \cite{Jordan1, Jordan2}. Inspired by these considerations Brans and Dicke published in the early sixties a theory in which the gravitational constant is replaced by the reciprocal of a scalar field \cite{Brans1}. Further developments going in the same direction can be inferred from \cite{Narlikar1, Isern1, Uzan2}. The model that we present in this article originates from purely bottom-up considerations. It is however worth mentioning that many theoretical approaches, such as models with extra dimensions, string theory or scalar–tensor models of quintessence \cite{Peebles1,Steinhardt1,Lykkas1} contain a built–in mechanism for a possible time variation of the couplings \cite{Dvali3, Dvali4, Dvali5,Parikh1,Damour1,Uzan1,Lykkas1}. This phenomenon, usually referred to as the running of the coupling constants, is well known from quantum field theory and has been extensively studied by using renormalization group techniques \cite{PeskinSchroeder, ReuterSaueressig1, Shankar1}. The main difference between the standard general relativistic theory and our nonlocally modified theory lies in the way in which the energy-momentum tensor source term is translated into spacetime curvature. In the standard theory of gravity this translation is ensured by the gravitational coupling constant $G$, whereas in our modified approach the coupling between the energy source term and the gravitational field will be in the truest sense of the word more differentiated. The covariant d'Alembert operator is sensitive to the characteristic wavelength of the gravitating system under consideration $1/\sqrt{-\Box_g} \sim \lambda_c$. We will see that our precise model will be constructed in such a way that the long-distance modification is almost inessential for processes varying in spacetime much faster than $1/\sqrt{\Lambda}$ and large for slower phenomena at wavelengths $\sim \sqrt{\Lambda}$ and larger. In this regard spatially extended processes varying very slowly in time, with a small characteristic frequency $\nu_c\sim 1/\lambda_c$, will produce a less stronger gravitational field than smaller fast moving objects like solar-system planets or even earth sized objects. The latter possess rather small characteristic wavelengths and will therefore couple to the gravitational field in almost the usual way. Deviations from the purely general relativistic results will be discussed in the second half of this article in the context of astrophysical binary-systems. Cosmologically extended processes with a small characteristic frequency will effectively decouple from the gravitational field. John Wheeler's famous statement about the mutual influence of matter and spacetime curvature remains of course true, the precise form of the coupling differs however according to the dynamical nature of the gravitating object under consideration. Indeed promoting Newton's constant $G$ to a differential operator $G_{\Lambda}(\Box_g)$ allows for an interpolation between the Planckian value of the gravitational constant and its long distance magnitude \cite{Barvinsky1,Barvinsky2},
\begin{eqnarray*}
G_P>G_\Lambda(\Box_g)>G_{L}.
\end{eqnarray*}
Thus the differential operator acts like a high-pass filter with a macroscopic distance filter scale $\sqrt{\Lambda}$. In this way sources characterized by characteristic wavelengths much smaller than the filter scale  ($\lambda_c\ll\sqrt{\Lambda}$) pass undisturbed through the filter and gravitate normally, whereas sources characterized by wavelengths larger than the filter scale are effectively filtered out \cite{Dvali1,Dvali2}. In a more quantitative way we can see how this filter mechanism works by introducing the dimensionless parameter $z\,=\,-\Lambda \Box_g\sim \Lambda/\lambda_c^2$,
\begin{eqnarray*}
G(z)\rightarrow G, \ |z|\gg 1 \ (\lambda_c \ll 1),\quad \quad G(z)\rightarrow 0, \ \  |z|\ll 1 \ (\lambda_c \gg 1).
\end{eqnarray*}
For small and fast moving objects with large values of $|z|$ (small characteristic wavelengths) the covariant coupling operator will essentially reduce to Newton's constant $G$, whereas for slowly varying processes characterized by small values of $|z|$ (large characteristic wavelengts) the coupling will be much smaller.  Despite the fact that the equations of motion $\eqref{NonlocalEinstein}$ are themselves generally covariant, they cannot, for nontrivial $G_\Lambda(\Box_g)$, be represented as a metric variational derivative of a diffeomorphism invariant action. The solution to this problem was suggested in \cite{Barvinsky1,Barvinsky2, Modesto1} by considering equation $\eqref{NonlocalEinstein}$ only as a first, linear in the curvature, approximation for the correct equations of motion. Further technical details, regarding this particular issue, can be withdrawn from \cite{Alain1,Barvinsky3,Barvinsky4, Barvinsky5}. In the context of the cosmological constant problem we aim to briefly summarize the main features of the differential coupling model outlined for the first time in \cite{Alain1}, 
\begin{equation*}
G_{\Lambda}(\Box)\,=\,\mathcal{G}_{\kappa}(\Box_g) \ \mathcal{F}_\Lambda(\Box_g).
\end{equation*}
We recall that $\mathcal{G}_{\kappa}=\frac{G}{1-\sigma e^{\kappa\Box_g}}$ is a purely ultraviolet (UV) modification term and $\mathcal{F}_\Lambda=\frac{\Lambda \Box_g}{\Lambda \Box_g-1}$ is the nonlocal infrared (IR) contribution. We also remind that this particular model contains all of the {\it degravitation} properties mentioned earlier in this section \cite{Alain1}. We observe that, in the limit of vanishing wavelengths or infinitely large frequencies, we obtain Einstein's theory of general relativity as the UV-term reduces to the Newtonian coupling constant ($\lim_{z\rightarrow +\infty} \mathcal{G}_{\kappa}(z)=G$) and the IR-term goes to one ($\lim_{z\rightarrow +\infty}\mathcal{F}_\Lambda=1$). The IR-degravitation essentially comes from $\lim_{z\rightarrow 0}\mathcal{F}_\Lambda(z)=0$ while the UV-term $\lim_{z\rightarrow 0}\mathcal{G}_\kappa(z)=\frac{G}{1-\sigma}$ taken alone does not vanish in this limit. We will encounter this Newtonian coupling constant, $G(\sigma)=1/(1-\sigma)$, again in chapter five when we work out the effective Newtonian potential. The dimensionless UV-parameter $\sigma$ is a priori not fixed, however in order to make the infrared degravitation mechanism work properly $\sigma$ should be different from one. In addition we will slightly restrain the general character of our theory by assuming that $|\sigma|<1$ should be rather small and here again we will see in chapter five that this particular assumption is indeed well motivated from a phenomenological point of view. The second UV-parameter $\kappa$ s well as the IR-degravitation parameter $\Lambda$ are both of dimension length squared. The constant factor $\sqrt{\Lambda}$ is the cosmological scale at which the infrared degravitation process sets in. In the context of the cosmological constant problem this parameter needs to be typically of the order of the horizon size of the present visible Universe $\sqrt{\Lambda} \sim  10^{30} m$ \cite{Dvali1,Barvinsky1, Barvinsky2, Dvali2}. Moreover we assume that $\sqrt{\kappa}\ll\sqrt{\Lambda}$, so that we can perform a formal series-expansion $\mathcal{G}_\kappa(z)=\sum_{n=0}^{+\infty}\sigma^n e^{-n\frac{\kappa}{\Lambda}z} $ in the UV-regime ($|z|\ll1$). The parameter $\kappa$, although named differently, was encountered in the context of various nonlocal modified theories of gravity which originate from the pursuit of constructing a UV-complete theory of quantum gravity or coming from models of noncommutative geometry \cite{Modesto1,Modesto2, Spallucci1, Sakellariadou1}. Finally it should be observed that the standard Einstein field equations are recovered in the limit of vanishing UV parameters and infinitely large IR parameter ($\lim_{\sigma,\kappa\rightarrow 0}\lim_{\Lambda\rightarrow +\infty}G_{\Lambda}(\Box_g)=G$).
\subsection{The vacuum energy and the degravitation mechanism:}
Shortly after the final publication of the theory of general relativity, Einstein tried to apply his new theory to the whole Universe by implicitly assuming that the latter is homogeneous on cosmological length scales \cite{Einstein5}. This assumption, which is usually referred to as the cosmological principle, claims that on scales of the order of $10^8$-$10^9$ light years all positions of the Universe are essentially the same \cite{Weinberg1, Weinberg2}. We will return to this idea in the next subsection where we work out a succinct cosmological model in the context of our nonlocally modified theory of gravity. Although Einstein's guiding principle was that the Universe must be static, no such static solutions of his original equations could be found. He therefore introduced an additional term to his field equations, the cosmological constant, which he later considered as an unnecessary complication to his initial field equations \cite{Einstein5, Weinberg1, Weinberg2}. However from a microscopic point of view it is not so straightforward to discard such a term, because anything that contributes to the energy density of the vacuum acts just like a cosmological constant. Indeed from a quantum point of view the vacuum is a very complex state in the sense that it is constantly permeated by fluctuating quantum fields of different origins. In agreement to Heisenberg's energy-time uncertainty principle one important contribution to the vacuum energy comes from the spontaneous creation of virtual particle-antiparticle pairs which annihilate shortly after \cite{Weinberg1}. Even though there is some freedom in the precise computation of the vacuum energy density, the most reasonable theoretical estimates range around a value of $\rho_{th}\approx 10^{111} J/m^3$ \cite{Carroll1}. Towards the end of the past century two independent research groups, the {\it High-Z Supernova Team} and the {\it Supernova Cosmology Project}, searched for distant type Ia supernovae in order to determine parameters that were supposed to provide information about the cosmological dynamics of the Universe. The two research groups were able to obtain a deeper understanding of the expansion history of the Universe by observing how the brightness of these supernovae varies with redshift. They initially expected to find signs that the expansion of the Universe is slowing down as the expansion rate is essentially determined by the energy-momentum density of the Universe. However in 1998 they published their results in two separate papers and came both independently from each other to the astonishing result that the opposite is true: the expansion of the Universe is accelerated. The supernovae results in combination with the Cosmic Microwave Background data \cite{Planck1}, interpreted in terms of the Standard Model of Cosmology ($\Lambda$CDM-model), allow for a precise determination of the vacuum energy density of the order of $\rho_{ob}\sim 10^{-9} J/m^3$ as it is observed in the Universe. These investigations together with the theoretically computed value for the vacuum energy are at the origin of the famous 120 orders of magnitude discrepancy $\rho_{th} \sim 10^{120} \rho_{ob}$ between the observational and theoretical estimations of the vacuum energy density \cite{Carroll1, Inverno,Weinberg1}. Most efforts in solving this problem have focused on the question why the vacuum energy is so small. However, since nobody has ever measured the energy of the vacuum by any means other than gravity, perhaps the right question to ask is why does the vacuum energy gravitates so little \cite{Dvali1,Barvinsky1,Dvali2, Barvinsky2}. In this regard our aim is not to question the theoretically computed value of the vacuum energy density, but we will rather try to see if we can find a mechanism by which the vacuum energy is effectively degravitated on cosmological scales. In order to sketch how the degravitation mechanism works in the context of our nonlocal coupling model we take up the concise but very illustrative description of the vacuum energy on macroscopic scales outlined in \cite{Alain1}. In this context we will assume that the Universe is essentially flat. This assumption, which is in good agreement to cosmological observations \cite{Planck1}, will permit us to replace the differential coupling operator by its flat spacetime counterpart. We further presume that the quantum vacuum energy can be modelled, on macroscopic scales, by an almost time independent Lorentz-invariant energy process of the form, $\langle T_{\alpha\beta}\rangle_{v}\,\simeq\, T_v \ \cos(\textbf{k}_c \cdot \textbf{x}) \ \eta_{\alpha\beta}$. $T_v$ is the average vacuum energy density and $\textbf{k}_c=1/\mathbf{\lambda}_c$ is the three dimensional characteristic wave vector $(|\mathbf{\lambda}_c|\gg 1)$. Moreover we suppose that the vacuum energy is homogeneously distributed throughout the whole universe so that the components of the wave vector $k_x=k_y=k_z\,\sim 1/\lambda_c$ can be considered equal in all three spatial directions. In this particular framework the effective coupling to the vacuum energy is $G_\Lambda(\Box) \ \langle T_{\alpha\beta} \rangle_{v}=\mathcal{G}(\kappa/\lambda_c^2) \ \mathcal{F}(\Lambda/\lambda_c^2) \ \langle T_{\alpha\beta} \rangle_{v}$, where $\mathcal{G}(\kappa/\lambda_c^2)=\frac{G}{1-\sigma e^{-\kappa/\lambda_c^2} }$ and $\mathcal{F}(\Lambda/\lambda_c^2)=\frac{\Lambda /\lambda_c^2}{1+\ \Lambda/\lambda_c^2}$ \cite{Kragler1}. We observe that energy processes with a characteristic wavelength, much larger than the macroscopic filter scale $\lambda_c \gg \sqrt{\Lambda}$ effectively decouple from the gravitational field $\lim_{\lambda_c \rightarrow+\infty} G_\Lambda(\Box) \langle T_{\alpha\beta} \rangle_{v}\,=\, 0$. For energy processes characterized by very large frequencies, $\lim_{\lambda_c\rightarrow 0}\mathcal{G}(\kappa/\lambda^2_c)=G$, $\lim_{\lambda_c\rightarrow 0}\mathcal{F}(\Lambda/\lambda^2_c)=1$ we essentially retrieve the standard Newtonian coupling constant $G$. 
\begin{figure}[h]
\begin{center}
\includegraphics[width=8.8cm,height=4.5cm]{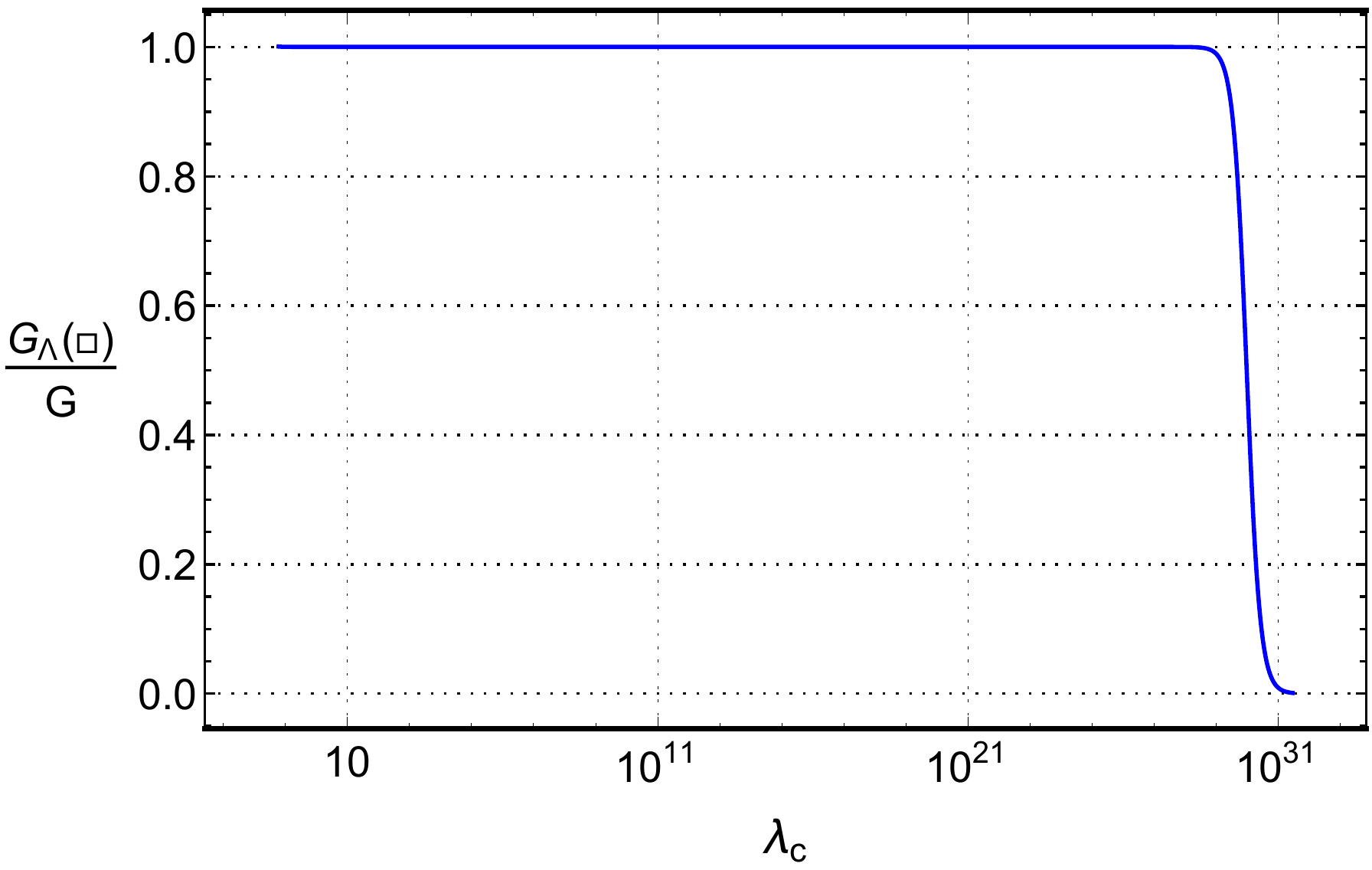} 
\end{center}
\caption{\label{Degravitation1}The function $\frac{G_\Lambda(\Box)}{G}=\frac{\mathcal{G}(\kappa/\lambda_c^2)\mathcal{F}(\Lambda/\lambda^2_c)}{G}$ is plotted against the characteristic wavelength $\lambda_c$ (m) for $\sigma=2 \ 10^{-4}$, $\kappa=5 \ 10^{-3}$ m$^2$ and $\Lambda=10^{60}$ m$^2$. A strong degravitational effect is observed for energy processes with a characteristic wavelength larger or equal to $\lambda_c=10^{29}$m.}
\end{figure}
The degravitation mechanism is illustrated in FIG. \ref{Degravitation1} where we plotted the function $\big[\frac{G_\Lambda(\Box)}{G}\langle T_{\alpha\beta}\rangle_v\big][\langle T_{\alpha\beta}\rangle_v]^{-1}=\frac{\mathcal{G}(\kappa/\lambda_c^2)\mathcal{F}(\Lambda/\lambda^2_c)}{G}$ against the characteristic wavelength for the following set of UV and IR parameters $\sigma=2\ 10^{-4}$, $\kappa=5 \ 10^{-3}$ m$^2$ and $\Lambda=10^{60}$ m$^2$. We deduce from FIG. \ref{Degravitation1} that in the context of our vacuum energy model we have for small characteristic wavelengths $G_\Lambda(\Box)\sim G$ while for large wavelengths of the order $\lambda_c=10^{29}$ m we observe a strong {\it degravitational} effect. In the remaining chapters of this article we will investigate in how far the relaxed Einstein equations are affected by the nonlocal ultraviolet term $\mathcal{G}_\kappa(\Box_g)$. In particular we will examine in chapter five the effective orbital dynamics of a binary-system and in the penultimate chapter of this article we will investigate in how far the barycentre of an $n$-body system deviates from the purely general relativistic result. However before we embark for these computations we aim to present in the next subsection a concise cosmological model based on the cosmological principle assumption.
\subsection{A succinct cosmological model:}
In this section we will analyse our nonlocally modified theory of gravity by constructing a concise cosmological model based on the cosmological principle assumption. The latter claims that when averaged over length-scales ($\sim 10^9$ light years) the matter distribution of the Universe is homogeneous and isotropic \cite{Weinberg2, Inverno}. The metric that goes along with this assumption is the famous Robertson-Walker metric,
\begin{equation*}
ds^2\,=\,-c^2dt^2+q_{ab}(t,\textbf{x}) \ dx^adx^b,
\end{equation*}
where $q_{ab}$ is a three-dimensional, diagonal matrix whose line element is commonly chosen to be $q_{ab} \ dx^adx^b=R^2(t)\big[\frac{dr^2}{1-kr^2}+r^2d\theta^2+r^2\sin^2\theta d\phi^2\big]$ \cite{Weinberg1, Weinberg2}. $R(t)$ is the cosmic scale factor, $r$, $\theta$, $\phi$ are dimensionless spherical coordinates and $k\in\{-1, \ 0, \ +1\}$ is the dimensionless curvature parameter describing an open, flat or closed Universe respectively. Moreover we will assume that the contents of the Universe are, on the average, at rest in the coordinate system $r$, $\theta$, $\phi$, so that the velocity field of the matter distribution simplifies to $u^\alpha=\gamma (c,\textbf{0})$ and where $\gamma^{-1}=\sqrt{-g_{\mu\nu}\frac{v^\mu v^\nu}{c^2}}$ is dimensionless relativistic factor. In this particular context the Universe's energy-momentum tensor will essentially reduce to a perfect fluid, $T^{\alpha\beta}=(c^2\rho+p) \ u^\alpha u^\beta/c^2+p \ g^{\alpha\beta}$, where $\rho$ is the matter density, $p$ is the pressure and $g_c^{\alpha\beta}$ is the inverse four-dimensional Robertson-Walker metric outlined above. In terms of this metric the generally covariant d'Alembert operator becomes\cite{PoissonWill, Woodard1, Weinberg2},
\begin{equation*}
\Box_R\,=\, -\partial_0^2-\frac{\dot{q}}{2q}\partial_0+\frac{1}{\sqrt{q}}\partial_a\Big(\sqrt{q} \ q^{ab}\partial_b\Big),
\end{equation*}
where $q=det(q_{ab})$ is the three-dimensional metric determinant. It should be noticed that $\frac{\dot{q}}{2q}=\frac{\dot{R}}{R}=3H$ can be rephrased in terms of the Hubble parameter \cite{Weinberg1,Weinberg2,Inverno}, which accounts for the expansion rate of the Universe. Further computational details are provided in the appendix-section related to this chapter. The present value of the cosmic scale factor, which is sometimes called the "radius of the Universe" \cite{Weinberg2}, is rather large and that the current cosmological pressure term is small ($p\approx 0$) compared to the value of the early Universe. In the framework of this first concise cosmological analysis we will therefore simplify the effective cosmological energy-momentum tensor to the following expression, $\mathcal{T}^{\alpha\beta}_\Lambda=\frac{G_\Lambda(\Box_R)}{G}T^{\alpha\beta}\approx\text{diag}(c^2\rho_\Lambda,\ q^{11}\ p_\Lambda,\ q^{22}\ p_\Lambda,\ q^{33}\ p_\Lambda)$, where $\rho_\Lambda=\frac{G_\Lambda(\Box_R)}{G}\rho$ is the effective matter density, $p_\Lambda=\frac{G_\Lambda(\Box_R)}{G}p$ is the fluid's effective pressure term and $T^{\alpha\beta}=\text{diag}(\rho c^2,\ q^{11}p,\ q^{22}p,\ q^{33}p)$is the diagonal energy-momentum tensor with, $q^{aa}=q_{aa}^{-1}$. It should be noticed that in the limit of vanishing UV-parameters ($\sigma,\kappa \rightarrow 0$) and infinitely large IR-parameter ($\Lambda\rightarrow +\infty$) we recover the standard energy-momentum tensor of the perfect fluid. We previously saw that the nonlocally modified Einstein field equations $G^{\alpha\beta}=\frac{8\pi}{c^4} \ G\ \mathcal{T}^{\alpha\beta}$ together with the contracted Bianchi identities $\nabla_\alpha G^{\alpha\beta}=0$ \cite{PoissonWill, Weinberg2, MisnerThroneWheeler} give rise to an effective energy-momentum conservation equation $\nabla_\alpha \mathcal{T}^{\alpha\beta}=0$. This allows us to derive an energy conservation equation which relates the effective matter density $\rho_\Lambda$ and effective pressure $p_\Lambda$ with the cosmic scale factor, $\dot{\rho}_\Lambda\,=\,-\frac{3\dot{R}}{R} \Big(\rho_\Lambda+\frac{p_\Lambda}{c^2}\Big)$, where $\dot{\rho}_\Lambda$ is the first temporal derivative of the effective matter density function. Further computational details can be withdrawn from the appendix-section related to this chapter. Another important relation that can be worked out in this particular context is the effective cosmic acceleration equation $\ddot{R}=-\frac{4\pi G}{3} \big(\rho_\Lambda+\frac{3p_\Lambda}{c^2}\big) R$. Combining the last two equations we obtain the effective Friedmann-Lema\^{i}tre equation,
\begin{equation*}
\dot{R}^2\,=\, \frac{8\pi G}{3}  \rho_\Lambda R^2-kc^2.
\end{equation*}
We observe that in the context of this succinct cosmological model, based essentially on the cosmological principle, we arrived at an equation which resembles the standard Friedmann-Lema\^{i}tre equation \cite{Weinberg1, Weinberg2, Inverno}. The nonlocal complexity is stored entirely inside the effective matter density $\rho_\Lambda=\frac{G_\Lambda(\Box_R)}{G}\rho$ and in the limit $\lim_{\sigma,\kappa\rightarrow 0}\lim_{\Lambda\rightarrow +\infty} \rho_\Lambda=\rho$ we recover the usual equation. Our next task is to study the inverse differential coupling operator $G^{-1}_\Lambda(\Box_R)$ acting on a generic function $f(R)$ depending only on the cosmic scale factor $R(t)$. In this context the cosmological Robertson-Walker d'Alembert operator reduces to $\Box_R=-\partial^2_0-3c^{-1}H\partial_0$, where $\frac{\dot{q}}{2q}=3H$ and $H=\frac{\dot{R}}{R}$ is the time-dependent Hubble parameter \cite{Weinberg2}. In addition we will use the fact that $\sqrt{\Lambda}\sim 10^{30}\ m$ is of the order of the horizon size of the present visible Universe \cite{Barvinsky1,Barvinsky2}, so that in good approximation the nonlocal IR term can be set to one ($\mathcal{F}_\Lambda(\Box_R)\approx 1)$. The leading order term of the remaining nonlocal coupling operator, acting on a general cosmic scale depending function will become in the sense of a post-Newtonian expansion, $\frac{\mathcal{G}(\Box_R)^{-1}}{G}f(R)=[1-\sigma e^{\kappa\Box_R}]f(R)[1-\sigma(1+\kappa\ \Box_R)]f(R)+\mathcal{O}(c^{-4})$, where the precise form of $\Box_R\propto c^{-2}$ for this particular situation was outlined above. This eventually allows us to see how the differential operator acts on the left hand side of the Friedmann-Lema\^{i}tre equation,
\begin{equation*}
\Big[1-\sigma+\frac{\epsilon}{c^2}\partial^2_t+\frac{3H\epsilon}{c}\partial_t\Big]\ \Big[\frac{\dot{R}^2}{R^2}+k\frac{c^2}{R^2}\Big]\,=\,\frac{8\pi G}{3} \rho,
\end{equation*}
at the 1.5 post-Newtonian order of accuracy and we remind that $\epsilon=\sigma \kappa$ is a parameter of dimension length square.  In a first step we will carry out separately the first and second order temporal derivatives,
\begin{gather*}
\frac{3\epsilon}{c^2}H\partial_t\Big[\frac{\dot{R}^2}{R^2}+k\frac{c^2}{R^2}\Big]\,=\,\frac{\epsilon}{c^2}\Big[-\frac{6 \dot{R}^4}{R^4}+\frac{6 \dot{R}^2 \ddot{R}}{R^3}-kc^2\frac{6\dot{R}^2}{R^4}\Big],\\
\frac{\epsilon}{c^2}\partial^2_t\Big[\frac{\dot{R}^2}{R^2}+k\frac{c^2}{R^2}\Big]\,=\,\frac{\epsilon}{c^2}\Big[\frac{6\dot{R}^4}{R^4}-\frac{10\dot{R}^2\ddot{R}}{R^3}+\frac{2\ddot{R}^2}{R^2}+\frac{2\dot{R}\ \dddot{R}}{R^2}+kc^2\Big(\frac{6\dot{R}^2}{R^4}-\frac{2\ddot{R}}{R^3}\Big)\Big].
\end{gather*}
We see that the leading order terms proportional to $\frac{\epsilon}{c^2}\frac{\dot{R}^4}{R^4}$ and $\frac{\kappa\dot{R}^2}{R^4}$ cancel out each other. The remaining contributions contain terms proportional to second and third order derivative terms of the cosmic scale factor. We will assume that the latter is a very slowly varying function and therefore the second and third order derivatives will, in good approximation, vanish ($\ddot{R}\approx 0$ and $\dddot{R}\approx 0$). Although this assumption is not true for the very early Universe it certainly applies for the more recent expansion history of the Universe \cite{Weinberg2}. By discarding higher order derivative terms of the of the cosmic scale factor we obtain the effective Friedmann-Lema\^{i}tre equation, $\dot{R}^2+kc^2=\frac{8\pi G}{3}\frac{R^2}{1-\sigma}\rho+\mathcal{O}[(\frac{\sqrt{\epsilon}\ddot{R})}{c^2}]$. We will, in the remaining part of this section, work out a solution for this equation in the context of a Universe in which the energy density is dominated by nonrelativistic matter with negligible pressure ($p\ll \rho c^2$). In this context the time dependent matter density function takes the form, $\rho=\rho_0(\frac{R_0}{R})^3$, where $\rho_0=\rho(t=0)$ and $R_0=R(t=0)$ are initial values for the matter density and the cosmic scale factor. With this the leading order term of the effective Friedmann-Lema\^{i}tre equation can be recast in the following form $\dot{R}=c\sqrt{\frac{R_\sigma}{R}-k}$, where $R_\sigma=\frac{8\pi G}{3(1-\sigma)}\frac{\rho_0R^3_0}{c^2}$ is the effective scale factor depending on the dimensionless UV parameter $\sigma$. This differential equation can be solved, for $k\neq 0$, by the ansatz $R(\eta)=\frac{R_\sigma}{k}\sin^2\big(\frac{\sqrt{k}\eta}{2}\big)=\frac{R_\sigma}{2k}\big(1-\cos(\sqrt{k}\eta)\big)$, where $\eta$ is a new dimensionless variable. After separation of the time and cosmic scale variables ($dR=R_\sigma/(2\sqrt{k})\ \sin(\sqrt{k}\eta)$) we obtain,
\begin{equation*}
t\,=\,\frac{R_\sigma}{2kc}\int_0^\eta d\tilde{\eta}\ \frac{\sin(\sqrt{k}\tilde{\eta})}{\sqrt{\frac{1}{\sin^2\big(\frac{\sqrt{k}\tilde{\eta}}{2}\big)-1}}}\,=\,\frac{R_\sigma}{2kc}\int_0^\eta d\tilde{\eta}\ [1-\cos(\sqrt{k}\tilde{\eta})]\,=\,\frac{R_\sigma}{2kc}\Big[\eta-\frac{\sin(\sqrt{k}\eta)}{\sqrt{k}}\Big].
\end{equation*}
It should be noticed that for $k=0$ the effective Friedmann-Lema\^{i}tre equation can be solved without recurring to the parametrisation of the cosmic scale factor as a straightforward computation reveals the following explicit relation between the time variable and the cosmic scale factor, $t=[c\sqrt{R_\sigma}]^{-1}\int_0^Rd\tilde{R}\ \sqrt{\tilde{R}}=\frac{2R^{3/2}}{3c\sqrt{R_\sigma}}$.
\begin{figure}[h] 
\begin{center}
\includegraphics[width=8.4cm,height=4.5cm]{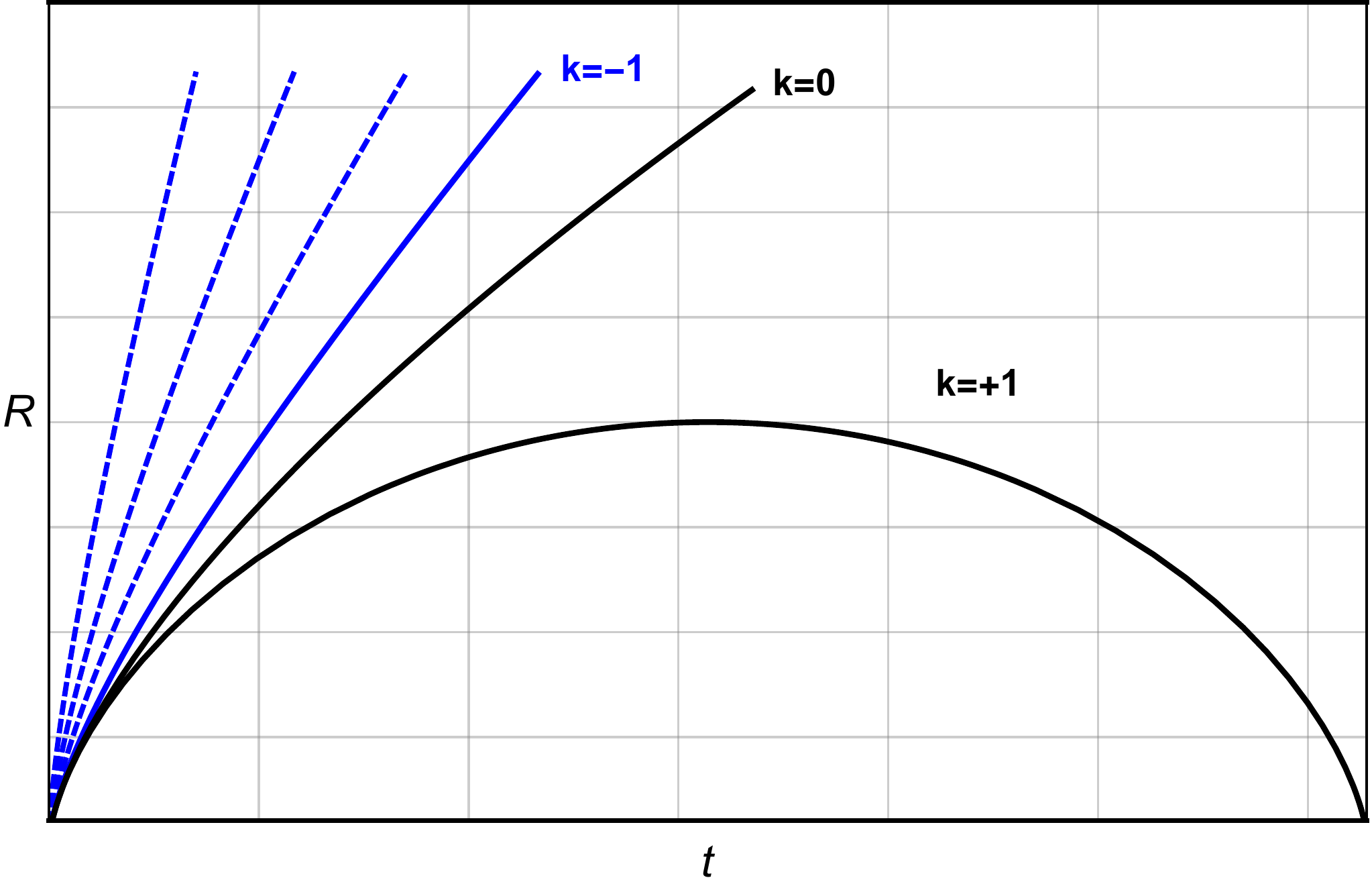} 
\end{center}
\caption{\label{effectiveFriedmann}Solid blue and black curves: The cosmic scale factor $R$ as a function of time, with curvature $k\in \{-1,\ 0, \ 1\}$, in the standard Frieddmann-Lema\^{i}tre model. Solid blue curve: The cosmic scale factor $R$ as a function of time for $k=-1$ and the parameter $\sigma= 0$ taking its minimal possible value in the the context of the effective Friedmann-Lema\^{i}rte model. Dashed blue curves: The cosmic scale factor $R$ as a function of time for $k=-1$ and three different different $\sigma$-values in between zero and one ($0<\sigma<1$).}
\end{figure}
We will close this section by providing the three possible solutions according to the three possible values for the curvature parameter $k\in\{-1,0,+1\}$, which correspond to an open, flat or closed Universe \cite{Weinberg1,Weinberg2},
\vspace{0.2cm}
\begin{center}
\begin{tabular}{c|c|c}
$k=-1$&$k=+1$&$k=0$\\ \hline
\parbox[0pt][3em][c]{0cm}{}
\quad\quad$R=\frac{R_\sigma}{2}\ [\cosh(\eta)-1]$\quad\quad\quad &\quad\quad\quad$R=\frac{R_\sigma}{2}\ [1-\cos(\eta)]$\quad\quad\quad &\quad\quad\quad $R=R_\sigma\big(\frac{t}{t_\sigma}\big)^{2/3}$\quad\quad \\
\quad\quad$t=\frac{R_\sigma}{2c}\ [\sinh(\eta)-\eta]$\quad\quad\quad & \quad\quad\quad $t=\frac{R_\sigma}{2c}\ [\eta-\sin(\eta)]$\quad\quad\quad&\quad\quad\quad $t_\sigma=\frac{2R_\sigma}{3c}$\quad\quad
\end{tabular}
\end{center}
\vspace{0.2cm}

Here we used $\cos(\sqrt{-1}\ \eta)=\cos(i\eta)=\cosh(\eta)$, $\sin(\sqrt{-1}\ \eta)=\sin(i \eta)=i\sinh(\eta)$ and $i^2=-1$ is the square of the imaginary unit. We will close this chapter with a graphical representation FIG. \ref{effectiveFriedmann} of the three different solutions for the effective Friedmann-Lema\^{i}tre equation. We observe that for increasing $\sigma$-values, in between zero and one (blue dashed curves), the cosmic scale factor grows faster than in the standard general relativistic case (solid blue curve), where the dimensionless UV parameter is set to zero ($\sigma=0$). As the plot depends on quantities like the initial matter density $\rho_0$ or the initial cosmic scale factor $R_0$ this concise cosmological model can only provide a qualitative description of the expansion dynamics of the Universe.
\section{An equation for the gravitational potentials:}
In this chapter we will briefly review the relaxed Einstein equations in the context of the Landau-Lifshitz formulation of the Einstein field equations \cite{LandauLifshitz,MisnerThroneWheeler,WillWiseman,PatiWill1,Blanchet1,Buonanno1,PoissonWill,PatiWill2},
\begin{equation*}
\partial_{\mu\nu} H^{\alpha\mu\beta\nu}\,=\, \frac{16\pi G}{c^4}(-g)\ \big(T^{\alpha\beta}+t^{\alpha\beta}_{LL}\big),
\end{equation*}
where $H^{\alpha\mu\beta\nu}\,\equiv\,\mathfrak{g}^{\alpha\beta} \mathfrak{g}^{\mu\nu}-\mathfrak{g}^{\alpha\nu}\mathfrak{g}^{\beta\nu}$ is a tensor density which possesses the same symmetries as the Riemann tensor. In the Landau-Lifshitz formulation of gravity the main variables are not the components of the metric tensor $g_{\alpha\beta}$ but those of the gothic inverse metric, $\mathfrak{g}^{\alpha\beta}\,\equiv\, \sqrt{-g} \ g^{\alpha\beta}$, where $g^{\alpha\beta}$ is the inverse metric and $g$ the metric determinant \cite{LandauLifshitz, MisnerThroneWheeler,WillWiseman,PatiWill1,PatiWill2,Blanchet1,Buonanno1,PoissonWill,Virbhadra2}. $T^{\alpha\beta}$ is the energy-momentum tensor of the matter source term and the Landau-Lifshitz pseudotensor,
\begin{equation*}
\begin{split}
(-g)t^{\alpha\beta}_{LL}\,=\,& \frac{c^4}{16 \pi G} \big[\partial_\lambda \mathfrak{g}^{\alpha\beta} \partial_\mu \mathfrak{g}^{\lambda\mu}-\partial_\lambda \mathfrak{g}^{\alpha\lambda}\partial_\mu \mathfrak{g}^{\beta\mu}+
\frac{1}{2} g^{\alpha\beta} g_{\lambda\mu} \partial_\rho \mathfrak{g}^{\lambda\nu}\partial_\nu \mathfrak{g}^{\mu\rho}
-g^{\alpha\lambda}g_{\mu\nu} \partial_\rho \mathfrak{g}^{\beta\nu}\partial_\lambda \mathfrak{g}^{\mu\rho}-
g^{\beta\lambda}g_{\mu\nu}\partial_\rho \mathfrak{g}^{\alpha\nu}\partial_\lambda \mathfrak{g}^{\mu\rho}\\
&\quad\quad\quad\quad\quad +g_{\lambda\mu}g^{\nu\rho}\partial_\nu \mathfrak{g}^{\alpha\lambda} \partial_\rho \mathfrak{g}^{\beta\mu}+
 \frac{1}{8} (2g^{\alpha\lambda} g^{\beta \mu}-g^{\alpha\beta}g^{\lambda\mu})(2 g_{\nu\rho} g_{\sigma\tau}-g_{\rho\sigma}g_{\nu\tau})
\partial_\lambda \mathfrak{g}^{\nu\tau}\partial_\mu \mathfrak{g}^{\rho\sigma}\big],
\end{split}
\end{equation*}
can be interpreted as an energy momentum pseudotensor for the gravitational field. By virtue of the antisymmetry of $H^{\alpha\mu\beta\nu}$ in the last pair of indices, we have that the equation $\partial_{\beta\mu\nu} H^{\alpha\mu\beta\nu}=0$ holds as an identity. This together with the equation of the Landau-Lifshitz formulation of general relativity implies that, $\partial_\beta \big[(-g)\big(T^{\alpha\beta}+t^{\alpha\beta}_{LL}\big)\big]=0$. It is conventional to choose a particular coordinate system and to impose the four conditions $\partial_\beta \mathfrak{g}^{\alpha\beta}=0$ on the gothic inverse metric, known as the harmonic coordinate conditions. It is also common practice to introduce the gravitational potentials defined by $h^{\alpha\beta}:=\eta^{\alpha\beta}-\mathfrak{g}^{\alpha\beta}$, where $\eta^{\alpha\beta}=diag(-,+,+,+)$ is the Minkowski metric \cite{Blanchet1,Blanchet3,Blanchet4,WillWiseman, PatiWill1, PatiWill2, Buonanno1}. In terms of the potentials the harmonic coordinate conditions read $\partial_\beta h^{\alpha\beta}=0$, and in this context they are usually referred to as the harmonic gauge conditions. It is straightforward to verify that the left-hand side of the Landau-Lifshitz formulation of the Einstein field equations reduces to $\partial_{\mu\nu}H^{\alpha\mu\beta\nu}=-\Box h^{\alpha\beta}+h^{\mu\nu}\partial_{\mu\nu}h^{\alpha\beta}-\partial_\mu h^{\alpha\nu}\partial_\nu h^{\beta\mu}$, where $\Box=\eta^{\mu\nu}\partial_{\mu\nu}$ is the flat-spacetime d'Alembert operator. The right-hand side of the field equations remains essentially unchanged, but the harmonic conditions do slightly simplify the form of the Landau-Lifshitz pseudotensor, namely the first two terms in $(-g)t^{\alpha\beta}_{LL}$ vanish. Isolating the wave operator on the left-hand side and putting the remaining terms on the other side, gives rise to the formal wave equation \cite{PoissonWill,WillWiseman,PatiWill1,PatiWill2,Blanchet1,Blanchet3,Blanchet4,Maggiore1,Buonanno1},
\begin{equation*}
\Box h^{\alpha\beta}\,=\,-\frac{16 \pi G}{c^4} \tau^{\alpha\beta},
\end{equation*}
where $\tau^{\alpha\beta}:=-\frac{16 \pi G}{c^4} \big[ \tau^{\alpha\beta}_m+\tau^{\alpha\beta}_{LL}+\tau^{\alpha\beta}_H\big]$ is defined as the effective energy-momentum pseudotensor composed by a matter $\tau_m^{\alpha\beta}=(-g) T^{\alpha\beta}$ contribution, the Landau-Lifshitz contribution $\tau^{\alpha\beta}_{LL}=(-g)t^{\alpha\beta}_{LL}$ and the harmonic gauge contribution, $\tau^{\alpha\beta}_H=(-g)t^{\alpha\beta}_H=\frac{c^4}{16\pi G} \big(\partial_\mu h^{\alpha\nu}\partial_\nu h^{\beta\mu}-h^{\mu\nu}\partial_{\mu\nu}h^{\alpha\beta}\big)$. It is easy to verify that because of the harmonic gauge condition this additional contribution is separately conserved, $\partial_\beta\big[(-g)t^{\alpha\beta}_H\big]=0$. This together with the conservation relation introduced previously leads to a conservation relation for the effective energy-momentum tensor $\partial_\beta \tau^{\alpha\beta}=0$. The wave equation taken by itself, independently of the harmonic gauge condition or the conservation condition, is known as the relaxed Einstein field equation \cite{PoissonWill, WillWiseman, PatiWill1, PatiWill2}. The formal retarded solution of the wave equation is known from the literature \cite{PoissonWill,WillWiseman,PatiWill1,PatiWill2, Blanchet1,Blanchet3,Blanchet4,Maggiore1, Buonanno1, Maggiore2} and was outlined in a more detailed way in our previous work \cite{Alain1},
\begin{equation*}
h^{\alpha\beta}(t,\textbf{x})\,=\,\frac{4G}{c^4} \int d\textbf{y} \ \frac{\tau^{\alpha\beta}(y^0-|\textbf{x}-\textbf{y}|,\textbf{y})}{|\textbf{x}-\textbf{y}|},
\end{equation*}
where the domain of integration extends over the past light cone of the field point $x=(ct,\textbf{x})$. 
In order to work out a solution to this particular integral we need to introduce the near and wave zones \cite{PoissonWill, WillWiseman,PatiWill1,PatiWill2,Blanchet1,Maggiore1} domains defined by $r\ll \lambda_c$ and $r\gg \lambda_c$ respectively. Thus the near zone is the region of three dimensional space in which $r=|\textbf{x}|$ is small compared with the characteristic wavelength $\lambda_c$ of the gravitational radiation produced by the source, while the wave zone is the region in which $r$ is large compared with this length scale. We introduce the arbitrarily selected radius $\mathcal{R}\lesssim\lambda_c$ to define the near-zone domain $\mathcal{M}:|\textbf{x}|<\mathcal{R}$. While the gravitational potentials, originating from these two different integration domains, will individually depend on the cutoff radius their sum is guaranteed to be $\mathcal{R}$-independent and we will therefore systematically discard such terms in the remaining part of this article \cite{PoissonWill,WillWiseman,PatiWill1}. It can be shown \cite{PoissonWill, WillWiseman, PatiWill1, PatiWill2} that the formal near zone solution to the wave equation, for a far away wave zone field point ($|\textbf{x}|\gg\lambda_c$) can be is given by,
\begin{equation*}
h^{ab}_\mathcal{N}=\frac{4G}{c^4r}\sum_{l=0}^{+\infty}\frac{n_L}{l!}\Big(\frac{d}{du}\Big)^l\int_{\mathcal{M}} d\textbf{y} \ \tau^{ab}(u,\textbf{y}) \ y^L+\mathcal{O}(r^{-2}).
\end{equation*}
This result was derived by expanding the ratio $\frac{\tau^{\alpha\beta}(t-|\textbf{x}-\textbf{y}|/c,\textbf{y})}{|\textbf{x}-\textbf{y}|}=\frac{1}{r} \ \sum_{l=0}^\infty \frac{y^L}{l!}  \ n_L \ \Big(\frac{\partial}{\partial u}\Big)^l \ \tau^{\alpha\beta}(u,\textbf{y})+\mathcal{O}(1/r^2)$ in terms of the retarded time $u=c\tau$, the radial unit vectors $\textbf{n}=\frac{\textbf{x}}{r}$ and $y^Ln_L=y^{j1}\cdots y^{jl}n_{j1}\cdots n_{jl}$. The far away wave zone is characterized by the fact that only leading order terms, propotional to $1/r$ need to be retained. The post-Minkowskian theory is an approximation method that will not only reproduce the predictions of Newtonian theory but is a method that can be pushed systematically to higher and higher order to produce an increasingly accurate description of a weak gravitational field $||h^{\alpha\beta}||<1$. The gravitational potentials can be determined by using a formal expansion of the form $h^{\alpha\beta}=Gk_1^{\alpha\beta}+G^2k_2^{\alpha\beta}+G^3k_3^{\alpha\beta}+...$ \cite{PoissonWill,WillWiseman, PatiWill1,PatiWill2, Blanchet1}. As the spacetime deviates only moderately from Minkowski spacetime, we can assemble its metric from the gravitational potentials,
\begin{equation*}
g_{\alpha\beta}=\eta_{\alpha\beta}+h_{\alpha\beta}-\frac{1}{2}h\eta_{\alpha\beta}+h_{\alpha\mu}h^{\mu}_\beta-\frac{1}{2}hh_{\alpha\beta}+\Big(\frac{1}{8}h^2-\frac{1}{4}h^{\mu\nu}h_{\mu\nu}\Big)\eta_{\alpha\beta}+\mathcal{O}(G^3),
\end{equation*}
where the indices on $h^{\alpha\beta}$ are lowered with the Minkowski metric $h_{\alpha\beta}=\eta_{\alpha\mu}\eta_{\beta\nu}h^{\mu\nu}$, the metric determinant is given by $(-g)=1-h+\frac{1}{2}h^2-\frac{1}{2}h^{\mu\nu}h_{\mu\nu}+\mathcal{O}(G^3)$ and $h=\eta_{\mu\nu}h^{\mu\nu}$. In what follows we will assume that the matter distribution of the source is deeply situated within the near zone $r_c\ll \lambda_c$, where we remind that $r_c$ is the characteristic length scale of the source. It is straightforward to observe that this equation is tantamount to a slow motion condition $v_c\ll c$ for the matter source term. The post-Newtonian theory (PN) is an approximation method to GR that incorporates the weak-field and slow-motion conditions. The dimensionless expansion parameter in this approximation procedure is $(Gm_c)/(c^2r_c)=v^2_c/c^2$, where $m_c$ is the characteristic mass of the system under consideration. The material source term is modelled by a collection of $n$-fluid balls with negligible pressure, $T^{\alpha \beta}=\rho\ u^\alpha u^\beta$, where $\rho\big(m_A,\textbf{r}_A(t)\big)=\frac{\rho^*}{\sqrt{-g}\gamma_A}$ is the energy-density and $u^\alpha=\gamma_A (c,\textbf{v}_A)$ is the relativistic four-velocity of the fluid ball with mass $m_A$ and individual trajectory $r_A(t)$. Taking into account that for point masses we have $\rho^*=\sum_{A=1}^Nm_A\ \delta\big(\textbf{x}-\textbf{r}_A(t)\big)$, $\frac{1}{\sqrt{-g}}=1-\frac{1}{2}h^{00}+\mathcal{O}(c^{-4})$ and $\gamma_A^{-1}=\big[-g_{\mu\nu}\frac{v^\mu_Av^\nu_A}{c^2}\big]^{1/2}=1-\frac{1}{2}\frac{\textbf{v}_A^2}{c^2}-\frac{1}{4}h^{00}+\mathcal{O}(c^{-4})$, we obtain for the time-time component of  the 1.5 post-Newtonian matter energy-momentum pseudotensor, 
\begin{equation*}
c^{-2}(-g)T^{00}=\sum_A m_A\ \Big[1+\frac{1}{c^2}(\frac{\textbf{v}^2}{2}+3U)\Big]\ \delta(\textbf{x}-\textbf{r}_A)+\mathcal{O}(c^{-4}).
\end{equation*}
$U$ is the Newtonian potential of a $n$-body system with point masses $m_A$ and $h^{00}=\frac{4}{c^2}U+\mathcal{O}(c^{-4})$ is the corresponding gravitational potential at the 1.5 post-Newtonian order of accuracy. Another important quantity, that will be frequently used in the last chapter of this article, is the time-time component of the Landau-Lifshitz tensor, 
\begin{equation*}
c^{-2}(-g)t^{00}_{LL}\,=\,-\frac{7}{8\pi G c^2}\ \big[\partial_pU\partial^pU\big]+\mathcal{O}(c^{-4}).
\end{equation*}
worked out to the same degree of accuracy \cite{PoissonWill,WillWiseman,PatiWill1,PatiWill2,Virbhadra1} as the matter contribution. Further details on the derivation of this important quantity can be withdrawn from the second appendix-section of \cite{Alain1}. Using some of the previously outlined developments we see that the harmonic gauge contribution is beyond the 1.5 post-Newtonian order of accuracy $c^{-2}\tau_H^{00}=\mathcal{O}(c^{-4})$. The slow-motion condition gives rise to a hierarchy between the components of the energy-momentum tensor $T^{0b}/T^{00}\sim v_c/c$ and $T^{ab}/T^{00}\sim (v_c/c)^2$, where we used the approximate relations $T^{00}\approx \rho\ c^2$, $T^{0b}\approx\rho\ v^b c$, $T^{ab}\approx \rho\ v^a v^b$ and $\textbf{v}$ is the velocity vector of the fluid balls. A glance at the relaxed Einstein equations reveals that this hierarchy is inherited by the gravitational potentials $h^{0b}/h^{00}\sim v_c/c$, $h^{ab}/h^{00}\sim (v_c/c)^2$. Taking into account the factor $c^{-4}$ in the field equations, we have for the potentials $h^{00}=\mathcal{O}(c^{-2})$, $h^{0b}=\mathcal{O}(c^{-3})$ and $h^{ab}=\mathcal{O}(c^{-4})$. It should be mentioned that this particular notation should only be considered as a powerful mnemonic to judge the importance of various terms inside a post-Newtonian expansion. The real dimensionless post-Newtonian expansion parameter however is rather $(Gm_c)/(c^2r_c)=v^2_c/c^2$. It should be mentioned that the approach which is used in this article to determine the gravitational potentials is usually called the Direct Integration of the Relaxed Einstein equations or DIRE approach for short. An alternative method, based on a formal multipolar expansion of the potential outside the source was presented in \cite{Blanchet1,Blanchet5, BlanchetDamour1}. To conclude we would like to point out that further developments, about the concepts outlined in this chapter, can be withdrawn from a large number of outstanding articles \cite{BlanchetDamourIyerWillWiseman,DamourJaranowskiSchaefer1,BlanchetDamourIyer,BlanchetDamourEsposito-FareseIyer1,BlanchetDamourEsposito-FareseIyer2,DamourJaranowskiSchaefer2}.

\section{The effective wave equation:}
In this section we will work out the effective wave equation and present a solution for a far away wave zone field point. We saw in \cite{Alain1} that the effective  relaxed Einstein equations originate from the quest of rewriting the wave equation, containing the nonlocally modified energy-momentum tensor $\mathcal{T}^{\alpha\beta}= \frac{G_\Lambda(\Box)}{G} \ T^{\alpha\beta}$, in such a way that it can be solved most readily. This can be achieved by spreading out some of the differential complexity inside the nonlocally modified energy-momentum tensor to both sides of the differential equation. However before we can come to the actual derivation of the modified wave equation we first need to carefully prepare the grounds by setting in place a couple of important preliminary results.
\subsection{The nonlocally modified energy-momentum tensor:}
The major difference between the nonlocally modified theory of gravity outlined in this article and the standard theory of gravity lies in the way in which the energy (matter or field energy) couples to the gravitational field. In the purely Einsteinian theory the time-dependent energy distribution in space is translated via the constant coupling $G$ into spacetime curvature. In our particular model however the coupling-strength itself varies according to the characteristic wavelength $\lambda_c$ of the source term under consideration. From a strictly formal point of view however, the modified field equations can be formulated in a very similar way to Einstein's field equations,
\begin{equation*}
G^{\alpha\beta}\,=\, \frac{8\pi}{ c^4}\ G \ \mathcal{T}^{\alpha\beta}.
\end{equation*}
$G^{\alpha\beta}$ is the usual Einstein tensor and $\mathcal{T}^{\alpha\beta}$ is the modified energy-momentum tensor outlined in the introduction of this chapter. We see that this formulation is possible only because the nonlocal modification can be put entirely into the source term $\mathcal{T}^{\alpha\beta}$, leaving in this way the geometry ($G^{\alpha\beta}$) unaffected. In this regard we can easily see that, by virtue of the contracted Bianchi identities $\nabla_\beta G^{\alpha\beta}=0$, the effective energy-momentum tensor is conserved $\nabla_\beta \mathcal{T}^{\alpha\beta}=0$. This allows us to use the Landau-Lifshitz formalism introduced previously by simply replacing the energy-momentum tensor $T^{\alpha\beta}$, inside the relaxed Einstein field equations, through its nonlocal counterpart, 
\begin{equation*}
\Box h^{\alpha\beta}\,=\,-\frac{16\pi G}{c^4} \ (-g) \ \Big[\mathcal{T}^{\alpha\beta}+t^{\alpha\beta}_{LL}+t^{\alpha\beta}_H\Big].
\end{equation*}
Instead of trying to integrate out by brute force the nonlocally modified relaxed Einstein field equations, we rather intend to bring part of the differential complexity, stored inside the effective energy momentum tensor, to the left-hand-side of the field equation. These efforts will finally bring us to an equation that will be more convenient to solve. Loosely speaking we aim to separate inside the nonlocal covariant differential coupling the flat spacetime contribution from the the curved one. In this way we can rephrase the relaxed Einstein equations in a form that we will eventually call the effective relaxed Einstein equations. This new equation will have the advantage that the nonlocal complexity will be distributed to both sides of the equation and hence it will be easier to work out its solutions to the desired post-Newtonian order of accuracy. In this context we aim to rewrite the covariant d'Alembert operator $\Box_g$ in terms of a flat spacetime contribution $\Box$ plus an additional piece $w$ depending on the gravitational potentials $h^{\alpha\beta}$. The starting point for the splitting of the differential operator $\Box_g=\nabla_\alpha\nabla^\alpha$ is the well known relation \cite{PoissonWill,Woodard1, Weinberg1, Maggiore2},
\begin{equation*}
\Box_g\,=\,\frac{1}{\sqrt{-g}}\partial_\mu\big(\sqrt{-g}g^{\mu\nu}\partial_\nu)\,=\,\Box+w(h,\partial),
\end{equation*}
where $\Box=\partial^\alpha\partial_\alpha$ is the flat spacetime d'Alembert operator and the differential operator function $w(h,\partial)\,=\,-h^{\mu\nu}\partial_\mu\partial_\nu+\tilde{w}(h)\Box-\tilde{w}(h)h^{\mu\nu}\partial_{\mu}\partial_\nu+\mathcal{O}(G^4)$ is composed by the four-dimensional spacetime derivatives $\partial_\beta$ and the potential function $\tilde{w}(h)= \frac{h}{2}-\frac{h^2}{8}+\frac{h^{\rho\sigma}h_{\rho\sigma}}{4}+\mathcal{O}(G^3)$. We remind that the actual expansion parameter in a typical situation involving a characteristic mass $m_c$ confined to a region of characteristic size $r_c$ is the dimensionless quantity $Gm_c/(c^2 r_c)$. The result above was derived by employing the post-Minkoskian expansion of the metric $g_{\alpha\beta}$ in terms of the gravitational potentials \cite{PoissonWill, Will1, PatiWill1, PatiWill2,Blanchet1} outlined in the previous chapter. Further computational details can be found in the appendix relative to this chapter. With this result at hand we are ready to split the nonlocal gravitational coupling operator $G(\Box_g)$ into a flat spacetime contribution $G(\Box)$ multiplied by a piece $\mathcal{H}(\Box,w)$ that may contain correction terms originating from a possible curvature of spacetime,
\begin{equation*}
\mathcal{T}^{\alpha\beta}\,=\,G(\Box) \ \mathcal{H}(\Box,w) \ T^{\alpha\beta},
\end{equation*}
For astrophysical processes confined to a rather small volume of space $r_c\ll \sqrt{\Lambda}$ we can reduce the nonlocal coupling operator $G_{\Lambda}(\Box_g)$ to its ultraviolet component $G(\Box_g)=G\ \big[1-\sigma e^{\kappa\Box_g}\big]^{-1}$ only. Using the relation for the general covariant d'Alembert operator we can split the differential UV coupling into two separate contributions \cite{Efimov1, Efimov2,Namsrai1, Spallucci1}, 
\begin{equation*}
G(\Box)\,=\,\frac{1}{1-\sigma e^{\kappa\Box}},\quad \ \mathcal{H}(w,\Box)\,=\,1+\sigma \frac{e^{\kappa\Box}}{1-\sigma e^{\kappa\Box}} \sum_{n=1}^{+\infty} \frac{\kappa^n}{n!} w^n+...
\end{equation*}
The price to pay to obtain such a concise result is to assume that the modulus of the dimensionless parameter $\sigma$ has to be smaller than one ($|\sigma|<1$). We will see later in this article that this assumption will be well confirmed from a phenomenological point of view, when we work out the modified Newtonian potential or analyse the perihelion precession of Mercury. Here again the reader interested in the computational details is referred to the appendix of this chapter or to our previous work \cite{Alain1}. It will turn out that the splitting of the nonlocal coupling operator, into two independent pieces, will be of serious use when it comes to the integration of the relaxed Einstein equations. For later purposes we introduce the effective curvature energy-momentum tensor,
\begin{equation*}
\mathcal{B}^{\alpha\beta}\,=\,\mathcal{H}(\Box,w) \ T^{\alpha\beta}.
\end{equation*}
It is understood that a nonlocal theory involves infinitely many terms, however in the context of a post-Newtonian expansion, the newly introduced curvature energy-momentum tensor $\mathcal{B}^{\alpha\beta}$, can be truncated at a certain order of accuracy. In this sense the first four leading terms (appendix) of the effective curvature energy-momentum tensor are,

\begin{minipage}{0.6\textwidth}
\begin{equation*}
\begin{split}
\mathcal{B}^{\alpha\beta}_1\,&=\,\Big[\frac{\tau^{\alpha\beta}_m}{(-g)}\Big],\\
\mathcal{B}^{\alpha\beta}_2\,&=\,\epsilon e^{\kappa\Box}\Big[\frac{w}{1-\sigma e^{\kappa \Box}}\Big] \Big[\frac{\tau^{\alpha\beta}_m}{(-g)}\Big],
\end{split}
\end{equation*}
\end{minipage}
\hspace{-2.9cm}
\begin{minipage}{0.4\textwidth}
\begin{equation*}
\begin{split}
\mathcal{B}^{\alpha\beta}_3\,&=\,\epsilon\frac{\kappa}{2} e^{\kappa\Box}\Big[\frac{w^2}{1-\sigma e^{\kappa \Box}}\Big] \Big[\frac{\tau^{\alpha\beta}_m}{(-g)}\Big],\\
\mathcal{B}^{\alpha\beta}_4\,&=\,\epsilon\frac{\kappa^2}{3!} e^{\kappa\Box}\Big[\frac{w^3}{1-\sigma e^{\kappa \Box}}\Big] \Big[\frac{\tau^{\alpha\beta}_m}{(-g)}\Big].
\end{split}
\end{equation*}
\end{minipage}
$\newline$

For clarity reasons we introduced the parameter $\epsilon=\kappa \sigma$ of dimension length squared. To conclude this subchapter we would like to point out that the leading term in the curvature energy-momentum tensor can be reduced to the matter source term, $\mathcal{B}^{\alpha\beta}_1=T^{\alpha\beta}$.  
\subsection{The effective relaxed Einstein equations:}
The main purpose of this chapter is to work out the nonlocally modified wave equation which essentially originates from the quest of sharing out some of the differential complexity of the nonlocal coupling operator $G(\Box_g)$ to both sides of the relaxed Einstein equations. It was shown in the previous subsection that it is possible to split the nonlocal coupling operator, acting on the matter source term $T^{\alpha\beta}$, into a flat spacetime contribution $G(\Box)$ multiplied by a highly nonlinear differential term $\mathcal{H}(w,\partial)$. In order to remove some of the differential complexity from the effective energy-momentum tensor $\mathcal{T}^{\alpha\beta}=G(\Box) \ \mathcal{H}(w,\partial) \ T^{\alpha\beta}$ we will apply the flat spacetime inverse coupling operator $G^{-1}(\Box)$ to both sides of the relaxed Einstein field equations, $G^{-1}(\Box) \ \Box h^{\alpha\beta}\,=\,-\frac{16 \pi G}{c^4} \ G^{-1}(\Box) \big[(-g)\mathcal{T}^{\alpha\beta}+\tau_{LL}^{\alpha\beta}+\tau_H^{\alpha\beta}\big]$. We will see in this chapter that it is precisely this formal operation which will eventually lead to the effective wave equation,
\begin{equation*}
\Box_{c} \ h^{\alpha\beta}(x)\,=\, -\frac{16 \pi G}{c^4}N^{\alpha\beta}(x).
\end{equation*}
where $\Box_{c}$ is the effective d'Alembert operator $\Box_{c}=\big[1-\sigma e^{\kappa\Delta}\big] \ \Box$. $N^{\alpha\beta}$ is the effective energy-momentum pseudotensor, $N^{\alpha\beta}=G^{-1}(\Box) \big[(-g)\mathcal{T}^{\alpha\beta}+\tau_{LL}^{\alpha\beta}+\tilde{\tau}_H^{\alpha\beta}\big]$ \cite{Alain1}. $\tilde{\tau}^{\alpha\beta}_m=(-g)\mathcal{T}^{\alpha\beta}$ is the effective matter pseudotensor, $\tau_{LL}^{\alpha\beta}=(-g)t_{LL}^{\alpha\beta}$ is the Landau-Lifshitz pseudotensor and $\tilde{\tau}_H^{\alpha\beta}=(-g)t^{\alpha\beta}_H+G(\Box)\mathcal{O}^{\alpha\beta}(h)$ is the effective harmonic gauge pseudotensor where $\mathcal{O}^{\alpha\beta}(h)=-\sigma\sum_{n=1}^{+\infty}\frac{\kappa^n}{n!}\partial^{2n}_0 e^{\kappa\Delta} \Box h^{\alpha\beta}$ is the iterative post-Newtonian potential correction contribution. This term is added to the right-hand-side of the wave equation very much like the harmonic gauge contribution is added to the right-hand-side in the standard relaxed Einstein equation \cite{PoissonWill, Will1, PatiWill1, PatiWill2, Blanchet1}. It should be noticed that the modified d'Alembert operator $\Box_{c}$ is of the same post-Newtonian order than the standard d'Alembert operator, $\Box_c=\mathcal{O}(c^{-2})$ and reduces to the usual one in the limit of vanishing UV modification parameters $\lim_{\sigma,\kappa \rightarrow 0}\ \Box_{c}\,=\, \Box$. In the same limit the effective pseudotensor $N^{\alpha\beta}$ reduces to the purely general relativistic one, $\lim_{\sigma,\kappa \rightarrow 0}  \ N^{\alpha\beta}=\tau^{\alpha\beta}=\tau^{\alpha\beta}_m+\tau^{\alpha\beta}_{LL}+\tau^{\alpha\beta}_H$. At the level of the wave equations, these two properties can be summarized by the following relation,
\begin{equation*}
\Box_c \ h^{\alpha\beta}(x)\,=\, -\frac{16 \pi G}{c^4}N^{\alpha\beta}(x) \ \ \underset{\sigma,\kappa\rightarrow 0}{\Longrightarrow} \ \ \Box \ h^{\alpha\beta}(x)\,=\, -\frac{16 \pi G}{c^4}\tau^{\alpha\beta}(x).
\end{equation*}
In order to solve this equation we will use, in analogy to the standard wave equation, the following ansatz, $h^{\alpha\beta}(x)\,=\,-\frac{16 \pi G_N}{c^4} \int d^4y \ G(x-y) \ N^{\alpha\beta}(y)$ together with the identity for the effective Green function, $\Box_{c} G(x-y)\,=\,\delta(x-y)$, to solve for the potentials $h^{\alpha\beta}$ of the modified wave equation. Following the usual procedure \cite{PoissonWill,Maggiore1,Buonanno1} we obtain the nonlocally modified Green function in momentum space,
\begin{equation*}
\begin{split}
G(k)\,=\,\frac{1}{(k^0)^2-|\textbf{k}|^2} \ \frac{1}{1-\sigma e^{-\kappa \textbf{k}^2}}\,=\,\frac{\sum_{n=0}^{+\infty} \sigma^n \ e^{-n\kappa \textbf{k}^2}}{(k^0)^2-\textbf{k}^2},
\end{split}
\end{equation*} 
where we remind that by assumption we have $|\sigma|<1$. It should be noticed that the leading term ($n=0$) of this infinite sum of contributions will give rise to the usual Green function. Further computational details can be found in the appendix related to this chapter. These considerations finally permit us to work out an expression for the retarded Green function, $G_r(x-y)=G_r^{GR}+G_r^{NL}$, where $G_r^{GR}=\frac{-1}{4\pi}\frac{\delta(x^0-|\textbf{x}-\textbf{y}|-y^0)}{|\textbf{x}-\textbf{y}|}$ is the well known retarded Green function and
$G_r^{NL}=\frac{-1}{4\pi}\frac{1}{|\textbf{x}-\textbf{y}|}\sum_{n=1}^{+\infty}\frac{\sigma^n}{2\sqrt{\pi n\kappa}}e^{-\frac{(x^0-|\textbf{x}-\textbf{y}|-y^0)^2}{4\kappa}}$ is the nonlocal correction term. In this way we are able to recover in the limit of vanishing modification parameters the usual retarded Green function, $\lim_{\sigma,\kappa \rightarrow 0} \ G_r(x-y)=G_r^{GR}$. In addition it should be pointed out that we have, by virtue of the exponential representation of the dirac distribution, $\lim_{\kappa\rightarrow 0} \frac{1}{2\sqrt{\pi n\kappa}}e^{-\frac{(x^0-|\textbf{x}-\textbf{y}|-y^0)}{4n\kappa}}\,=\, \delta(x^0-|\textbf{x}-\textbf{y}|-y^0)$. In analogy to the purely general relativistic case, we can write down the formal solution to the modfied wave equation,
\begin{equation*}
h^{\alpha\beta}(x)\,=\,  \frac{4 \ G}{c^4}  \int d\textbf{y} \ \frac{N^{\alpha\beta}(x^0-|\textbf{x}-\textbf{y}|,\textbf{y})}{|\textbf{x}-\textbf{y}|}.
\end{equation*}
The retarded effective pseudotensor can be decomposed into two independent pieces according to the two contributions coming from the retarded Green function, $N^{\alpha\beta}(x^0-|\textbf{x}-\textbf{y}|,\textbf{y})=\mathcal{D} N^{\alpha\beta}(y^0,\textbf{y})+\sum_{n=1}^{+\infty}\sigma^n \mathcal{E}_n N^{\alpha\beta}(y^0,\textbf{y})$, where for later convenience we introduced the following two retardation integral operators, $\mathcal{D}= \int dy^0 \ \delta(x^0-|\textbf{x}-\textbf{y}|-y^0)$ and $\mathcal{E}_n=\int dy^0 \ \frac{1}{2\sqrt{\pi n \kappa}} e^{-\frac{(x^0-|\textbf{x}-\textbf{y}|-y^0)^2}{4n\kappa}}$. We will see in chapter five in how far the effective Green function will modify the Newtonian potential, a quantity which is frequently used in post-Newtonian developments.
\subsection{A particular solution:}
In this subsection we will derive a general solution for the gravitational potentials for a far away wave zone field point ($|\mathbf{x}|\gg \lambda_c$). Moreover we will consider in this article only the near zone energy-momentum contribution to the gravitational potentials $h^{ab}_\mathcal{N}$. In order to determine the precise form of the spatial components of the near zone potentials we need to expand the ratio inside the formal solution \cite{PoissonWill, Will1, PatiWill1} in terms of a power series,
\begin{equation*}
\begin{split}
 \frac{N^{ab}(x^0-|\textbf{x}-\textbf{y}|,\textbf{y})}{|\textbf{x}-\textbf{y}|}\,
=\,\frac{1}{r} \ \sum_{l=0}^\infty \frac{y^L}{l!}  \ n_L \ \Big(\frac{\partial}{\partial u}\Big)^l \ N^{ab}(u,\textbf{y})+\mathcal{O}(r^{-2}),
\end{split}
\end{equation*}
where $u=c\tau$ and $\tau=t-r/c$ is the retarded time. The distance from the matter source term's center of mass to the far away field point is given by $r=|\textbf{x}|$ and its derivative with respect to spatial coordinates is $\frac{\partial r}{\partial x^a}=n^a$ where $n^a=\frac{x^a}{r}$ is the $a$-th component of the unit radial vector. We remind that the far away wave zone is characterized by the fact that only leading order terms $1/r$ need to be retained and $y^Ln_L=y^{j1}\cdots y^{jl}n_{j1}\cdots n_{jl}$. Additional technical details can be found in the appendix relative to this subsection. By introducing the far away wave zone expansion of the effective energy-momentum distance ratio into the formal solution for the potentials we finally obtain the near zone contribution to the gravitational potentials for a far away wave zone field point in terms of the retarded derivatives,
\begin{equation*}
\begin{split}
h^{ab}_{\mathcal{N}}(x)\,&=\, \frac{4 G}{c^4 r} \sum_{l=0}^\infty \frac{n_L}{l!} \Big(\frac{\partial}{\partial u}\Big)^l \Big[ \int_{\mathcal{M}} d\textbf{y} \ N^{ab}(u,\textbf{y}) \ y^L\Big]+\mathcal{O}(r^{-2}),
\end{split}
\end{equation*}
We recall that $\mathcal{M}$ is the three-dimensional near zone integration domain (sphere) defined by $|\textbf{x}| <\mathcal{R}\leq \lambda_c$. Further computational details can be withdrawn from the appendix related to this section as well as from \cite{Alain1}. The effective energy momentum pseudotensor, $\partial_\beta N^{\alpha\beta}=0$ is conserved because we can store the complete differential operator complexity inside the modified energy-momentum tensor $\mathcal{T}^{\alpha\beta}=G(\Box_g)T^{\alpha\beta}$. We saw in the second chapter that no matter what the precise form of the energy-momentum tensor is, we have the following conservation relation, $\partial_\beta \big[(-g)(\mathcal{T}^{\alpha\beta}+t_{LL}^{\alpha\beta})\big]=0$. It should be noticed that similarly to the harmonic gauge contribution $\partial_\beta t_H^{\alpha\beta}=0$ the iterative potential contribution is separately conserved $\partial_\beta \mathcal{O}^{\alpha\beta}(h)=0$ because of the harmonic gauge condition. As the linear differential operator with constant coefficients $G^{-1}(\Box)$, commutes with the partial derivative operator ($[G^{-1}(\Box),\partial_\beta]=0$), we can immediately conclude for the conservation of the effective energy-momentum pseudotensor, $\partial_\beta N^{\alpha\beta}=G^{-1}(\Box) \ \partial_\beta \big[(-g)\mathcal{T}^{\alpha\beta}+\tau^{\alpha\beta}_{LL}+\tilde{\tau}^{\alpha\beta}_H\big]=0$. It was shown in \cite{Alain1} that the solution for the near zone gravitational potentials, outlined above, can be rephrased in terms of the nonlocally modified radiative multipole moments,
\begin{equation*}
h^{ab}_{\mathcal{N}}\,=\, \frac{2G}{c^4 r} \frac{\partial^2}{\partial \tau^2} \Big[Q^{ab}+Q^{abc} \ n_c+Q^{abcd} \ n_c n_d+\frac{1}{3}Q^{abcde} \ n_c n_d n_e+[l\geq 4]\Big]+\frac{2 G}{c^4 r}\left[P^{ab}+P^{abc} n_c\right]+\mathcal{O}(r^{-2}).
\end{equation*}
It should be noticed that in analogy to the purely general relativistic case \cite{PoissonWill, Maggiore1, Will1, PatiWill1, PatiWill2, Buonanno1, Blanchet1} the leading order term is proportional to the second derivative in $\tau$ of the radiative quadrupole moment. 
This result was derived by making use of the conservation of the effective energy-momentum pseudotensor \cite{Alain1}. The precise form of the effective radiative multipole moments is given in the appendix-section related to this chapter. It can be shown that the surface terms $P^{ab}$ and $P^{abc}$, outlined in \cite{Alain1}, will give rise to $\mathcal{R}$-dependent contributions only. These terms will eventually cancel out with contributions coming from the wave zone as was shown in \cite{PatiWill1,PatiWill2}. 

\section{A brief review of the effective energy-momentum pseudotensor:}
In the previous chapter we transformed the original wave equation, in which all the nonlocal complexity was contained inside the nonlocally modified energy-momentum tensor $\mathcal{T}^{\alpha\beta}$, into an effective wave equation that is much easier to solve. This effort gave rise to a new pseudotensorial quantity, the effective energy-momentum pseudotensor $N^{\alpha\beta}$. This chapter is devoted to the analysis of this important quantity by reviewing the matter, field and harmonic gauge contributions one after the other. We will study these three terms $N^{\alpha\beta}_m$, $N^{\alpha\beta}_{LL}$, $N^{\alpha\beta}_H$ separately and extract all the relevant contributions that lie within the 1.5 post-Newtonian order of accuracy.
\subsection{The effective matter pseudotensor:}
From the previous chapter we recall the precise expression for the matter contribution to the effective energy-momentum pseudotensor,
\begin{equation*} 
N_{m}^{\alpha\beta}\,=\,G^{-1}(\Box) \big[(-g) \ \mathcal{T}^{\alpha\beta}\big]\,=\,[G(\Box)]^{-1} \big[(-g) \ G(\Box) \ \mathcal{B}^{\alpha\beta}\big].
\end{equation*}
In order to extract from this expression all the relevant pieces that lie within the order of accuracy that we aim to work at in this article, we essentially need to address two different tasks. In a first step we have to review the leading terms of $\mathcal{B}^{\alpha\beta}$ and see in how far they may contribute to the 1.5 post-Newtonian order of accuracy. In a second step we have to analyze how the differential operator $G^{-1}(\Box)$ acts on the product of the metric determinant $(-g)$ multiplied by the nonlocally modified energy-momentum tensor $\mathcal{T}^{\alpha\beta}=G(\Box) \ \mathcal{B}^{\alpha\beta}$. Although this formal operation will lead to additional terms, the annihilation of the operator $G(\Box)$ with is inverse counterpart will substantially simplify the differential structure of the original energy-momentum tensor $\mathcal{T}^{\alpha\beta}$. We need to introduce a couple of preliminary results before we come to the two duties mentioned earlier. From a technical point of view we need to introduce the operators of instantaneous potentials \cite{Blanchet1, Blanchet3, Blanchet4},
$ \Box^{-1}[ \bar{\tau}]=\sum_{k=0}^{+\infty} \Big(\frac{\partial}{c\partial t}\Big)^{2k} \ \Delta^{-k-1}[\bar{\tau}]$. This operator is instantaneous in the sense that it does not involve any integration over time. However one should be aware that unlike the inverse retarded d'Alembert operator, this instantaneous operator will be defined only when acting on a post-Newtonian series $\bar{\tau}$. Another important computational tool which we borrow from \cite{Blanchet1, Blanchet3, Blanchet4} are the generalized iterated Poisson integrals, $\Delta^{-k-1}[\bar{\tau}_m](\textbf{x},t)=-\frac{1}{4\pi} \int d\textbf{y} \ \frac{|\textbf{x}-\textbf{y}|^{2k-1}}{(2k)!} \ \bar{\tau}_m(\textbf{y},t)$, where $\bar{\tau}_m$ is the $m$-th post-Newtonian coefficient of the energy-momentum source term $\bar{\tau}=\sum_{m=-2}^{+\infty} \bar{\tau}_m/c^{m}$. An additional important result that needs to be mentioned is the generalized regularization prescription\footnote{The author would like to thank Professor E. Poisson (University of Guelph) for useful comments regarding this particular issue.}, $\big[\nabla^m \frac{1}{|\textbf{x}-\textbf{r}_A|}\big] \ \big[\nabla^n \delta(\textbf{x}-\textbf{r}_A)\big]\equiv 0, \ \forall n,m\in \mathbb{N}$. The need for this kind of regularization prescription merely comes from the fact that inside a post-Newtonian expansion the nonlocality of the modified Einstein equations will lead to additional derivatives which will act on the Newtonian potentials. It is easy to see that in the limit $m=0$ and $n=0$ we recover the well known regularization prescription \cite{PoissonWill, Blanchet1, Blanchet2}. We are now ready to come to the first of the two tasks mentioned in the beginning of this subsection. In order to extract from $\mathcal{B}^{\alpha\beta}= \mathcal{H}(w,\Box)\ \big[\tau_m^{\alpha\beta}/(-g)\big]$ the contributions that lie within the 1.5 post-Newtonian order of accuracy, we need first to have a closer look at the differential curvature operator $\mathcal{H}(w,\Box)$. From the previous chapter we know that it is essentially composed by the potential operator function $w(h,\partial)$ and the flat spacetime d'Alembert operator,
\begin{equation*}
w(h,\partial)\,=\,-h^{\mu\nu} \partial_{\mu\nu}+\tilde{w}(h)\Box-\tilde{w}(h) h^{\mu\nu}\partial_{\mu\nu}\,=\,-\frac{h^{00}}{2}\Delta+\mathcal{O}(c^{-4}).
\end{equation*}
We see that to the desired order of precision, the potential operator function $w(h,\partial)$ reduces to one single contribution, composed by the potential $h^{00}=\mathcal{O}(c^{-2})$ \cite{PoissonWill, WillWiseman, PatiWill1} and the flat spacetime Laplace operator $\Delta$. Additional computational details can be found in  \cite{Alain1} as well as in the appendix-section related to this chapter. With this in mind we can have a closer look at the leading two contributions of the curvature energy-momentum tensor $\mathcal{B}^{\alpha\beta}$,
\begin{equation*}
\begin{split}
\mathcal{B}^{\alpha\beta}_{1}\,&=\,\tau^{\alpha\beta}_{m}(c^{-3})-\tau^{\alpha\beta}_m(c^0) \ h^{00}+\mathcal{O}(c^{-4}),\\
B^{\alpha\beta}_{2}\,&=\,-\frac{\epsilon}{2}\sum_A m_A v^\alpha_A v^\beta_A \ \Big[\sum_{n=0}^\infty \sigma^n e^{(n+1)\kappa \Delta} \Big] \ \Big[h^{00}\Delta \delta(\textbf{y}-\textbf{r}_A)\Big]+\mathcal{O}(c^{-4}),
\end{split}
\end{equation*}
The terms $\mathcal{B}^{\alpha\beta}_3$ and $\mathcal{B}^{\alpha\beta}_4$ are beyond the order of accuracy at which we aim to work at in this article because they are proportional to $\omega^2=\mathcal{O}(c^{-4})$ and $\omega^3=\mathcal{O}(c^{-6})$ respectively \cite{Alain1}. Moreover it should be mentioned that $\tau_m(c^0)$ is the matter pseudotensor at the leading order of accuracy. We will see later in this chapter that  $\mathcal{B}^{\alpha\beta}_1$ will generate the usual 1.5 post-Newtonian matter source term as the second piece of the latter will precisely cancel out with another contribution. This allows us to come to the second task, namely to look at the differential operation, $\big[1-\sigma e^{\kappa\Box}\big] \ \big[(-g)\mathcal{T}^{\alpha\beta}\big]$, mentioned in the introduction of this section. A very detailed derivation for this was outlined in \cite{Alain1} and we content ourselves here to present the main results,
\begin{equation*}
N_{m}^{\alpha\beta}\,=\,\mathcal{B}^{\alpha\beta}+\mathcal{B}^{\alpha\beta}h^{00}-\sigma D^{\alpha\beta}+\mathcal{O}(c^{-4}),
\end{equation*}
where we remind that $\mathcal{T}^{\alpha\beta}=G(\Box) \mathcal{B}^{\alpha\beta}$ and $\mathcal{B}^{\alpha\beta}=\mathcal{B}^{\alpha\beta}_{1}+\mathcal{B}^{\alpha\beta}_{2}+\mathcal{O}(c^{-4})$. It is understood that there are numerous additional terms which we do not list here because they are beyond the degree of precision of this article. The two leading contributions of $N^{\alpha\beta}_m$ give rise to the usual 1.5 post-Newtonian contribution \cite{PoissonWill,WillWiseman,PatiWill1,PatiWill2},
\begin{equation*}
\mathcal{B}^{\alpha\beta}_{1}(c^{-3})+\mathcal{B}_{1}^{\alpha\beta}(c^{-1})h^{00}\,=\,\sum_A m_A v_A^\alpha v^\beta_A \Big[1+\frac{\textbf{v}_A^2}{2c^2}+\frac{3V}{c^2}\Big] \ \delta(\textbf{x}-\textbf{r}_A)+\mathcal{O}(c^{-4}),
\end{equation*}
where $V=\frac{U}{1-\sigma}$ is the effective Newtonian potential. The additional tensor contribution \cite{Alain1} at the 1.5 post-Newtonian order of accuracy is,
\begin{equation*}
 D^{\alpha\beta}\,=\,\sum_A m_A v_A^\alpha v^\beta_A \ \mathcal{S}(\sigma,\kappa) \ \big[ \nabla^{2p+2n-m} \delta(\textbf{x}-\textbf{r}_A)\big]\big[ \nabla^m h^{00}\big] +\mathcal{O}(c^{-4}),
\end{equation*}
where for clarity reasons we introduced $\mathcal{S}(\sigma,\kappa)\,=\,\sum_{n=1}^{\infty} \frac{\kappa^n}{n!} \sum_{m=1}^{2n} \binom{2n}{m} \ \sum_{s=0}^{+\infty} \sigma^s \ \sum_{p=0}^{+\infty} \frac{(s\kappa)^p}{p!}$ to summarize the four sums inside $D^{\alpha\beta}$ (appendix). It should be noticed that the first two sums originate from the inverse differential operator $G^{-1}(\Box)$ while the last two sums originate from the extraction of the 1.5 post-Newtonian contribution of the modified energy-momentum tensor $\mathcal{T}^{\alpha\beta}=G(\Box) \mathcal{B}^{\alpha\beta}$ and $\binom{2n}{m}=\frac{(2n)!}{(2n-m)!m!}$ is the binomial coefficient. 
Although we cannot review the derivation in such a detailed way as we did in \cite{Alain1}, we will however, for reasons of completeness, present the most important intermediate computational results in the appendix-section related to this chapter.

\subsection{The effective harmonic gauge and Landau-Lifshitz pseudotensors:}
We will briefly review the time-time-component of the effective Landau-Lifshitz and harmonic gauge pseudotensors outlined in a more detailed way in our previous work \cite{Alain1},
\begin{gather*}
N_H^{\alpha\beta}\,=\,G^{-1}(\Box) \ \tilde{\tau}^{\alpha\beta}_H\,=\,G^{-1}(\Box) \ \tau^{\alpha\beta}_H+\mathcal{O}^{\alpha\beta},\\
N^{00}_{LL}\,=\,G^{-1}(\Box) \ \tau^{00}_{LL}\,=\, \Big[ \big(1-\sigma\big)\tau^{00}_{LL} -\epsilon \Delta \tau^{00}_{LL}-\sigma\sum_{m=2}\frac{\kappa^m}{m!}\Delta^m \tau^{00}_{LL}\Big]+\mathcal{O}(c^{-4}),
\end{gather*}
where $\tau_H=(-g)t_H^{\alpha\beta}$ is the standard harmonic gauge pseudotensor contribution, $\mathcal{O}^{\alpha\beta}(h)=-\sigma\sum_{n=1}^{+\infty}\frac{\kappa^n}{n!}\partial^{2n}_0 e^{\kappa\Delta} \Box h^{\alpha\beta}$ is the iterative potential contribution and $\tau^{00}_{LL}=\frac{-7}{8\pi G} \partial_jV\partial^jV+\mathcal{O}(c^{-2})$ \cite{PoissonWill, WillWiseman, PatiWill1}. We know from \cite{Alain1, PoissonWill, WillWiseman, PatiWill1} that $\frac{16\pi G}{c^4} N^{00}_H\,=\,\mathcal{O}(c^{-6})$ is beyond the 1.5 post-Newtonian order of accuracy. Moreover we recall that, because $\lim_{\kappa,\sigma\rightarrow 0} G^{-1}(\Box)=1$ and $\lim_{\sigma,\kappa\rightarrow 0}\mathcal{O}^{\alpha\beta}=0$, we recover the standard harmonic gauge contribution in the limit of vanishing UV parameters. Concerning the effective Lanadau-Lifshitz pseudotensor, we observe that from the leading term we will be able to eventually retrieve the standard post-Newtonian field contribution.

\section{The effective orbital dynamics of a two-body system:}
The purpose of this section is to outline in how far our nonlocally modifed theory of gravity affects the orbital dynamics of a binary-system in the context of Newtonian gravity as well as in the context of linearised general relativity. We begin by working out the effective Newtonian potential which we use in order to solve the famous Kepler problem. We continue with the relativistic Kepler problem and we derive from the perihelion precession of Mercury an upper bound for the dimensionless UV parameter $\sigma$. We conclude this chapter by computing the total amount of gravitational energy released by a binary-system moving along circular orbits. We observe that for all of the three different situations we recover, in the limit of a vanishing UV parameter $\sigma$, either the Newtonian or the linearised GR result.
\subsection{The effective Newtonian potential:}
It is known from the penultimate chapter that the effective retarded Green function is composed by the standard retarded Green function $G_r^{GR}$ together with a nonlocal correction term $G_r^{NL}$ which disappears in the limit of vanishing UV parameters. In addition we obtained a formal solution to the modified wave equation which can be decomposed into two different independent terms according to the two contributions coming from the retarded Green function. Moreover we saw in the previous chapter that the leading order contribution of the effective energy-momentum pseudotensor reduces to $N^{00}=\sum_A m_A  c^2 \ \delta(\textbf{x}-\textbf{r}_A)+\mathcal{O}(c^{-1})$. This allows us to derive the modified gravitational potential $h^{00}(x)=\frac{4}{c^2} V(x)$, where $V(x)=\sum_A  \ \frac{G\tilde{m}_A}{|\textbf{x}-\textbf{r}_A|}= \frac{U(\textbf{x})}{1-\sigma}$ is the effective Newtonian potential. $U(x)$ is the standard Newtonian potential term and $\tilde{m}_A= \frac{m_A}{1-\sigma}$ is the effective mass of the body $A$. Additional computational details regarding the precise derivation of this result can be found in the appendix-section related to this chapter. It should be mentioned that in \cite{Alain1} we reduced the modified retarded Green function to the two leading order terms only and therefore obtained a slightly different result for the effective mass. The modified Newtonian potential worked out in this article is however more accurate and we will therefore stick to this result in the remaining part of this article. It should be noticed that the usual Newtonian potential is recovered in the limit of vanishing $\sigma$. Experimental results \cite{Chiaverini1, Kapner1} from deviation measurements of the Newtonian law at small length scales ($\sim 75 \mu m$) suggest that the dimensionless correction constant needs to be of the order $\sigma \lesssim 10^{-3}$. We see that this experimental bound confirms our theoretical assumption made previously for $\sigma$ to be a small dimensionless parameter.

\subsection{The effective Kepler orbits:}
\begin{figure}[h]
\begin{center}
\includegraphics[width=8.7cm,height=5cm]{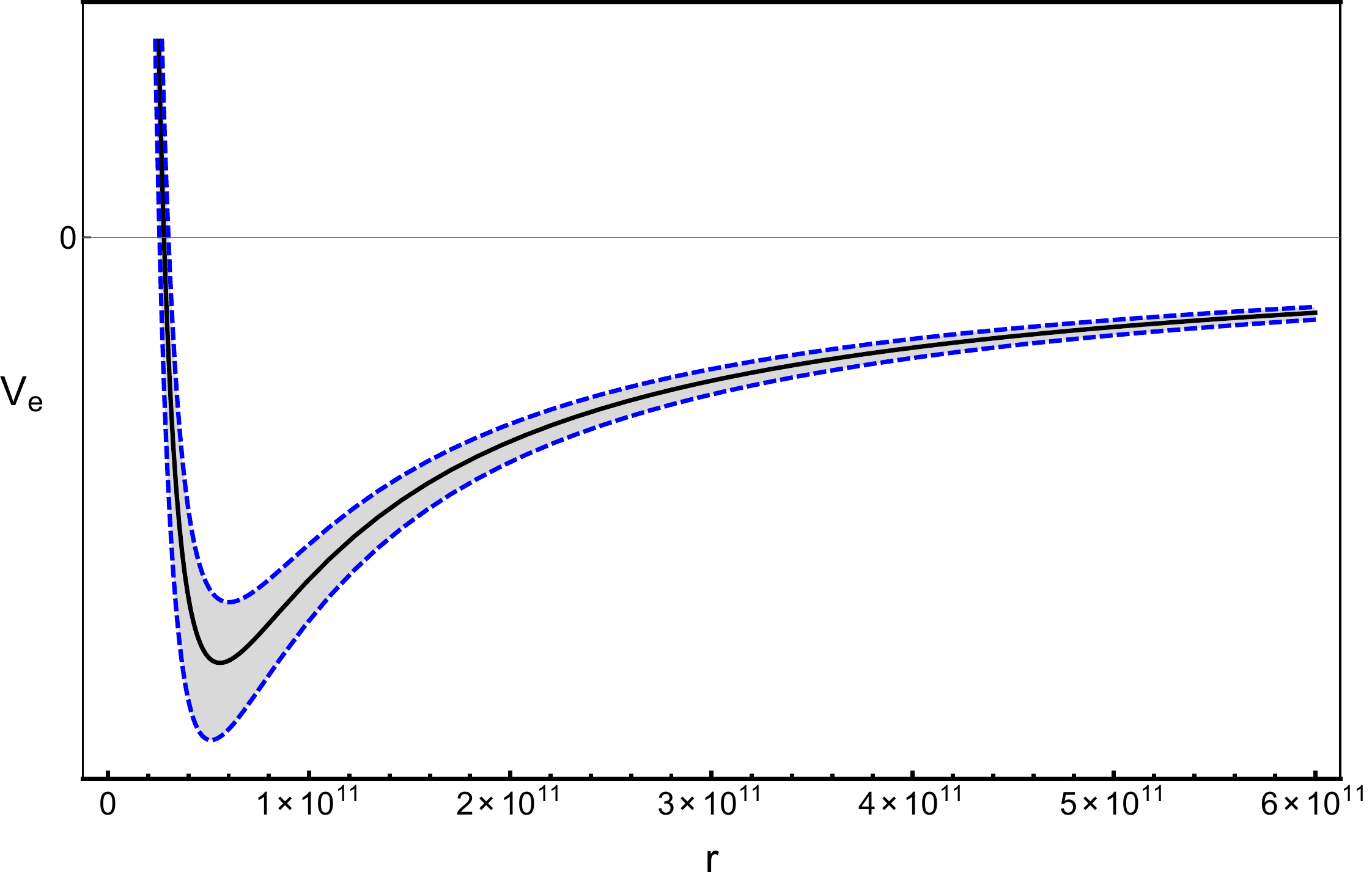}
\end{center}
\caption{\label{KeplerPotential}The effective potential $V_e(r,\sigma)$ is plotted against the relative separation $r$ of the two-body system (Sun-Mercury) for three different $\sigma$-values: $0$ (black curve), $-0.08$ (upper blue dashed curve) and $+0.08$ (lower blue dased curve).}
\end{figure}

The Kepler problem consists in determining the motion of two bodies subjected to their mutual gravitational attraction by assuming that each body by itself can be taken to be spherically symmetric. Although this is one of the simplest problems of celestial mechanics it is also one of the most relevant ones because it provides to a good first approximation the motion of any planet around the Sun ignoring the effects of other planets \cite{PoissonWill}. Moreover it is one of the few systems that can be solved exactly and completely in terms of simple functions. We will analyse in how far the effective Newtonian potential will affect the Kepler problem by deriving the effective spatial solution for a two-body system composed by the Sun and its closest planet Mercury. In a first time we will however keep the problem generic and work out the solution for an arbitrary two body-system with masses $m_1$ and $m_2$. Their respective equations of motion are given in terms of the polar coordinates by, $m_1 \ddot{\textbf{r}}_1=-G(\sigma)\frac{m_1 m_2}{r^2} \textbf{e}_r$ and $m_2 \ddot{\textbf{r}}_2=-m_1\ddot{\textbf{r}}_1$, where $\textbf{e}_r=\textbf{r}/r$ and $\textbf{e}_\phi=\dot{\textbf{e}}_r$ are time-dependent unit polar vectors. $G(\sigma)=\frac{G}{1-\sigma}$ is the effective leading order Newtonian coupling constant containing the dimensionless UV parameter $\sigma$. Using the relative separation vector between body one and body two, $\textbf{r}=\textbf{r}_1-\textbf{r}_2$, as well as the position of the Newtonian barycentre $m\textbf{R}=m_1 \textbf{r}_1+m_2\textbf{r}_2$, where $m=m_1+m_2$ is the total mass of the two-body system, we can deduce the position vectors in the center-of-mass frame ($\textbf{R}=\textbf{0}$) to be $\textbf{r}_1=+\frac{m_2}{m} \textbf{r}$ and $\textbf{r}_2=-\frac{m_1}{m}\textbf{r}$. It should be noticed that in contrary to the previous chapters $r=|\textbf{r}_1-\textbf{r}_2|$ is the distance between the two bodies with respective masses $m_1$ and $m_2$ and should not be confused with $|\textbf{x}|$ which is the distance between the source and the observer (detector). By combining these results it is straightforward to work out, in the context of an effective one-body description, the equation of motion of a fictitious body of reduced mass $\mu=_1m_2/m$ and its solution in terms of the polar angle and the dimensionless ultraviolet (UV) parameter,
\begin{equation*}
\mu \ddot{\textbf{r}}\,=\, -G(\sigma)\frac{m_1m_2}{r^2}\textbf{e}_r, \quad \quad \ r(\phi,\sigma)=\frac{p(\sigma)}{1+e\cos(\phi)}.
\end{equation*} 
It is no surprise to observe that the effective $\sigma$-depending solution has the same general shape as the standard Kepler solution and we remind that $e$ is the orbit's eccentricity, $\phi$ the polar angle, $p(\sigma)=\frac{h^2}{G(\sigma)\ m}$ is a quantity of dimension length commonly known as the orbit's {\it semi-latus rectum} \cite{PoissonWill} and $h=l/\mu$ is the binary-system's reduced angular momentum. Further computational details are provided in the appendix-section related to this chapter. The energy conservation relation of the binary-system can be obtained from the equation of motion, $\epsilon=\frac{\dot{r}^2}{2}+\frac{h^2}{2 r^2}-\frac{G(\sigma) \ m}{r}$, in which $\epsilon=E/\mu$ is a constant of motion also known as the reduced energy and $E$ is the total energy of the two-body system. It is instructive to rewrite the last equation in the form $\frac{\mu}{2}\dot{r}^2=E-V_e(r)$, in which we introduced the effective potential defined by,
\begin{equation*}
V_e(r,\sigma)\,=\,\mu \Big(\frac{h^2}{2 r^2}-\frac{G(\sigma)\ m}{r}\Big).
\end{equation*}
This particular form allows us to explore the qualitative features of Keplerian motion without having to perform additional calculations. The potential consists of an 
attractive (negative) gravitational well and a repulsive (positive) centrifugal barrier rising to infinity as $r$ approaches $0$. We outlined in Fig \ref{KeplerPotential} in the context of an effective one-body description \cite{PoissonWill} the effective potential of a fictitious body with reduced mass $\mu=M_{\odot}M_{\mercury}/(M_{\odot}+M_{\mercury})$ and reduced angular momentum $h\approx2.75\cdot 10^{15} \frac{m^2}{s}$. $M_{\odot}\approx 2\cdot 10^{30} \ kg$ is the solar mass and $M_{\mercury}\approx 3\cdot 10^{23}\ kg$ is the mass of the planet Mercury. A turning point occurs when the first temporal derivative of the relative separation between the two bodies vanishes ($\dot{r}=0$), that is when the effective potential equals the total energy of the two-body system ($E=V_e(r)$). At such points the radial velocity changes sign and the motion changes from incoming to outgoing or the other way around. If the fictitious body (effective one-body description) has a positive energy ($E>0$) then there is a single turning point at some innermost radius $r_{min}$ and the motion takes place for $r\geq r_{min}$. The particle starts at infinity with a negative radial velocity ($\dot{r}=-\sqrt{2 \epsilon}$) and a vanishing angular velocity $\dot{\phi}$. As $r$ decreases the angular velocity increases to obey conservation of angular momentum and $\dot{r}$ becomes increasingly negative until the body has reached the position of the minimum of the gravitational well. While the angular velocity continues to increase $\dot{r}$ becomes from now on decreasingly negative until it finally vanishes. This is when the body reaches its turning point at $r=r_{min}$ before $\dot{r}$ turns positive and the particle begins its way back to infinity. Such an orbit, known as a hyperbola, is not bound to the gravitating center as the total energy is dominated by (positive) kinetic energy instead of (negative) gravitational potential energy. The limiting case of an unbound orbit corresponds to parabolic motion where $E=0$. Here the body begins from rest at infinity, proceeds to a single turning point at $r=r_{min}$ and returns to a state of rest at infinity. For the case where $E<0$ the gravitational potential energy dominates over kinetic energy and Fig. \ref{KeplerPotential} reveals that there are now two turning points at $r=r_{min}$ and $r=r_{max}$. In this case the orbital motion is bound to the gravitating center and takes place between the innermost and outermost radii. This situation is known as elliptic motion and a special case occurs when $E$ is made equal to the minimum value of the effective potential. In this case the turning points merge to a single radius $r_0$ and motion proceeds on a circular orbit with fixed radius $r_0$. 
It should be reminded that the effective one-body description outlined in this subsection is in fact a fictitious representation of the relative orbit. However due to the position vectors of the two bodies expressed in terms of the relative separation vector $\textbf{r}$, their respective motion is merely a scaled version of the relative orbital motion and can thus be described in the same language. In the limit of small mass ratios it becomes increasingly true that $\textbf{r}_1\rightarrow \textbf{r}$ and $\textbf{r}_2\rightarrow \textbf{0}$ and in this particular case $m_1$ becomes a test mass in the field of $m_2\rightarrow m$. We observe that for $|\sigma|\neq0$ (blue dased curves) the general shape of the two-body dynamics remains essentially the same as for the standard non-modified Newtonian case (black curve). However for a gravitationally bound two-body system ($E<0$) the exact form of the effective potential $V_e(r,\sigma)$ is slightly altered according to the precise value of the UV parameter. For negative $\sigma$-values the gravitational well is less deep and for a given negative energy $E$ the two turning points appear to be closer one to the other. For positive $\sigma$-values the opposite is true and the orbit's semi-major axes therefore slightly increases. For clarity reasons we used in Fig. \ref{KeplerPotential} $\sigma$-values that are one or two orders of magnitude larger than those allowed by Newtonian-potential experiments used to constrain the gravitational constant value $G$ \cite{Chiaverini1, Kapner1}. Nevertheless for smaller and in this sense more realistic $\sigma$-parameters, the general behaviour of the modified orbits remains essentially  the same. In the next subsection we will have a closer look at the relativistic Kepler problem which will eventually bring us to the famous perihelion precession of Mercury.
\subsection{The perihelion precession of Mercury:}
In this section we discuss the nonlocally modified Einstein field equations for a spherically symmetric spacetime and we derive from the perihelion precession of Mercury an upper bound for the dimensionless UV parameter $\sigma$. We will see that the $\sigma$-value inferred from experimental data obtained in the context of a verification of Newton's law at small distances perfectly agrees with the result obtained t the end of this section. Spherical symmetry encourages the use of spherical polar coordinates $(r,\theta,\phi)$ in terms of which the metric of flat spacetime takes the form of $ds^2=-d(ct)^2+dr^2+r^2(d\theta^2+\sin^2\theta d\phi^2)$. Generalizing to curved spacetime, we assert that the metric of any spherically symmetric spacetime can always be written in the form, $ds^2=-e^{-2\Phi/c^2} d(ct)^2+e^{2\Lambda/c^2}dr^2+r^2(d\theta^2+\sin^2\theta d\phi^2)$, where $\Phi(t,r)$ and $\Lambda(t,r)$ are arbitrary functions of the coordinates $t$ and $r$ \cite{PoissonWill,Inverno}. We assume that we are dealing with a single isolated body, so that the spacetime becomes asymptotically flat in the limit $r\rightarrow \infty$. This leads to the boundary conditions for the functions which need to disappear $\lim_{r\rightarrow \infty} \Phi(t,r)=0$ $\lim_{r\rightarrow \infty} \Lambda(t,r)=0$ in order to allow the metric to reduce to the Minkowski metric for $r\rightarrow \infty$. In place of $\Lambda(t,r)$ it is helpful to employ instead a relativistic mass-energy function $m(t,r)$ defined by $e^{-2\Lambda/c^2}=1-\frac{2 [G(\Box) m(t,r)]}{c^2 r}:=f(t,r)$. It should be noticed that to this order of accuracy the generally covariant d'Alembert operator can be replaced by its flat spacetime counterpart. This particular substitution produces a substantial simplification to the field equations and we obtain, after a rather lengthy but essentially identical derivation compared to the purely general relativistic one, the following Einstein tensor components, $G^0_{\ 0}=-\frac{2}{c^2 r^2}\partial_r[G(\Box)m(t,r)]$, $G^0_{\ r}=-\frac{2}{c^3 r^2} e^{2\Phi/c^2} f^{-1} \partial_t[(G(\Box)m(t,r)]$ and $G^r_{\ r}=-\frac{2}{c^2 r} f\partial_r\Phi-\frac{2 [G(\Box) m(t,r)]}{c^2 r^3}$. In vacuum the effective energy-momentum tensor vanishes and we infer after, inserting the first two Einstein tensor components mentioned above into the effective Einstein field equations, that the relativistic mass-energy function is a constant, $m(r,t)=m$. This particular situation affects the coupling of the Newtonian operator to the mass-energy function (constant), $G(\Box)m=G\sum_{n=0}^{+\infty}\sigma^n m=G(\sigma) m$ ,where after making use of the geometric series ($|\sigma|<1$) we obtain the familiar result $G(\sigma)=\frac{G}{1-\sigma}$. With this assignment the equation for the function $\Phi$ integrates to $\partial_r\Phi=\big[1-\frac{2G(\sigma) m}{c^2 r}\big]^{-1}\frac{G(\sigma) m}{r^2}\Rightarrow\Phi=-\frac{c^2}{2}\text{ln}\big[1-\frac{2G(\sigma)m}{c^2 r}\big]+h(t)$ in which $h(t)$ is an arbitrary function of integration which eventually vanishes due to the boundary condition $\lim_{r\rightarrow\infty}\Phi(t,r)=0$ mentioned earlier \cite{PoissonWill,Inverno}. With this we arrive at $e^{-2\Phi/c^2}=1-\frac{2G(\sigma)m}{c^2 r}$ and the effective Schwarzschild metric becomes to this order of accuracy,
\begin{equation*}
ds^2\,=\,-\Big(1-\frac{R^\sigma_s}{r}\Big)\ d(ct)^2+\Big(1-\frac{R_s^\sigma}{r}\Big)^{-1} dr^2+r^2(d\theta^2+\sin^2\theta\ d\phi^2),
\end{equation*}
where $R^\sigma_s=\frac{2G(\sigma)m}{c^2}$ is the $\sigma$-dependent effective Schwarzschild radius. The effective Schwarzschild metric leads to, $c^2=-\big(1-\frac{R_s^\sigma}{r}\big) (\dot{x}^0)^2+\big(1-\frac{R_s^\sigma}{r}\big)^{-1} \dot{r}^2+r^2\dot{\theta}^2+r^2\sin^2\theta \dot{\phi}^2$, where $x^0=ct$ is the temporal length and an overdot indicates differentiation with respect to the proper time $\tau$. From the geodesic equation, $\frac{d^2 x^\mu}{d\tau^2}+\Gamma^\mu_{\alpha\beta}\frac{dx^\alpha}{d\tau}\frac{dx^\beta}{d\tau}=0$, we obtain in the context of the Schwarzschild-metric ($\Gamma^0_{00}=\Gamma^0_{11}=\Gamma^0_{22}=\Gamma^0_{33}=0$) the following equation, $\big(1-\frac{R_s^\sigma}{r}\big) \dot{x}^0\,=\,k$, where $k$ is a constant of integration. In addition we recall the angular momentum conservation relation, $r^2 \dot{\phi}\,=\,h$, where $h$ is the reduced angular momentum introduced previously. By inserting these two results, together with the previously obtained relation between the radial velocity and the reduced angular momentum $\frac{dr}{dt}\,=\,-h\frac{du}{d\phi}$ inside the equation outlined above ($\theta=\pi/2$) we get, $k^2 \ \big(1-R_s^\sigma u\big)^{-1}-\big(1-R_s^\sigma u\big)^{-1} h^2 \ \big(\frac{du}{d\phi}\big)^2-h^2 u^2=-c^2$. After differentiating this last relation with respect to $\phi$ and performing some simple algebra we obtain the following nonlinear differential equation for the variable $u=\frac{1}{r}$, $\frac{d^2 u}{d \phi^2}+u=\frac{c^2}{h^2} \frac{R_s^\sigma}{2}+\frac{3}{2} R_s^\sigma u^2$. This allows us to perform another substitution, $u=\frac{v}{\alpha}$, where $\alpha\gg R_s^\sigma$ can be chosen to be much larger than the effective Schwarzscild radius Multiplying the equation above by $\alpha$ which is of dimension length leaves us with, $\frac{d^2v}{d\phi^2}+v=\lambda+\epsilon\ v^2$, where $\lambda=\frac{c^2}{h^2}\frac{R^\sigma_s\alpha}{2}\geq 1$ is much larger than $\epsilon=\frac{3}{2}\frac{R^\sigma_s}{\alpha}\ll 1$ and $v$ is a dimensionless variable. For the Sun-Mercury binary system discussed previously we approximately have $R_s^\sigma\approx 3\cdot 10^3\ m$, $\lambda \approx 1$ and $\epsilon\approx 10^{-8}$ for $\alpha=10^8 R_s^\sigma$ and $h\approx2.75\cdot 10^{15}\frac{m^2}{s}$. We see that the leading term of this equation is the usual differential equation, which we already encountered in the previous subsection when we worked out a solution to the non-relativistic Kepler problem, followed by an additional factor containing a small dimensionless parameter $\epsilon$. In this regard we will expect that the solution will reduce, to leading order to the one of the classical Kepler problem. We will see that the precise expression for $\alpha$ will not matter so that we can choose the latter in a way such that the nonlinear differential equation outlined above can be solved using perturbation methods. In this sense we will choose the following ansatz, $v=v_0+\epsilon v_1+\mathcal{O}(\epsilon^2)$ and systematically skip terms of the order $\mathcal{O}(\epsilon^2)$ or smaller \cite{Inverno,JordanSmith}. By introducing this particular ansatz into the equation we obtain upon corrections of the order $\mathcal{O}(\epsilon^2)$ a set of two coupled differential equations, $v''_0+v_0=\lambda$ and $v''_1+v_1=v_0^2$, where we observe that, according to the previous subsection, the first equation gives rise to a solution of the form, $v_0=\lambda \ (1+e \cos\phi)$. Plugging the leading order solution into the second equation we obtain $v''_1+v_1=\lambda^2\big(1+\frac{e^2}{2}\big)+2\lambda^2 e \cos(\phi)+\frac{\lambda^2 e^2}{2} \cos(2\phi)$, where we made use of the following trigonometric identity $\cos^2(\phi)=\frac{1}{2}[1+\cos(2\phi)]$. We choose the solution to be of the from $v_1=A+B \phi \sin \phi+C \sin 2\phi$ and find by comparison the constants to be, $A=\lambda^2\big(1+\frac{e^2}{2}\big)$, $B\,=\,\lambda^2 e$, $C\,=\,-\frac{\lambda^2 e^2}{6}$. With this we obtain, $v=v_0+\epsilon \lambda^2 \Big(1+ \frac{e^2}{2}\Big)+\epsilon\lambda^2 e \phi \sin\phi-\epsilon\frac{\lambda^2 e^2}{6}\cos(2\phi)+\mathcal{O}(\epsilon^2)$ and we observe that the leading term as well as the third term are the dominant quantities of the solution. By skipping the other two terms we obtain in good approximation the general solution to be $v\approx\lambda \big[1+e\cos[\phi(1-\epsilon \lambda)]\big]+\mathcal{O}(\epsilon^2)$, where we used the Taylor expansion result, $1+e \cos\phi+\epsilon \lambda e \phi \sin\phi=1+e\cos[\phi(1-\epsilon \lambda)]+\mathcal{O}(\epsilon^2)$. The final solution ($r=\frac{\alpha}{v}$) for the relative separation of the two-body system as function of the angle $\phi$, which is of course independent from the arbitrarily chosen $\alpha$-value, 
\begin{equation*}
r(\phi,\sigma)\,\approx\frac{p(\sigma)}{1+e \cos[\phi(1-\Delta(\sigma))]},
\end{equation*}
where we remind that $p(\sigma)=\frac{h^2}{G(\sigma)\ m}$ is the orbit's {\it semi-latus rectum} and $\Delta(\sigma)=\lambda\epsilon=\frac{3}{4}\frac{c^2}{h^2}(R^\sigma_s)^2$ is a $\sigma$-dependent angular shift. We notice that the orbit remains approximately the one of an ellipse and the trajectory remains periodic with a period this time of $\frac{2\pi}{1-\Delta(\sigma)}\approx 2\pi [1+\Delta(\sigma)]$. In simple words the planet will essentially move on an elliptic orbit with an axis which will, in contrast to the nonrelativistic Kepler problem, be shifted between two points of closest approach by an angle of $\Delta\phi=2\pi \Delta(\sigma)$. For Mercury the shift, which is commonly known as perihelion precession or perihelion advance,is about $\Delta\phi=42.98\pm 0.04$ arcseconds per century. Using the relation between the reduced angular momentum, the orbit's semi-major axis $a$ with eccentricity $e$, $h(\sigma)=\sqrt{G(\sigma) (1-e^2) a}$, and the definition fro the effective Schwarzschild radius we see that $\Delta\phi(\sigma)=\frac{\Delta\phi}{1-\sigma}$, where the standard angular shift is given by $\Delta\phi=42.98$ \cite{PoissonWill,Inverno,Weinberg2}. This allows us to work out bounds for the dimensionless UV parameter $|\sigma|\leq 9.3\cdot 10^{-4}$. It should be noticed that this result agrees with the $\sigma$-value inferred from Newtonian potential experiments designed to measure the Newtonian coupling parameter \cite{Chiaverini1, Kapner1}.
\subsection{The energy released by a binary-system:}
\begin{figure}[h]
\begin{center}
\includegraphics[width=9cm,height=5.1cm]{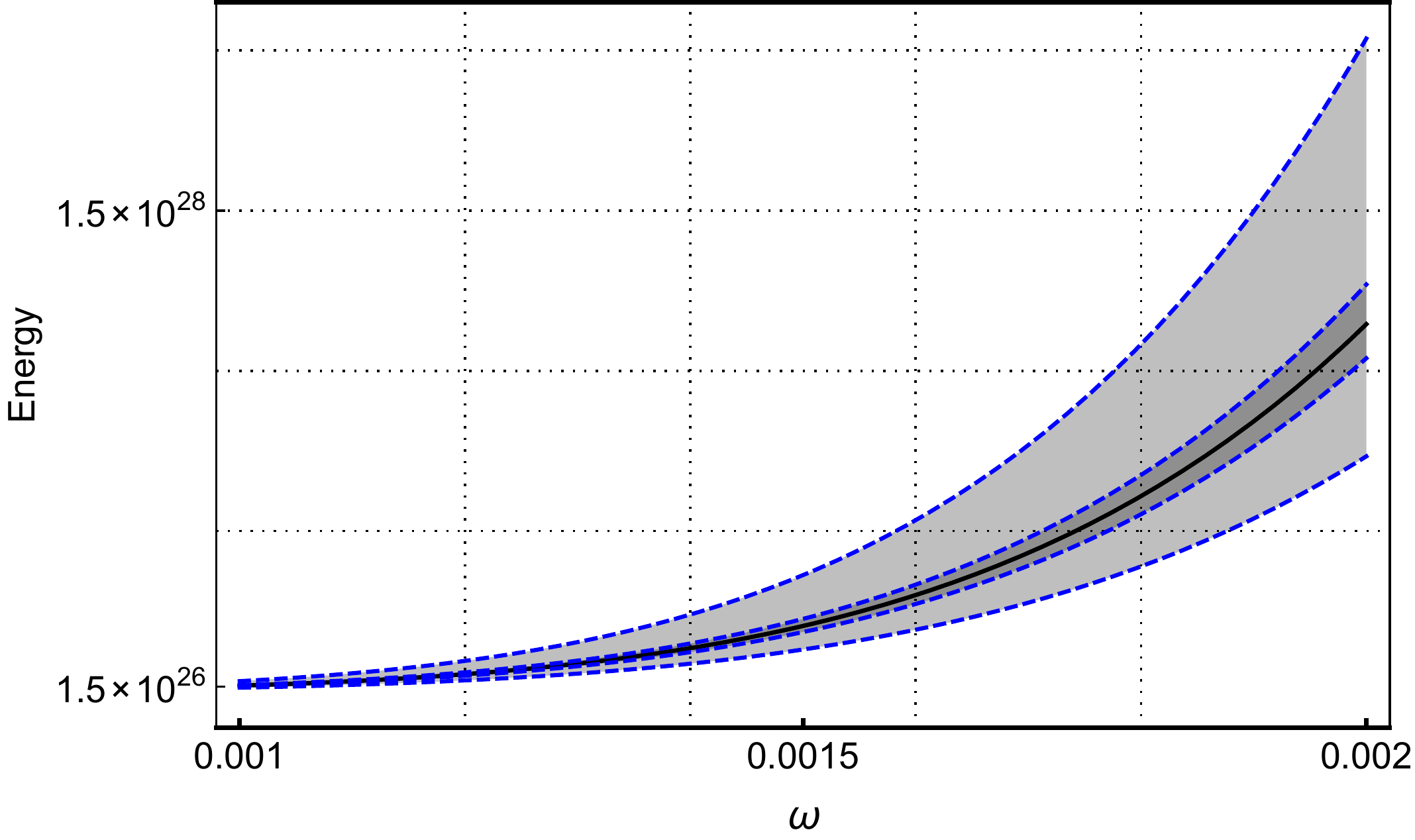} 
\end{center}
\caption{\label{QuadrupoleFormula}Energy (Joule) released by a binary-system ($m_1=1.34 M_{\odot}$, $m_2=1.25 M_{\odot}$ and $R=8.8 \cdot 10^{5} \ km$) for different UV parameters (blue dashed curves) as well as for the purely general relativistic case (black curve) as a function of the orbital frequency $\omega$ (Hz). The outer blue dashed curves correspond to $|\sigma|=0.25$ and $\sqrt{\kappa}=10^{-6}\ m$ and the inner ones correspond to $|\sigma|=0.05$ and $\sqrt{\kappa}=10^{-6}\ m$.}
\end{figure}

The purpose of this chapter is to work out the effective quadrupole formula for a binary-system evolving on circular orbits and to analyse in how far we observe a deviation from the linearised general relativistic result. We saw in the previous section that to leading order the effective energy-momentum pseudotensor reduces to $N^{\alpha\beta}=\sum_A m_A v^\alpha v^\beta\delta(\textbf{y}-\textbf{r}_A)$. This allows us to derive the retarded radiative quadrupole moment,
\begin{equation*}
Q^{ab}\,=\,c^{-2}\int_\mathcal{M}d\textbf{y}\ \sum_{n=0}^{+\infty}\sigma^n \mathcal{E}_n N^{00}(y^0,\textbf{y}) \ y^ay^b,
\end{equation*} 
where we remind that $\mathcal{E}_0=\mathcal{D}=\int dy^0 \ \delta(x^0-|\textbf{x}-\textbf{y}|-y^0)$ and $\mathcal{E}_n=\frac{1}{2\sqrt{\pi n \kappa}}\int dy^0  e^{-\frac{(x^0-|\textbf{x}-\textbf{y}|-y^0)^2}{4n\kappa}}$ are the retardation integrals and $\mathcal{M}$ is the near zone domain. It is common practice to resume the orbital dynamics of a two body-system by the motion of a fictitious body of reduced mass $\mu=\eta m$ with position vector $\textbf{r}=r [\cos(\omega t), \sin(\omega t),0]$ and orbital velocity $\omega$. We remind that in this particular context the position vectors of the single bodies are $\textbf{r}_1=\frac{m_2}{m}\textbf{r}$, $\textbf{r}_2=-\frac{m_1}{m}\textbf{r}$, where $m=m_1+m_2$ is the sum of the bodies respective masses and $\eta=\frac{m_1 m_2}{m^2}$ is a dimensionless parameter. It should be noticed that in contrary to the previous chapters $r=|\textbf{r}_1-\textbf{r}_2|$ is the distance between the two bodies with respective masses $m_1$ and $m_2$ and should not be confused with $|\textbf{x}|$ which is the distance between the source and the observer (detector). With this we can derive the quadrupole matrix for a binary system evolving on circular orbits,
\begin{equation*}
Q^{\langle ab\rangle}\,=\,\frac{\eta \tilde{m}}{6} r^2 \ \begin{pmatrix}
\ 1+3\ Z(\sigma,\kappa,\omega) \ \cos(2\omega \tau) & 3\ Z(\sigma,\kappa,\omega)\ \sin(2\omega\tau)& \ 0 \ \\ 
\ 3\ Z(\sigma,\kappa,\omega)\ \sin(2\omega\tau) & 1-3\ Z(\sigma,\kappa,\omega) \ \cos(2\omega \tau)& \ 0 \ \\
\ 0 & 0 &\ 0 \
\end{pmatrix},
\end{equation*}
where $Z(\sigma,\kappa,\omega)=(1-\sigma)/(1-\sigma e^{-4\kappa\frac{\omega^2}{c^2}})$ is a dimensionless function depending on the UV parameters, the orbital velocity and $\tilde{m}=(m_1+ m_2)/(1-\sigma)$ the effective mass of the binary-system. Further computational details can be withdrawn from the appendix-section related to this chapter. The energy released by the two-body system, as a function of the orbital frequency, is obtained from the effective quadrupole formula $P(\omega)\,=\,\frac{32 }{5 c^5}\ G(\sigma,\kappa, \omega)\  \mu^2 \ \ r^4  \ \omega^6$, where $G(\sigma,\kappa,\omega)=G/\big(1-\sigma e^{-4\kappa \frac{\omega^2}{c^2}}\big)^2$ is the effective frequency depending Newtonian coupling. We observe that in the limit of vanishing UV parameters ($\lim_{\sigma,\kappa\rightarrow 0}G(\sigma,\kappa,\omega)=G$) we recover the Newtonian constant. The amount of energy released by a binary-system, evolving on circular orbits, is visualized by Fig. \ref{QuadrupoleFormula} for mass and radial separation parameters taken from the Double Pulsar system ($m_1=1.34 M_{\odot}$, $m_2=1.25 M_{\odot}$ and $r=8.8 \cdot 10^{5} \ km$). The latter (J0737-3039) is composed by two massive neutron stars orbiting their common center-of-mass on almost circular orbits ($e\approx 0.088$) \cite{DoublePulsar}. We observe a rather strong deviation from the purely general relativistic result for values of the order of $ |\sigma| \geq 10^{-2}$ or larger. However for smaller ($|\sigma|\sim 0.0001$) and in this sense more realistic values (previous subsection) the effective curves approach the linearised general relativistic energy emission curve (black solid curve).   
At this order of accuracy the precise value of the dimension length-squared parameter $\kappa$ seems to be rather unimportant and it is easy to observe that in the limit of vanishing UV parameters we recover the usual linearised quadrupole formula. We will see later, when we work out the effective barycentre at the 1.5 post-Newtonian order, that stronger modifications to the effective coupling parameter $G$ will only set in beyond leading order. A generic feature of this kind of nonlocal modifications to the Einstein field equations is that the higher the post-Newtonian accuracy is the more complicated the effective Newtonian coupling parameter $G$ becomes. The 1.5 post-Newtonian quadrupole radiation is currently being investigated and will be outlined in the near future \cite{Alain2}.
\section{The effective barycentre:}
In this section we review the different contributions that make up the effective barycentre at the 1.5 post-Newtonian order of accuracy. In a first step we will work out the various contributions of the effective barycentre for a generic many body system before we finally rephrase the obtained results in terms of the characteristic notation for a binary-system. The total near-zone barycentre of a N-body system \cite{MisnerThroneWheeler, PoissonWill, WillWiseman,PatiWill1,PatiWill2}is composed by the matter and field energy confined in the region of space $\mathcal{M}:|\textbf{x}|<\mathcal{R}$ such that,
\begin{equation*}
M\boldsymbol{R}\,=\,c^{-2}\int_\mathcal{M}d\textbf{x}\ \big[N^{00}_m+N^{00}_{LL}\big] \textbf{x}=M_m\boldsymbol{R}+M_{LL}\boldsymbol{R},
\end{equation*}
where $N^{00}_m$ and $N_{LL}^{00}$ are the effective nonlocally modified matter and field (Landau-Lifshitz) pseudotensors respectively. We remind that the harmonic gauge contribution, $N^{00}_H=\mathcal{O}(c^{-4})$, is beyond the order of accuracy at which we aim to work at in this article.
\subsection{The matter contribution:}
We will work through the various matter-contributions first and systematically retain all the terms that are within the 1.5 post-Newtonian order of accuracy. The first integral essentially leads to the general relativistic matter contribution \cite{PoissonWill,WillWiseman,PatiWill1,PatiWill2},
\begin{equation*}
M_{\mathcal{B}_1+\mathcal{B}h}\boldsymbol{R}\,=\,c^{-2} \int_{\mathcal{M}}d\textbf{x} \ \textbf{x} \ \big[\mathcal{B}^{00}_{1}+\mathcal{B}h^{00}\big]\,=\,[M_m\boldsymbol{R}]^{GR}+3\sigma\frac{G}{c^2}\sum_A\sum_{B\neq A}\frac{m_A\tilde{m}_B}{r_{AB}} \boldsymbol{r}_A+\mathcal{O}(c^{-4}),
\end{equation*} 
where $\tilde{m}_B=(1-\sigma)^{-1}m_B$ is the effective mass of body $B$. We recover the standard matter piece $[M_m\boldsymbol{R}]^{GR}=\sum_Am_A\ \big[1+\frac{1}{2c^2}v^2_A+\frac{3G}{c^2}\sum_{B\neq A}\frac{m_B}{r_{AB}}\big]\ \textbf{r}_A$ as well as an additional term which merely originates from the effective Newtonian potential introduced previously. Further computational details are outlined in the appendix-section related to this chapter. The next two terms which could contribute to the 1.5 PN order contain infinitely many derivatives. Similarly to the discussion for the corresponding near-zone mass terms \cite{Alain1} we see that only the lowest order differential terms are able to provide a non-vanishing contribution,
\begin{equation*}
M_{\mathcal{B}_2} \boldsymbol{R}\,=\, c^{-2} \int_{\mathcal{M}}d\textbf{x} \ \textbf{x} \ \mathcal{B}^{00}_{2}\,=\,  \frac{4 \epsilon}{1-\sigma}\frac{G}{c^2}\sum_A\sum_{B\neq A} \frac{m_A\tilde{m}_B}{r_{AB}^3}\textbf{r}_{AB}+\mathcal{O}(c^{-4}).
\end{equation*}
A careful analysis shows that a similar reasoning applies for the derivative term $\sigma D^{\alpha\beta}$ worked out in the penultimate chapter,
\begin{equation*}
\sigma M_{D}\boldsymbol{R}\,=\,\sigma \int_{\mathcal{M}} d\textbf{x} \ D^{00} \ \textbf{x}\,=\,\frac{24\epsilon}{1-\sigma}\frac{G}{c^2} \sum_A \sum_{B\neq A}  \frac{m_A\tilde{m}_B}{r_{AB}^3}\textbf{r}_{AB}+\mathcal{O}(c^{-4}).
\end{equation*}
This contribution is less straightforward than the previous one in the sense that one has to distinguish many different cases according to the parameters in the sum $\mathcal{S}(\sigma,\kappa)$ contained within $D^{\alpha\beta}$ (appendix). Partial integration was used and surface terms were discarded for the same reasons as in the previous subsection. However it should be noticed that eventually both terms vanish because they are proportional to $\textbf{r}_{AB}$, so that we finally have: $M_{\mathcal{B}_2}\boldsymbol{R}=\sigma M_{D}\boldsymbol{R}=\textbf{0}+\mathcal{O}(c^{-4})$. Additional computational details can be found in the appendix-section related to this chapter. With this we have reviewed all the different matter contributions and we can finally write down the total near-zone matter centre-of-mass for a many-body system,
\begin{equation*}
M_m\boldsymbol{R}\,=\, [M_m\boldsymbol{R}]^{GR}+[M_m\boldsymbol{R}]^{NL}+\mathcal{O}(c^{-4}), 
\end{equation*}
where $[M_m\boldsymbol{R}]^{NL}=3\sigma\frac{G}{c^2}\sum_A\sum_{B\neq A}\frac{m_A\tilde{m}_B}{r_{AB}} \textbf{r}_A$. It should be noticed that nonlocal corrections disappear in the limit of vanishing UV parameters $\lim_{\sigma,\kappa\rightarrow 0}M_m\boldsymbol{R}= [M_m\boldsymbol{R}]^{GR}$.
\subsection{The field contribution:}
The next task is to work out the near-zone field contribution to the effective barycentre, $M_{LL}\boldsymbol{R}=c^{-2}\int_{\mathcal{M}}d\textbf{x} \ N^{00}_{LL} \ \textbf{x}$, where we recall \cite{Alain1} from the penultimate chapter the precise form of the effective Landau-Lifshitz pseudotensor, $N^{00}_{LL}=(1-\sigma)\tau^{00}_{LL}-\epsilon \Delta \tau^{00}_{LL}-\sigma \sum_{m=2}^{+\infty}\frac{\kappa^m}{m!}\Delta^m\tau^{00}_{LL}$, together with $\tau^{00}_{LL}=-\frac{7}{8\pi G}\partial_pV\partial^pV$. The first term \cite{PoissonWill,WillWiseman,PatiWill1,PatiWill2} gives essentially rise to the standard 1.5 post-Newtonian term,
\begin{equation*}
-\frac{7}{8\pi G c^2}\int_{\mathcal{M}} d\textbf{x} \ \textbf{x} \ \partial_pV\partial^pV\,=\,-\frac{7G}{2c^2} \sum_A\sum_{B\neq A} \frac{\tilde{m}_A\tilde{m}_B}{r_{AB}}\textbf{r}_A\,=\,(1-\sigma)^{-2}\ [M_{LL}\boldsymbol{R}]^{GR},
\end{equation*}
where we remind that $\tilde{m}_A=(1-\sigma)^{-1}m_A$ is the effective mass of body $A$. We observe that the first integral is proportional to the usual 1.5 post-Newtonian near-zone field contribution $[M_{LL}\boldsymbol{R}]^{GR}$. Additional computational details about the derivation of this integral are provided in the appendix-section related to this chapter. The second term of the effective Landau-Lifshitz pseudotensor $N^{00}_{LL}$ is less straightforward and therefore needs a more careful investigation. It should be noticed that we have $\Delta \tau^{00}_{LL}=\frac{-7}{4\pi G}\big[(\partial_p\partial_m\partial^m V)\ \partial_pV+(\partial_m\partial_pV)\ (\partial^m\partial^pV)\big]$, where we remind that $\partial_m\partial^m=\Delta$ is the Laplace-operator. We will review these two terms separately and we see that the first one vanishes after integration over the near-zone domain, 
\begin{equation*}
-\frac{7\epsilon}{4\pi c^2G} \int_{\mathcal{M}}d\textbf{x} \ \textbf{x} \ (\partial_p\Delta V) (\partial^p V)\,=\,-\frac{7\epsilon}{2}\frac{G}{c^2}  \sum_A\sum_{B\neq A} \tilde{m}_A \tilde{m}_B \frac{\textbf{r}_{AB}}{r^3_{AB}}\,=\,0,
\end{equation*}
because we have $\textbf{r}_{AB}=-\textbf{r}_{BA}$. Surface terms originating from partial integration are proportional to $\delta(\mathcal{R}-r_B)$ and will therefore disappear in the near-zone $\mathcal{M}:|\textbf{x}|<\mathcal{R}$ too. In addition we used $\Delta |\textbf{x}-\textbf{r}_A|^{-1}=-4\pi\ \delta(\textbf{x}-\textbf{r}_A)$ as well as $(\partial_p \textbf{x})\ (r_B-x)^p\ |\textbf{x}-\textbf{r}_B|^{-3}=(\textbf{r}_B-\textbf{x})\ |\textbf{x}-\textbf{r}_B|^3$ in the derivation of this result and we remind that $\epsilon=\sigma \kappa$ is of dimension length squared. Additional computational details are presented in the corresponding appendix-section. The last contribution is the most demanding one and full computational details are provided in the appendix related to this chapter,
\begin{equation*}
-\frac{7\epsilon}{4\pi c^2}\int_{\mathcal{M}}d\textbf{x} \ \textbf{x} \ (\partial_m\partial_pV) \ (\partial^m\partial^pV)\,=\,-\frac{831\epsilon}{10}\frac{G}{c^2}\sum_A\sum_{B\neq A} \tilde{m}_A\tilde{m}_B \  \frac{\textbf{r}_B}{r^3_{AB}}.
\end{equation*}
It should be mentioned that all the non-vanishing field contributions to the center-of-mass are of first post-Newtonian order and that besides the standard general relativistic term $[M_{LL}]^{GR}$ all additional terms disappear in the limit of vanishing UV parameters. Most of the contributions in the remaining piece of the effective Landau-Lifshitz pseudotensor, containing infinitely many derivative terms $\sum_{m=2}^{+\infty}\frac{\kappa^m}{m!}\Delta^m \tau^{00}_{LL}$, will not contribute to the effective barycentre at the 1.5 PN order of accuracy. After multiple partial integration they will be proportional to $\sum_A\sum_{B\neq A}\tilde{m}_A \tilde{m}_B \ \nabla^q\delta(\textbf{r}_A-\textbf{r}_B)\,=\,0,\ \forall q\geq 0, \ \text{or}\ \nabla^q \textbf{x}\,=\,0, \ \forall q\geq 2$ or to both terms at the same time. A similar situation was already encountered in the previous subsection when we worked out the matter-contribution to the center-of-mass as well as in \cite{Alain1} where we determined the total effective near-zone mass at the 1.5 post-Newtonian order of accuracy. Surface terms, coming from (multiple) partial integration, are proportional to $\delta(\mathcal{R}-r_A)$ and will eventually vanish in the near-zone defined by $\mathcal{M}:|\textbf{x}|<\mathcal{R}$. However each derivative order $m\geq 2$ will produce a term proportional to $(\partial_{p_1}\cdots \partial_{p_m} V) \ (\partial^{p_1}\cdots\partial^{p_m}V)$. We can infer from the analysis of the first two contributions ($\tau^{00}_{LL}$ and $\Delta \tau^{00}_{LL}$) that the forthcoming term will be proportional to $\frac{\kappa^2}{2!}\Delta^2\tau^{00}_{LL}\propto \big(\frac{\kappa}{r^2_{AB}}\big)^2$, where we remind that the UV parameter $\kappa$ is of dimension length squared. Assuming that the bodies $A$ and $B$ are separated by astrophysical distances such that $\kappa\ll r^2_{AB}$, we see that this term is smaller than the previous one by a factor $\frac{\kappa}{r^2_{AB}}\ll 1$. The next term, originating from $\frac{\kappa^3}{3!}\Delta^3\tau^{00}_{LL}$, will even be smaller than the leading term by a factor $\big(\frac{\kappa}{r^2_{AB}}\big)^2\ll 1$ this time. In principle it is possible to evaluate these remaining terms to all possible orders using the computational techniques (appendix) outlined in this article. In the context of the present post-Newtonian analysis we will however truncate the result at this level, not including terms of the order $\mathcal{O}\big(\frac{\kappa^2}{r^4_{AB}}\big)$ and refer the reader to future developments \cite{Alain2}. Additional computational details on this particular issue are provided in the appendix-section related to this chapter. This allows us to write down the total near-zone field contribution of the center-of-mass at the 1.5 post-Newtonian order of accuracy,
\begin{equation*}
M_{LL}\boldsymbol{R}\,=\,\frac{[M_{LL}\boldsymbol{R}]^{GR}}{(1-\sigma)^2} \ -\frac{831\epsilon}{10}\frac{G}{c^2}  \sum_A\sum_{B\neq A} m_A m_B  \frac{\textbf{r}_B}{r^3_{AB}}+\mathcal{O}(c^{-4},\kappa^{2}),
\end{equation*}
where $[M_{LL}\boldsymbol{R}]^{GR}=-\frac{7G}{2c^2}\sum_A\sum_{B\neq A} \frac{m_Am_B}{r_{AB}}\textbf{r}_A $. We introduced the notation $\mathcal{O}(c^{4},\kappa^{2})$ to indicate that, according to the discussion outlined above, a truncation has been performed discarding terms proportional to $\kappa^2$ divided by the fourth power of the relative separation of the two bodies $A$ and $B$. In analogy to the notation $\mathcal{O}(c^{-4})$, which is only a convenient mnemonic to judge the importance of various terms inside a post-Newtonian expansion, the real dimensionless expansion parameter is rather $(\kappa^2/r^4_c)\ (Gm_c)/c^2 r_c)=(\kappa^2/r^4_c)\ (v^2_c/c^2)$, where we remind that $m_c$, $r_c$ and $v_c$ are respectively the characteristic mass, the characteristic scale and the characteristic velocity of the gravitational system.
\subsection{The 1.5 post-Newtonian barycentre:}
Combining the near-zone matter and field contributions, we obtain the effective 1.5 post-Newtonian barycentre for a generic many-body system,
\begin{equation*}
M\boldsymbol{R}\,=\,[M\boldsymbol{R}]^{GR}-\frac{\sigma}{(1-\sigma)^2}\frac{G}{c^2}\sum_A\sum_{A\neq B}\frac{m_Am_B}{r_{AB}}\ \Big[4-\frac{\sigma}{2}+\frac{831}{10}\frac{\kappa}{r^2_{AB}}\Big]\ \textbf{r}_A+\mathcal{O}(c^{-4},\kappa^2),
\end{equation*}
where $[M\boldsymbol{R}]^{GR}=\sum_Am_A\textbf{r}_A+c^{-2}\sum_A\frac{m_Av^2_A}{2}\textbf{r}_A-c^{-2}\sum_{A,B\neq A} \frac{Gm_Am_B}{2r_{AB}}\textbf{r}_A$ is the standard 1.5 post-Newtonian barycentre \cite{PoissonWill,WillWiseman,PatiWill1,PatiWill2}. We observe that the modification term, $[M\boldsymbol{R}]^{NL}=-\frac{\sigma}{(1-\sigma)^2}\frac{G}{c^2}\sum_A\sum_{A\neq B}\frac{m_Am_B}{r_{AB}}\ \big[4-\frac{\sigma}{2}+\frac{831}{10}\frac{\kappa}{r^2_{AB}}\big]\ \textbf{r}_A$, is of first post-Newtonian order. Moreover it should be noticed that in the limit of vanishing UV parameters, $\lim_{\sigma\rightarrow 0,\kappa\rightarrow 0} M \boldsymbol{R}=[M\boldsymbol{R}]^{GR}+\mathcal{O}(c^{-4})$ we recover the purely general relativistic center-of-mass. At this stage we would like to reduce the general framework outlined above to a spinless two-body system with masses $m_1$, $m_2$ and respective position vectors $\textbf{r}_1$ and $\textbf{r}_2$, 
\begin{equation*}
[M\boldsymbol{R}]_2\,=\,m_1\Big[1+\frac{1}{2c^2}\Big(v^2_1-\frac{G(\sigma,\kappa,r)\ m_2}{r}\ \big]\Big)\Big]\textbf{r}_1+m_2\Big[1+\frac{1}{2c^2}\Big(v^2_2-\frac{G(\sigma,\kappa,r)\ m_1}{r}\ \big]\Big)\Big]\textbf{r}_2+\mathcal{O}(c^{-4},\kappa^2),
\end{equation*}
where $G(\sigma,\kappa,r)=G\big[1-\frac{\sigma}{(1-\sigma)^2} \big(8-\sigma+\frac{831}{5}\frac{\kappa}{r^2}\big)\big]$ is the effective Newtonian coupling for a binary-system at this particular order of accuracy. In the limit of vanishing UV parameters the effective coupling reduces to the standard Newtonian coupling $\lim_{\sigma,\kappa\rightarrow 0}G(\sigma,\kappa,r)=G$ and we recover the usual 1.5 post-Newtonian barycentre of a two-body system \cite{PoissonWill,WillWiseman,PatiWill1,PatiWill2}. We that the post-Newtonian relations outlined in this chapter apply to fluid bodies from one another, so that their mutual gravitational interaction is weak. The description of the motion simplifies when the coordinate system is attached to the barycentre. In this regard we impose the condition $\boldsymbol{R}=0$ and together with the separation vector $\textbf{r}=\textbf{r}_1-\textbf{r}_2$ we eventually obtain the effective 1.5 post-Newtonian position vectors for a binary-system,
\begin{equation*}
\begin{split}
\textbf{r}_1\,=&\,+\frac{m_2}{m}\textbf{r}+\frac{\eta \gamma}{2c^2}\Big[v^2-\frac{G(\sigma,\kappa, r)\ m}{r} \Big]+\mathcal{O}(c^{-4},\kappa^2),\\
\textbf{r}_2\,=&\,-\frac{m_1}{m}\textbf{r}+\frac{\eta \gamma}{2c^2}\Big[v^2-\frac{G(\sigma,\kappa,r)\ m}{r}\Big]+\mathcal{O}(c^{-4},\kappa^2).
\end{split}
\end{equation*}
For clarity reasons we introduced the following two dimensionless quantities $\eta=\frac{m_1m_2}{m^2}$, $\gamma=\frac{m_1-m_2}{m}$ and we remind that $\kappa$ is of dimension length-squared and $m=m_1+m_2$ is the Newtonian mass of the binary-system. Further computational details can be withdrawn from the appendix-section related to this chapter. Here again we see that in the limit of vanishing UV parameters ($\sigma, \kappa \rightarrow 0$) the effective position vectors reduce to the standard 1.5 post-Newtonian position vectors \cite{PoissonWill,WillWiseman,PatiWill1,PatiWill2}. 

\section{Conclusion:}
In this manuscript we suggested a specific model of a nonlocally modified theory of gravity in which Newton's constant $G$ is promoted to a differential operator $G_\Lambda(\Box_g)$. This particular theory of gravity as well as the accompanying degravitation mechanism were presented for the first time in \cite{Alain1}. In this work we briefly reviewed its basic characteristic features and pursued additional phenomenological investigations in the context of binary-systems. We reminded that even though the nonlocal equations of motion are themselves generally covariant, they cannot (for nontrivial $G_\Lambda(\Box_g)$) be presented as a metric variational derivative of a diffeomorphism invariant action unless you assume that they are only a first, linear in the curvature, approximation for the complete equations of motion \cite{Barvinsky1, Barvinsky2}. We also recalled that the generic idea of a differential coupling was obviously formulated for the first time in \cite{Dvali1,Barvinsky1,Dvali2,Barvinsky2} in order to find a solution to the cosmological constant problem \cite{Weinberg1}. It should however be noticed that the concept of a varying coupling constant of gravitation goes back to early works of Dirac \cite{Dirac1} and Jordan \cite{Jordan1, Jordan2}. Inspired by these considerations Brans and Dicke published in the early sixties a theory in which the gravitational constant is replaced by the reciprocal of a scalar field \cite{Brans1}. In this article we presented a precise nonlocal model for infrared {\it degravitation} in which $G_\Lambda(\Box_g)$ acts like a high-pass filter with a macroscopic distance filter scale $\sqrt{\Lambda}$. In this way sources characterized by characteristic wavelengths much smaller than the filter scale ($\lambda_c\ll\sqrt{\Lambda}$) pass rather unhindered through the filter and gravitate almost in the ordinary way, whereas sources characterized by wavelengths larger than the filter scale are effectively filtered out \cite{Dvali1,Dvali2}. In the first chapter we quickly reviewed the main features of the cosmological constant problem and proposed a precise differential coupling model by which we can observe an effective {\it degravitation} of the vacuum energy on cosmological scales. We finished chapter one by presenting a concise cosmological model in which we determined the effective Friedmann-Lema\^{i}tre equation. For reasons of completeness we recalled in the second chapter the relaxed Einstein equations in the context of purely Einsteinian gravity and we briefly introduced the post-Newtonian theory as well as various related concepts that were used in the subsequent chapters. In chapter three we derived the effective relaxed Einstein equations and showed that in the limit of vanishing ultraviolet parameters and an infinitely large infrared parameter $\sqrt{\Lambda}$ we recover the standard equation for the gravitational potentials. It should be mentioned that further conceptual and computational details about this precise model can be withdrawn from \cite{Alain1}. The effective energy-momentum pseudotensor $N^{\alpha\beta}$ forms the main body of chapter four in which we worked out separately its matter, field and harmonic gauge contributions up to the 1.5 post-Newtonian order of accuracy. In chapter five we worked out the effective orbital dynamics of a binary-system and we provided an upper bound for the UV parameter $\sigma$. In the penultimate chapter of this article we gathered the most important results in order to determine the effective barycentre of a generic $n$-body system at the 1.5 post-Newtonian order of accuracy. We closed this chapter by computing the effective 1.5 post-Newtonian position vectors for a binary-system and we compared our results to those obtained in the context of the standard theory of gravity. We observe that in the limit of vanishing ultraviolet parameters we are always able to recover the well known 1.5 post-Newtonian corresponding results. 
\begin{acknowledgments}
A. D. would like to thank Professor Malte Henkel  (University of Lorraine) for stimulating discussions concerning the relativistic Kepler problem and Professor Eric Poisson (University of Guelph) for useful comments concerning the generalized regularization prescription. The author gratefully acknowledges support by the Ministry for Higher Education and Research of the G.-D. of Luxembourg (MESR-Cedies).
\end{acknowledgments}

\appendix
\section{Introduction:}
\subsection{A succinct cosmological model:}
The first task is to work out the d'Alembert operator in the context of the Robertson-Walker metric, \cite{PoissonWill,Woodard1,Esposito1},
\begin{equation*}
\Box_{R}\,=\,\frac{1}{\sqrt{-g}} \partial_\mu\big[\sqrt{-g} \ g^{\mu\nu} \partial_\nu\big]\,=\,\frac{1}{\sqrt{q}} \partial_\mu\big[\sqrt{q} \ g^{\mu\nu} \partial_\nu\big]\,=\,-\partial_0^2-\frac{1}{2c}\frac{\dot{q}}{q}\partial_0+\frac{1}{\sqrt{q}}\partial_a \big[\sqrt{q} \ q^{ab} \partial_b\big],
\end{equation*}
where $g^{\alpha\beta}=\text{diag}(-1,q^{ab})$ is the Robertson-Walker metric, $q=det(q_{ab})$ is the spatial-metric determinant and $\partial_\mu[\sqrt{q}\ g^{\mu\nu}\partial_\nu]=\partial_0[-\sqrt{q}\ \partial_0]+\sqrt{q}^{-1}\partial_a[\sqrt{q}\ q^{ab}\ \partial_b]$. In the remaining part of this appendix-subsection we will work with the Robertson-Walker metric outlined in the main part of the article, $g_{\alpha\beta}=\text{diag}(-1,\ \frac{R^2}{1-kr^2},\ R^{2} r^2, \ R^2 r^2\sin^2\theta)$, so that for definiteness, we have $q_{11}=\frac{R^2}{1-kr^2}$, $q_{22}=R^2 r^2$ and $q_{33}=R^2 r^2 \sin^2\theta$ and the cosmic scale factor $R$ is a time dependent function only. In agreement to the cosmological principle, which basically claims that all positions of the Universe are essentially the same on length scales of the order of $10^8-10^9$ light years \cite{Weinberg2,Inverno}, the energy-momentum tensor that we employ will be the perfect fluid, $T^{\alpha\beta}=(c^2\rho+p) \ u^\alpha u^\beta/c^2+p \ g^{\alpha\beta}$, where $\rho$ is the matter density, $p$ is the pressure and $u^\alpha$ is the velocity field \cite{PoissonWill, Weinberg2}. Moreover we will assume that the matter density and the pressure will be time dependent functions only and that the contents of the Universe are, on the average, at rest in the coordinate-system $r$, $\theta$, $\phi$ \cite{Weinberg2}. From this we infer that the velocity field has to be $u^\alpha=\gamma (c,\textbf{0})$, where we remind that $\gamma^{-1}=\sqrt{-g_{\mu\nu}\frac{v^\mu v^\nu}{c^2}}$ is dimensionless relativistic factor. In this particular context the energy-momentum tensor of the perfect fluid becomes, $T^{\alpha\beta}=\text{diag}(\rho c^2,\ q^{11}p,\ q^{22}p,\ q^{33}p)$, where $q^{aa}=q_{aa}^{-1}$. It should be noticed that the present value of the cosmic scale factor, which is sometimes called the "radius of the Universe" \cite{Weinberg2}, is rather large ($R^{-2}\ll 1$) and that the current cosmological pressure term is quite small ($p\approx 0$) compared to the value of the early Universe. In the framework of this first concise cosmological analysis we will therefore simplify the effective cosmological energy-momentum tensor to the following expression, $\mathcal{T}^{\alpha\beta}_\Lambda=\frac{G_\Lambda(\Box_R)}{G}T^{\alpha\beta}\approx\text{diag}(c^2\rho_\Lambda,\ q^{11}\ p_\Lambda,\ q^{22}\ p_\Lambda,\ q^{33}\ p_\Lambda)$, where $\rho_\Lambda=\frac{G_\Lambda(\Box_R)}{G}\rho$ is the effective matter density and $p_\Lambda=\frac{G_\Lambda(\Box_R)}{G}p$ is the fluid's effective pressure term. It is obvious that in the limit of vanishing UV-parameters ($\sigma,\kappa \rightarrow 0$) and infinitely large IR-parameter ($\Lambda\rightarrow +\infty$) we recover the standard energy-momentum tensor of the perfect fluid introduced above. We previously saw that the nonlocally modified Einstein field equations $G^{\alpha\beta}=\frac{8\pi}{c^4} \ G\ \mathcal{T}^{\alpha\beta}$ together with the contracted Bianchi identities $\nabla_\alpha G^{\alpha\beta}=0$ \cite{PoissonWill, Weinberg2, MisnerThroneWheeler} give rise to an effective energy-momentum conservation equation $\nabla_\alpha \mathcal{T}^{\alpha\beta}=0$. This allows us to derive an energy conservation equation which relates the effective matter density $\rho_\Lambda$ and effective pressure $p_\Lambda$ with the cosmic scale factor,
\begin{equation*}
\begin{split}
\nabla_\alpha \mathcal{T}^\alpha_{\ \ 0}\,=&\,\partial_\alpha \mathcal{T}_{\ \ 0}^{\alpha}+\Gamma_{\alpha\lambda}^{\alpha}\mathcal{T}_{\ \ 0}^\lambda-\Gamma^\lambda_{\alpha 0} \mathcal{T}_{\ \ \lambda}^{\alpha}\\
=&\,\partial_\alpha (-c^2\rho_\Lambda) \delta^{\alpha 0}+\Gamma_{\lambda\alpha}^\alpha(-c^2\rho_\Lambda)\delta^{\lambda 0}-\Gamma^\lambda_{0\alpha} \mathcal{T}^{\alpha\beta} g_{\beta\lambda}\,=\,-c\dot{\rho}_\Lambda-c \rho_\Lambda \frac{3\dot{R}}{R}-p_\Lambda \frac{3\dot{R}}{cR},
\end{split}
\end{equation*}
where $\mathcal{T}^\alpha_{\ \ 0}=-\mathcal{T}^{\alpha 0}=-c^2 \rho_\Lambda \delta^{\alpha 0}$, $\Gamma^\alpha_{0\alpha}=\frac{1}{2} \ g^{\alpha\alpha} \ \partial_0 g_{\alpha\alpha}=\frac{3\dot{R}}{cR}$ and $\Gamma^\lambda_{0\alpha}T^{\alpha\beta}g_{\lambda\beta}=\frac{1}{2} g^{\lambda \epsilon}(\partial_0 g_{\epsilon\alpha})g_{\lambda\beta} \mathcal{T}^{\beta \alpha}=\frac{3 \dot{R}}{cR} p_\Lambda$. From this we can easily deduce the energy conservation equation ($\nabla_\alpha\mathcal{T}^\alpha_{\ \ 0}=0$) outlined in the main part of the article. We extensively used the fact that for the Robertson-Walker metric as well as for the effective energy-momentum tensor all off-diagonal elements are vanishing. Our next task is to work out the cosmological acceleration equation. By contracting the effective Einstein field equations we obtain a relation between the Ricci-scalar and the effective energy-momentum tensor $R=-8\pi G/c^4 \ \mathcal{T}$. This allows us to rewrite down the nonlocally modified Einstein field equations,
\begin{equation*}
R^{\alpha}_{\  \gamma}+\frac{1}{2} \delta^{\alpha}_{\  \gamma}\frac{8\pi G}{c^4}\mathcal{T}\,=\,\frac{8\pi G}{c^4} \mathcal{T}^{\alpha}_{\ \ \gamma}\Leftrightarrow R^{\alpha}_{\  \gamma}\,=\, \frac{8\pi G}{c^4} \Big(\mathcal{T}^\alpha_{\ \ \gamma}-\frac{1}{2} \delta^{\alpha}_{\  \gamma} \mathcal{T}\Big),
\end{equation*}
where we remind that $\mathcal{T}^{\alpha\beta}=\frac{G_\Lambda(\Box_R)}{G} T^{\alpha\beta}$ is the effective energy-momentum tensor. The time-time component for the Ricci tensor is,
\begin{equation*}
R^0_{\ 0}=\,\frac{8\pi G}{c^4} \Big(\mathcal{T}^0_{\ \ 0}-\frac{1}{2} \delta^{0}_{\ 0} \mathcal{T}\Big)\,=\,\frac{8\pi G}{c^4}\Big(-\mathcal{T}^{00}+\frac{1}{2} (\mathcal{T}^{00}-\mathcal{T}^{aa} g_{aa})\Big)\,=\,-\frac{4\pi G}{c^4} \Big(\rho_\Lambda c^2+3p_\Lambda\Big).
\end{equation*}
On the other hand we have,
\begin{equation*}
R^0_{\ 0}\,=\,-\Big[\partial_\alpha\Gamma^\alpha_{00}-\partial_0 \Gamma^{\alpha}_{0\alpha}+\Gamma^\alpha_{\mu\alpha}\Gamma^\mu_{00}-\Gamma^\alpha_{\mu 0}\Gamma^\mu_{0\alpha}\Big]\,=\,-\Big[-\partial_0\Gamma^\alpha_{0\alpha}-\Gamma_{\mu0}^\alpha\Gamma_{0\alpha}^\mu\Big]\,=\, \frac{3}{c^2} \frac{\ddot{R}R-(\dot{R})^2}{R^2}+\frac{3}{c^2} \Big(\frac{\dot{R}}{R}\Big)^2\,=\, \frac{3\ddot{R}}{c^2 R}.
\end{equation*}
Here we used $\Gamma^\alpha_{00}=g^{\alpha\nu}\partial_0g_{\nu 0}-\frac{1}{2}g^{\alpha\nu}\partial_\nu g_{00}=0$ as well as  $\Gamma^\alpha_{0\beta}=\frac{1}{2}g^{\alpha\nu}(\partial_0 g_{\beta\nu})=\frac{\dot{R}}{cR} \ \delta^a_{\ b}$, where $a$, $b$ are spatial indices. Combining the last two equations we obtain the cosmic acceleration equation $\ddot{R}=-\frac{4\pi G}{3} \big(\rho_\Lambda+\frac{3p_\Lambda}{c^2}\big) R$. This equation and the previously derived equation for the energy conservation, $\dot{\rho}_\Lambda=-\frac{3\dot{R}}{R}\big(\rho_\Lambda+\frac{p_\Lambda}{c^2}\big)$, allow us to work out the effective Friedmann-Lema\^{i}tre equation,
\begin{equation*}
\ddot{R}\,=\,\frac{4\pi G}{3\dot{R}} \big(2\rho_\Lambda R \dot{R}+\dot{\rho}_\Lambda R^2\big)\Leftrightarrow
\partial_t (\dot{R}^2)\,=\,\frac{8\pi G}{3} \ \partial_t (\rho_\Lambda R^2)\Rightarrow
\dot{R}^2\,=\, \frac{8\pi G}{3}  \rho_\Lambda R^2-kc^2,
\end{equation*}
where $k$ is the dimensionless curvature parameter which can take in the context of the Robertson-Walker Universe the values $\{-1,0,+1\}$. Our next task is to study the inverse differential coupling operator $G^{-1}_\Lambda(\Box_R)$ acting on a generic function $f(R)$ depending only on the cosmic scale factor $R(t)$. In this context the cosmological Robertson-Walker d'Alembert operator reduces to $\Box_R=-\partial^2_0-3c^{-1}H\partial_0$, where $\frac{\dot{q}}{2q}=3H$ and $H=\frac{\dot{R}}{R}$ is the time-dependent Hubble parameter \cite{Weinberg2}. In addition we will use the fact that $\sqrt{\Lambda}\sim 10^{30}\ m$ is of the order of the horizon size of the present visible Universe \cite{Barvinsky1,Barvinsky2}, so that in good approximation the nonlocal IR-term can be set to one ($\mathcal{F}_\Lambda(\Box_R)\approx 1)$. The leading order term of the remaining nonlocal coupling operator, acting on a general cosmic scale depending function will become in the sense of a post-Newtonian expansion, $\frac{\mathcal{G}(\Box_R)^{-1}}{G}f(R)=[1-\sigma e^{\kappa\Box_R}]f(R)[1-\sigma(1+\kappa\ \Box_R)]f(R)+\mathcal{O}(c^{-4})$, where the precise form of $\Box_R\propto c^{-2}$ for this particular situation was outlined above. This eventually allows us to see how the differential operator acts on the left hand side of the Friedmann-Lema\^{i}tre equation,
\begin{equation*}
\Big[1-\sigma+\frac{\epsilon}{c^2}\partial^2_t+\frac{3H\epsilon}{c}\partial_t\Big]\ \Big[\frac{\dot{R}^2}{R^2}+k\frac{c^2}{R^2}\Big]\,=\,\frac{8\pi G}{3} \rho,
\end{equation*}
at the 1.5 post-Newtonian order of accuracy and we remind that $\epsilon=\sigma \kappa$ is a parameter of dimension length square. We see that in the limit of vanishing UV-parameters this nonlinear differential equation essentially reduces to the standard Friedmann-Lema\^{i}tre equation \cite{Weinberg2}. In a first step we will carry out separately the first and second order temporal derivatives,
\begin{gather*}
\frac{3\epsilon}{c^2}H\partial_t\Big[\frac{\dot{R}^2}{R^2}+k\frac{c^2}{R^2}\Big]\,=\,\frac{\epsilon}{c^2}\Big[-\frac{6 \dot{R}^4}{R^4}+\frac{6 \dot{R}^2 \ddot{R}}{R^3}-kc^2\frac{6\dot{R}^2}{R^4}\Big],\\
\frac{\epsilon}{c^2}\partial^2_t\Big[\frac{\dot{R}^2}{R^2}+k\frac{c^2}{R^2}\Big]\,=\,\frac{\epsilon}{c^2}\Big[\frac{6\dot{R}^4}{R^4}-\frac{10\dot{R}^2\ddot{R}}{R^3}+\frac{2\ddot{R}^2}{R^2}+\frac{2\dot{R}\ \dddot{R}}{R^2}+kc^2\Big(\frac{6\dot{R}^2}{R^4}-\frac{2\ddot{R}}{R^3}\Big)\Big].
\end{gather*}
We see that the leading order terms proportional to $\frac{\epsilon}{c^2}\frac{\dot{R}^4}{R^4}$ and $\frac{\kappa\dot{R}^2}{R^4}$ cancel out each other. The remaining contributions contain terms proportional to second and third order derivative terms of the cosmic scale factor. We will assume that the latter is a very slowly varying function and therefore the second and third order derivatives will, in good approximation, vanish ($\ddot{R}\approx 0$ and $\dddot{R}\approx 0$). Although this assumption is not true for the very early Universe it certainly applies for the more recent expansion history of the Universe \cite{Weinberg2}. 
\section{The effective wave equation:}
\subsection{The nonlocally modified energy-momentum tensor:}
We have for an arbitrary contravariant rank two tensor $f^{\alpha\beta}(x)$ \cite{PoissonWill, Weinberg2, Woodard1, Maggiore2},
\begin{equation*}
\begin{split}
\Box_g f^{\alpha\beta}(x)\,
=\,\frac{1}{\sqrt{-g}}\partial_\mu\Big[(\eta^{\mu\nu}-h^{\mu\nu})\partial_\nu f^{\alpha\beta}(x)\Big]\,
&=\,\frac{1}{\sqrt{-g}}\Big[\Box f^{\alpha\beta}(x)-h^{\mu\nu}\partial_\mu\partial_\nu f^{\alpha\beta}(x)\Big]\\
&=\,\Big[1-\frac{h}{2}+\frac{h^2}{8}-\frac{h^{\rho\sigma}h_{\rho\sigma}}{4}+\mathcal{O}(G^3)\Big]^{-1}\Big[\Box f^{\alpha\beta}(x)-h^{\mu\nu}\partial_\mu\partial_\nu f^{\alpha\beta}(x)\Big]\\
&=\, \big[\Box-h^{\mu\nu}\partial_\mu\partial_\nu+\tilde{w}(h)\Box-\tilde{w}(h)h^{\mu\nu}\partial_{\mu}\partial_\nu\big] f^{\alpha\beta}(x)\\
\end{split}
\end{equation*}
where the harmonic gauge conditions $\partial_\mu h^{\mu\nu}\,=\,0$ were used together with $\sqrt{-g}g^{\mu\nu}=\eta^{\mu\nu}-h^{\mu\nu}$ \cite{PoissonWill, Will1,PatiWill1,PatiWill2, Blanchet1} and the following definition $\tilde{w}(h)= \frac{h}{2}-\frac{h^2}{8}+\frac{h^{\rho\sigma}h_{\rho\sigma}}{4}+\mathcal{O}(G^3)$ was introduced for the potential function. In the quest of decomposing the effective energy-momentum tensor $\mathcal{T}^{\alpha\beta}=G(\Box) \ \mathcal{H}(w,\Box) \ T^{\alpha\beta}$ we should remind the important result for linear differential operators, $[A,B]=0\Rightarrow [\frac{1}{A},\frac{1}{B}]=0$, where $A$ and $B$ are supposed to be two linear differential operators. A derivation of this result can be found in the second appendix-section of our previous work \cite{Alain1} in which this precise theory of modified gravity was outlined for the first time. This result will be used in the splitting of the nonlocally modified energy-momentum tensor,
\begin{equation*}
\begin{split}
\mathcal{T}^{\alpha\beta}\,=\, G\big[\Box_g\big] \ T^{\alpha\beta}
&=\, \frac{1}{1-\sigma e^{\kappa \Box}} \ \Big[1-\sigma \frac{e^{\kappa \Box}}{1-\sigma e^{\kappa \Box}}\sum_{n=1}^\infty \frac{\kappa^n}{n!} w^n\Big]^{-1} \  \ T^{\alpha\beta}\\
 &=\,\Big[\frac{1}{1-\sigma e^{\kappa \Box}} \ \Big(1+\sigma \frac{e^{\kappa \Box}}{1-\sigma e^{\kappa \Box}} \sum_{n=1}^\infty \frac{\kappa^n}{n!} w^n+\mathcal{O}(\sigma^2)\Big)\Big]  \ T^{\alpha\beta}\\
&=\,G\big[\Box\big] \ \Big[\sum_{n=0}^{+\infty}\mathcal{B}^{\alpha\beta}_n+\mathcal{O}(\sigma^2)\Big],
\end{split}
\end{equation*}
where we used $1-\sigma e^{\kappa\Box_g}=1-\sigma e^{\kappa[\Box+\omega(h,\partial)]}=1-\sigma e^{\kappa\Box}\ \sum_{n=0}^{+\infty} \frac{\kappa^n}{n!}\omega^n$ \cite{Efimov1, Efimov2,Namsrai1, Spallucci1}. Moreover we needed to constrain the range for the modulus of the dimensionless parameter $\sigma$ which has to be smaller than one in order to make the perturbative expansion work. We will see later that this assumption will be confirmed when we work out the modified Newtonian potential or the perihelion precession of Mercury. We adopt the convention that differential operators appearing in the numerator act first ($[w,\Box]\neq 0$). 
\subsection{The effective relaxed Einstein equations:}
In this appendix-subsection we will present some additional computational details regarding the modified Green function outlined in the main part of this article. By substituting the Fourier representation of the modified Green function $G(x-y)=(2\pi)^{-4}\int dk\ G(k) e^{ik(x-y)}$, where $x=(ct,\textbf{x})$ and $k=(k^0,\textbf{k})$, inside Green the function condition $(1-\sigma e^{\kappa \Delta})\Box G(x-y)=\delta(x-y)$ we obtain a relation for the Green function in momentum-space, 
\begin{equation*}
G(k)\,=\,\frac{1}{(k^0)^2-|\textbf{k}|^2} \ \frac{1}{1-\sigma e^{-\kappa \textbf{k}^2}}\,=\,\frac{\sum_{n=0}^{+\infty} \sigma^n \ e^{-n\kappa \textbf{k}^2}}{(k^0)^2-\textbf{k}^2}\,=\,\frac{1}{(k^0)^2-|\textbf{k}|^2}+\sigma \ \frac{ \ e^{-\kappa|\textbf{k}|^2}}{(k^0)^2-|\textbf{k}|^2}+...
\end{equation*} 
The first term in this infinite expansion is the usual Green function followed by correction terms. We also remind that the modulus of the dimensionless parameter is assumed to be strictly smaller than one ($|\sigma|<1$). By making use of the residue theorem we can derive the modified Green function, $G=G^{GR}+G^{NL}$, in terms of its retarded and advanced contributions, $G^{GR}=\frac{-1}{4\pi} \frac{1}{|\textbf{x}-\textbf{y}|} \Big[\delta(x^0-|\textbf{x}-\textbf{y}|-y^0)-\delta(x^0+|\textbf{x}-\textbf{y}|-y^0)\Big]$ and $G^{NL}=\frac{-1}{4\pi} \frac{1}{|\textbf{x}-\textbf{y}|} \sum_{n=1}^{+\infty}\  \frac{\sigma^n}{2\sqrt{\pi n \kappa }}\Big[ e^{-\frac{x^0-|\textbf{x}-\textbf{y}|-y^0}{4n\kappa}}-e^{-\frac{x^0+|\textbf{x}-\textbf{y}|-y^0}{4n\kappa}}\Big]$. The precise derivation of the usual Green $G^{GR}$ function is well known from the literature \cite{PoissonWill, Appel} and will not be repeated here. The nonlocal correction term is obtained by a contour integral (negative sense) over a closed arc $\mathcal{A}$ in the complex plane with two poles on the real axis, 
\begin{equation*}
\begin{split}
G_\mathcal{A}^{NL}\,=&\,\frac{1}{(2\pi)^4} \int d\textbf{k} \oint dz \ \frac{e^{-iz(x^0-y^0)}}{(z-|\textbf{k}|) (z+|\textbf{k}|)}\ e^{-n\kappa |\textbf{k}|^2+i\textbf{k}(\textbf{x}-\textbf{y})}\\
=&\,\frac{-1}{(2\pi)^3} \int d\textbf{k} \ \frac{\sin[|\textbf{k}|(x^0-y^0)]}{|\textbf{k}|}\ e^{-n\kappa |\textbf{k}|^2+i\textbf{k}(\textbf{x}-\textbf{y})}\,
=\,\frac{(-2\pi^2)^{-1}}{|\textbf{x}-\textbf{y}|}\int d|\textbf{k}|\ \sin[|\textbf{k}|(x^0-y^0)]\sin[|\textbf{k}||\textbf{x}-\textbf{y}|]e^{-n\kappa |\textbf{k}|^2}, 
\end{split}
\end{equation*}
where we used $\int d \textbf{k}=\int_0^{2\pi}\int_0^\pi d\theta \ \sin\theta \int_0^{+\infty}d|\textbf{k}|\ |\textbf{k}|^2$. In the limit of infinite arc-radius $G_\mathcal{A}^{NL}$ reduces to $G^{NL}$ and the complex arc contribution vanishes very much like for the standard Green function. By evaluating the integral ($ \sin[|\textbf{k}|(x^0-y^0)]\sin[|\textbf{k}||\textbf{x}-\textbf{y}|]=\frac{1}{2} (\cos[|\textbf{k}|(x^0-|\textbf{x}-\textbf{y}|-y^0)]+\cos[|\textbf{k}|(x^0+|\textbf{x}-\textbf{y}|-y^0)])$) we finally obtain the nonlocal Green function contribution outlined above.

\subsection{A particular solution:}
In analogy to \cite{Alain1,PoissonWill, Will1, PatiWill1, PatiWill2}, we aim to expand the retarded effective pseudotensor in terms of a power series,
\begin{equation*}
 \frac{N^{\alpha\beta}(x^0-|\textbf{x}-\textbf{y}|,\textbf{y})}{|\textbf{x}-\textbf{y}|}\,=\, \sum_{l=0}^{\infty} \frac{(-1)^l}{l!} \textbf{y}^L\partial_L \Big[\frac{N^{\alpha\beta}(x^0-r,\textbf{y})}{r}\Big]\,=\,\frac{1}{r} \ \sum_{l=0}^\infty \frac{y^L}{l!}  \ n_L \ \Big(\frac{\partial}{\partial u}\Big)^l \ N^{\alpha\beta}(u,\textbf{y})+\mathcal{O}(r^{-2}),
\end{equation*}
where we used, $\partial_L N^{\alpha\beta}\,=\, \frac{\partial}{\partial x^{a1}}  \cdots  \frac{\partial}{\partial x^{al}} N^{\alpha\beta}= \Big(\frac{\partial}{\partial u}\Big)^l \ N^{\alpha\beta} \ \frac{\partial u }{\partial x^{a1}}  \cdots  \frac{\partial u}{\partial x^{al}}=(-1)^l \ \Big(\frac{\partial}{\partial u}\Big)^l \ N^{\alpha\beta}  \ n_{L}$, and where $\frac{\partial r}{\partial x^a}=\frac{x^a}{r}=n_a$ is the $a$-th componant of the radial unit vector and $u=c\tau=x^0-r$. The first  effective radiative multipole moments are, $Q^{ab}= c^{-2} \int_{\mathcal{M}} N^{00} y^a y^b d\textbf{y}$, $Q^{abc}= c^{-2} \int_{\mathcal{M}} (N^{0a} y^b y^c+N^{0b} y^a y^c- N^{0c} y^a y^b ) \ d\textbf{y}$. The surface terms $P^{ab}$, $P^{abc}$ as well as the remaining two radiative multipole moments are presented in \cite{Alain1}.
\section{A brief review of the effective energy-momentum pseudotensor:}
\subsection{The effective matter pseudotensor:}
We saw in \cite{Alain1} that the three leading contributions of the potential operator function $w(h,\partial)$ are of the following post-Newtonian orders, $h^{\mu\nu}\partial_{\mu\nu}=\mathcal{O}(c^{-4})$, $\tilde{w}(h)=\frac{h}{2}-\frac{h^2}{8}+\frac{h^{\rho\sigma}h_{\rho\sigma}}{4}+\mathcal{O}(G^3)=-\frac{h^{00}}{2}+\mathcal{O}(c^{-4})$, $\tilde{w}(h)h^{\mu\nu}\partial_{\mu\nu}=\mathcal{O}(c^{-6})$, where $h=\eta_{\alpha\beta}h^{\alpha\beta}$. The leading contribution of $\mathcal{B}^{\alpha\beta}$ gives rise to the usual 1.5 PN matter contribution,
\begin{equation*}
\begin{split}
\mathcal{B}^{\alpha\beta}_1\,=\,\frac{\tau^{\alpha\beta}_m}{(-g)}\,&=\,\Big[\tau^{\alpha\beta}_m(c^{-3})+\mathcal{O}(c^{-4})\Big]\Big[1-h^{00}+h^{aa}-\frac{h^2}{2}+\cdots\Big]\,=\,\tau^{\alpha\beta}_{m}(c^{-3})-\tau^{\alpha\beta}_m(c^0) \ h^{00}+\mathcal{O}(c^{-4}),
\end{split}
\end{equation*}
where $\tau^{\alpha\beta}_m$ is the standard matter pseudotensor introduced previously \cite{PoissonWill,WillWiseman,PatiWill1,PatiWill2,Blanchet1}.
The second contribution, $\mathcal{B}^{\alpha\beta}_2= \mathcal{B}^{\alpha\beta}_{2a}+\mathcal{B}^{\alpha\beta}_{2b}+\mathcal{B}^{\alpha\beta}_{2c}+\mathcal{O}(c^{-4})$, is more advanced and can be decomposed, at the 1.5 PN order of accuracy, into three  different contributions,
\begin{equation*}
\begin{split}
\mathcal{B}_{2a}^{\alpha\beta}\,=&\,-\frac{\epsilon}{2} \frac{1}{1-\sigma} \sum_A m_A v_A^\alpha v^\beta_A \  \Big[h^{00} \big(\Delta \delta(\textbf{y}-\textbf{r}_A)\big)\Big], \ \quad \mathcal{B}_{2b}^{\alpha\beta}\,=\, -\frac{\epsilon}{2} \frac{\kappa}{(1-\sigma)^2} \sum_A m_A v_A^\alpha v^\beta_A \Delta \Big[h^{00} \big(\Delta \delta(\textbf{y}-\textbf{r}_A)\big)\Big],\\ \mathcal{B}_{2c}^{\alpha\beta}\,= &\,-\frac{\epsilon}{2} \sum_A m_A v_A^\alpha v^\beta_A \   \sum_{n=0}^{+\infty} \sigma^n \ \sum_{m=2}^{+\infty} \frac{[(n+1)\kappa]^m}{m!} \Delta^m \Big[h^{00} \big(\Delta \delta(\textbf{y}-\textbf{r}_A)\big)\Big],
\end{split}
\end{equation*} 
where we used $\tau_m^{\alpha\beta}=\sum_A m_A v_A^\alpha v_A^\beta \delta(\textbf{y}-\textbf{r}_A)+\mathcal{O}(c^{-2})$ with $(-g)=1+h^{00}+\mathcal{O}(c^{-4})$ and $\omega=-\frac{h^{00}}{2}+\mathcal{O}(c^{-4})$. We will come back to this decomposition of $\mathcal{B}_2^{\alpha\beta}$ when we work out the effective barycentre. We used, according to the definition \cite{Spallucci1} for an exponential differential operator, $e^{(n+1)\kappa\Delta}=1+[n+1]\kappa \Delta+\sum_{m=2}^{+\infty} \frac{[(n+1)\kappa]^m}{m!} \Delta^m$ and we assumed that the modulus of the dimensionless parameter is smaller than one $|\sigma|<1$, so that we have, $\sum_{n=0}^{+\infty} \sigma^n=\frac{1}{1-\sigma}$ and $\sum_{n=0}^{+\infty} [n+1]\sigma^n=\frac{1}{(1-\sigma)^2}$. It should be mentioned that this particular assumption will be discussed in the next chapter. The exponential differential operator acting on the product of the nonlocally modified energy-momentum tensor and the metric determinant is,
\begin{equation*}
\begin{split}
\sigma e^{\kappa \Box} \Big[\mathcal{T}^{\alpha\beta} (-g)\Big]
&=\, \sigma \sum_{s=0}^{\infty} \frac{(-\kappa)^s}{s!}\sum_{n=0}^{\infty} \frac{\kappa^n}{n!} \sum_{m=0}^{2n} \binom{2n}{m} \sum_{p=0}^{2s} \binom{2s}{p}\Big[\partial^{2s-p}_0\Big(\nabla^{2n-m} \mathcal{T}^{\alpha\beta}\Big)\Big] \Big[\partial_0^p\Big(\nabla^m (-g)\Big)\Big]\\
&=\,\sigma [1+h^{00}] \ e^{\kappa\Box}\mathcal{T}^{\alpha\beta}+ \sigma \sum_{n=1}^{\infty} \frac{\kappa^n}{n!} \sum_{m=1}^{2n} \binom{2n}{m} \Big[ \Big(\nabla^{2n-m} \mathcal{T}^{\alpha\beta}\Big) \Big] \Big[ \nabla^m h^{00}\Big]+\mathcal{O}(c^{-4}),
\end{split}
\end{equation*}
where we used the generalized Leibniz product rule \cite{Alain1}. Further relations that were used in the derivation of this result are, $(-g)=1+h^{00}+\mathcal{O}(c^{-4})$, $\partial_0=\mathcal{O}(c^{-1})$ and $\binom{2n}{0}=\binom{2s}{0}=1$. In order to work out $D^{\alpha\beta}$ to the required order of precision we remind from \cite{Alain1} that,
\begin{equation*}
\mathcal{T}^{\alpha\beta}\,=\,G(\Box) \ \mathcal{H}(w,\Box) \ T^{\alpha\beta}\,=\,\sum_{s=0}^{+\infty} \sigma^s \sum_{p=0}^{+\infty} \frac{(s\kappa)^p}{p!} \Delta^p \big[\sum_A m_A v_A^\alpha v^\beta_A \ \delta(\textbf{x}-\textbf{r}_A) \big]+\mathcal{O}(c^{-2}),
\end{equation*}
where $G(\Box)=G(\Delta)+\mathcal{O}(c^{-2})$, $|\sigma|<1$, $\mathcal{H}(w,\Box)=1+\mathcal{O}(c^{-2})$ and $T^{\alpha\beta}=\sum_A m_A v^\alpha_A v^\beta_A \ \delta(\textbf{x}-\textbf{r}_A)+\mathcal{O}(c^{-2})$.
\section{The effective orbital dynamics of a two-body system:}
\subsection{The effective Newtonian potential:}
It is known from the penultimate chapter that the retarded Green function is composed by the standard retarded Green function $G_r^{GR}=\frac{-1}{4\pi}\frac{\delta(x^0-|\textbf{x}-\textbf{y}|-y^0)}{|\textbf{x}-\textbf{y}|}$ together with a nonlocal correction term $G_r^{NL}=\frac{-1}{4\pi}\frac{1}{|\textbf{x}-\textbf{y}|}\sum_{n=1}^{+\infty}\frac{\sigma^n}{2\sqrt{\pi n\kappa}}e^{-\frac{(x^0-|\textbf{x}-\textbf{y}|-y^0)^2}{4n\kappa}}$. Moreover we obtained a formal solution for the modified wave equation, $h^{\alpha\beta}(x)= \frac{4 \ G}{c^4}  \int d\textbf{y} \ \frac{N^{\alpha\beta}(x^0-|\textbf{x}-\textbf{y}|,\textbf{y})}{|\textbf{x}-\textbf{y}|}$. The retarded effective pseudotensor can be decomposed into two independent pieces according to the two contributions coming from the retarded Green function, $N^{\alpha\beta}(x^0-|\textbf{x}-\textbf{y}|,\textbf{y})=\mathcal{D} N^{\alpha\beta}(y^0,\textbf{y})+\sum_{n=1}^{+\infty}\sigma^n \mathcal{E}_n N^{\alpha\beta}(y^0,\textbf{y})$, where we chose for clarity reasons to introduce the following two integral operators, $\mathcal{D}=\mathcal{E}_0= \int dy^0 \ \delta(x^0-|\textbf{x}-\textbf{y}|-y^0)$ and $\mathcal{E}_n=\int dy^0 \ \frac{1}{2\sqrt{\pi n \kappa}} e^{-\frac{(x^0-|\textbf{x}-\textbf{y}|-y^0)^2}{4n\kappa}}$. We saw in the previous chapter that the leading order contribution to the effective energy-momentum pseudotensor reduces to $N^{00}=\sum_A m_A  c^2 \ \delta(\textbf{x}-\textbf{r}_A)+\mathcal{O}(c^{-1})$. This together with the general solution for the gravitational potentials allows us to derive the modified Newtonian potential,
\begin{equation*}
h^{00}\,=\,\frac{4 G}{c^4}\int_{\mathcal{M}}d \textbf{y}\  \int_{-\infty}^{+\infty} dy^0\ \sum_{n=0}^{+\infty} \frac{\sigma^n}{2\sqrt{\pi n \kappa}}  \ \frac{\sum_A m_A c^2 \ \delta(\textbf{y}-\textbf{r}_A)}{|\textbf{x}-\textbf{y}|}\ e^{-\frac{(x^0-|\textbf{x}-\textbf{y}|-y^0)^2}{4n\kappa}}\,=\,\frac{4 G}{c^2}\sum_A \frac{\tilde{m}_A}{|\textbf{x}-\textbf{r}_A|}.
\end{equation*}
Here we used the Gauss integral, the geometric series for the dimensionless parameter $|\sigma|<1$ and we remind that $\tilde{m}_A=\frac{m_A}{1-\sigma}$ is the effective mass of body $A$. 
\subsection{The effective Kepler orbits:}
The force acting on body one due to the effective Newtonian potential produced by the mass of body two is given by $\textbf{F}_{12}(\sigma)=-m_1\nabla_1V_{12}(\sigma)=-G(\sigma)\nabla_1\frac{m_1 m_2}{|\textbf{r}_1-\textbf{r}_2|}=-G(\sigma)\frac{m_1m_2}{r^2}\textbf{e}_r $, where $\nabla_1$ is the gradient operator with respect to the coordinates $\textbf{r}_1=(x_1,y_1,z_1)$ of body one and $r=|\textbf{r}_1-\textbf{r}_2|$ is the relative separation between the two bodies. We remind that $G(\sigma)=\frac{G}{1-\sigma}$ is the effective leading order Newtonian coupling constant containing the dimensionless UV parameter $\sigma$. From this vectorial relation it is uncomplicated to derive the equations of motion of the two bodies outlined in the main part of this article. In order to solve the effective one-body equation of motion we remind that the radial and polar unit vectors can be decomposed in terms of cartesian coordinates, $\textbf{e}_r=[\cos(\phi), \sin(\phi),0]$ and $\textbf{e}_\phi=[-\sin(\phi),\cos(\phi),0]$ and that in this context the first and second temporal derivatives of the radial vector $\textbf{r}=r \ \textbf{e}_r$ are respectively $\dot{\textbf{r}}=\dot{r} \ \textbf{e}_r+r \ \dot{\phi} \ \textbf{e}_\phi$ and $\ddot{\textbf{r}}=\big(\ddot{r}-r\dot{\phi}^2\big) \ \textbf{e}_r+\frac{1}{r}\frac{d}{dt}\big(r^2\dot{\phi}\big) \ \textbf{e}_\phi$.
These results can be used to split the motion of the fictitious particle of reduced mass $\mu=m_1m_2/(m_1+m_2)$ into two separate equations describing its radial and angular dynamics separately. In this regard we obtain a relation which accounts for the system's conservation of angular momentum as well as a relation which describes the dynamics of the relative separation of the binary-system,
\begin{equation*}
\frac{\mu}{r}\frac{d}{dt}\Big(r^2\dot{\phi}\Big)=0 \ \Rightarrow \ \mu r^2\dot{\phi}\,=\,l, \quad\quad \mu \ \Big(\ddot{r}-r\dot{\phi}^2\Big)=-\frac{G(\sigma)\ m_1 m_2}{r^2} \ \Rightarrow \ \ddot{r}+\frac{G(\sigma)\ m}{ r^2}-\frac{h^2}{r^3}=0,
\end{equation*}
where $l=h \mu$ is the angular momentum. We wish to rewrite the second relation, which is a homogeneous nonlinear second-order in time differential equation, in terms of the polar angle $\phi$. Using the derivative-chain-rule, $\dot{r}=\frac{h}{r^2} \frac{dr}{d\phi}$, $\ddot{r}=-\frac{2 h^2}{r^5}\big(\frac{dr}{d\phi}\big)^2+\frac{h^2}{r^4}\frac{d^2r}{d\phi^2}$, we obtain for the radial equation, $\frac{1}{r^2}\frac{d^2 r}{d\phi^2}-\frac{2}{r^3} \big(\frac{dr}{d\phi}\big)^2-\frac{1}{r}+\frac{G(\sigma)\ m}{ h^2}=0$. In order to work out the solution it turns out to be a convenient strategy to introduce the new variable $r=\frac{1}{u}$ as well as, $\frac{dr}{d\phi}=-\frac{1}{u^2}\frac{du}{d\phi}$ and $\frac{d^2r}{d\phi^2}=\frac{2}{u^3}\big(\frac{d u}{d\phi}\big)^2-\frac{1}{u^2}\frac{d^2u}{d\phi^2}$ in order to obtain the rather simple linear second-order in $\phi$ differential equation, $\frac{d^2 u}{d\phi^2}+u=\frac{G(\sigma) \ m}{h^2}$. Using the substitution $v=u-\frac{G(\sigma)\ m}{h^2}$ the latter becomes even simpler and we obtain the equation $\frac{d^2 v}{d\phi^2}+v=0$. The solution to this equation is $v=u_o \cos(\phi-\phi_o)$, where $u_o$ and $\phi_o$ are the constants of integration. We finally obtain, $u(\phi)=\frac{G(\sigma)\ m}{h^2} \ \big(1+e\cos(\phi)\big)$ which rephrased in terms of the initial separation variable becomes $r(\phi)=\frac{p(\sigma)}{1+e\cos(\phi)}$, where $u_o=\frac{eG(\sigma)\ m}{h^2}$ and where $p(\sigma)=\frac{h^2}{G(\sigma)\ m}$ is a quantity of dimension length commonly known as the orbit's {\it semi-latus rectum} \cite{PoissonWill}. The energy conservation relation can be obtained from the radial equation by integrating over time, $\frac{d}{dt}\big(\frac{\dot{r}^2}{2}+\frac{h^2}{2 r^2}-\frac{G(\sigma) \ m}{r}\big)=0\Rightarrow\epsilon=\frac{\dot{r}^2}{2}+\frac{h^2}{2 r^2}-\frac{G(\sigma) \ m}{r}$, where $\epsilon=E/\mu$ is the reduced energy of the two-body system.

\subsection{The energy released by a binary-system:}
The effective retarded radiative quadrupole moment for a two-body system is,
\begin{equation*}
\begin{split}
Q^{ab}\,=&\,c^{-2}\int_\mathcal{M} d\textbf{y}\ \Big[\mathcal{D}N^{00}(y^0,\textbf{y})+\sum_{n=1}^{+\infty}\sigma^n \mathcal{E}_n N^{00}(y^0,\textbf{y})\Big] \ y^ay^b\\
=&\,\eta\ m \Big[r^a(c\tau)\ r^b(c\tau)+\sum_{n=1}^{+\infty}\frac{\sigma^n}{2\sqrt{\pi n \kappa}} \int dy^0 \ e^{-\frac{(x^0-|\textbf{x}|-y^0)^2}{4n\kappa}}\ r^a(y^0)\ r^b(y^0)\Big],
\end{split}
\end{equation*}
where $\eta=\frac{m_1 m_2}{m^2}$ is a dimensionless parameter, $m=m_1+m_2$ is the sum of the bodies' respective masses and $\textbf{r}=r\ [\cos(\omega t), \sin(\omega t),0]$ is the position vector of the fictitious body with reduced mass $\mu=\eta m$ orbiting the binary-system's barycentre with a certain orbital velocity $\omega$. We remind that for a far distant source ($|\textbf{x}-\textbf{y}|\approx |\textbf{x}|$) the retardation integrals become in good approximation, $\mathcal{D}=\mathcal{E}_0= \int dy^0\ \delta(x^0-|\textbf{x}|-y^0)$ and $\mathcal{E}_n=\frac{1}{2\sqrt{\pi n \kappa}}\int dy^0 e^{-\frac{(x^0-|\textbf{x}|-y^0)^2}{4n \kappa}}$. We also remind that in contrary to the previous chapters $r=|\textbf{r}_1-\textbf{r}_2|$ is the relative separation of the two bodies with respective masses $m_1$ and $m_2$ and should not be confused with $|\textbf{x}|$ which is the distance from the source to the observer (detector). Using the relation between the quadrupole moment and the radiative quadrupole moment, $Q^{\langle ab\rangle}=Q^{ab}-\frac{\delta^{ab}}{3} Q^{cc}$, we can easily determine the non-vanishing components of the former as well as its third temporal derivatives \cite{PoissonWill,Maggiore1,Schutz1},

\begin{minipage}{0.5\textwidth}
\begin{equation*}
\begin{split}
Q^{\langle 11 \rangle}\,=&\,\frac{\eta \ \tilde{m}}{6}\ r^2\ \big[1 +3 Z(\sigma, \kappa, \omega)\ \cos(2\omega \tau)\big],\\
Q^{\langle 22 \rangle}\,=&\,\frac{\eta \ \tilde{m}}{6}\ r^2\ \big[1 -3 Z(\sigma, \kappa, \omega)\ \cos(2\omega \tau)\big],\\
Q^{\langle 12 \rangle}\,=&\,Q^{\langle 21 \rangle}\,=\,\frac{\eta \ \tilde{m}}{2}\ r^2\ Z(\sigma, \kappa, \omega) \ \sin(2\omega \tau), 
\end{split}
\end{equation*}
\end{minipage}
\begin{minipage}{0.5\textwidth}
\begin{equation*}
\begin{split}
\dddot{Q}^{\langle 11\rangle}\,=&\,+4\eta \ \tilde{m}\ r^2 \ Z(\sigma,\kappa,\omega) \ \omega^3 \ \sin(2\omega \tau),\\
\dddot{Q}^{\langle 22\rangle}\,=&\,-4\eta \ \tilde{m}\ r^2 \ Z(\sigma,\kappa,\omega) \ \omega^3 \ \sin(2\omega \tau),\\
\dddot{Q}^{\langle 12\rangle}\,=&\,-4\eta \ \tilde{m}\ r^2 \ Z(\sigma,\kappa,\omega) \ \omega^3 \ \cos(2\omega \tau).
\end{split}
\end{equation*}
\end{minipage}

$\newline$
where $Z(\sigma,\kappa,\omega)=(1-\sigma)/(1-\sigma e^{-4\kappa\frac{\omega^2}{c^2}})$ is a dimensionless function depending on the UV parameters as well as on the orbital velocity of the effective body with reduced mass $\mu=\eta m$. As the generic form of the quadrupole formula ($P=\frac{32 G}{5 c^5}\langle \dddot{Q}^{ab} \dddot{Q}^{ab}\rangle$) itself is not affected by the nonlocal modification to the Einstein field equations, we can easily deduce the amount of energy released by a binary-system using the results outlined above \cite{PoissonWill,Maggiore1,Buonanno1}.

\section{The effective barycentre:}
\subsection{The matter contribution:}
We provide additional computational steps in order to outline how the results presented in the main text were derived. We will therefore focus in this appendix-section mainly on technical issues and refer the reader for the notations and conceptual points to the main part of this article. In analogy to \cite{Alain1}, where we worked out the total near-zone mass, we observe that the first contribution to the center-of-mass of a many-body system essentially gives rise to the usual 1.5 post-Newtonian term,
\begin{equation*}
\begin{split}
M_{\mathcal{B}_1+\mathcal{B}h}\boldsymbol{R}\,=\,c^{-2}\int_{\mathcal{M}} d\textbf{x} \ \textbf{x} \ \big[\mathcal{B}^{00}_1+\mathcal{B}h^{00}\big]
=&\,c^{-2}\int_\mathcal{M}d\textbf{x}\ \textbf{x}\ \sum_A m_A \Big[1+\frac{v^2}{2c^2}+3\frac{V}{c^2}\Big]\delta(\textbf{x}-\textbf{r}_A) +\mathcal{O}(c^{-4})\\
=&\,[M_m\boldsymbol{R}]^{GR}+3\sigma\frac{G}{c^2}\sum_A\sum_{B\neq A}\frac{m_A\tilde{m}_B}{r_{AB}} \textbf{r}_A+\mathcal{O}(c^{-4}).
\end{split}
\end{equation*}
For reasons of completeness we remind that $\mathcal{B}_1^{00}+\mathcal{B}^{00}h^{00}=\sum_Am_A\big[1+c^{-2}\big(\frac{\textbf{v}^2}{2}+3V\big)\big]$, where $\mathcal{B}_1^{00}=\tau^{00}_m(c^{-3})-\tau_m^{00}h^{00}+\mathcal{O}(c^{-4})$, $\mathcal{B}^{00}h^{00}=\tau_m^{00}(c^{0})h^{00}+\mathcal{O}(c^{-4})$, $\tau^{00}_m=\sum_Am_A\big[1+\frac{\textbf{v}^2}{2c^2}+3\frac{V}{c^2}\big]\delta(\textbf{x}-\textbf{r}_A)$ and we used $\frac{1}{1-\sigma}=\sum_{n=0}^{+\infty}\sigma^n$ for $|\sigma|<1$. We also like to recall that $V=(1-\sigma)^{-1}U=G\sum_{A}\frac{\tilde{m}_A}{|\textbf{x}-\textbf{r}_A|}$ is the effective Newtonian potential, where $\tilde{m}_A=(1-\sigma)^{-1}m_A$ is the effective mass of body $A$ and $\frac{\delta(\textbf{x}-\textbf{r}_A)}{|\textbf{x}-\mathbf{r}_A|}\equiv 0$ is the standard regularization prescription for point masses \cite{PoissonWill, Blanchet1,Blanchet2}.
For the computation of the barycentre only the first term of $\mathcal{B}^{\alpha\beta}_2=\mathcal{B}_{2a}^{\alpha\beta}+\mathcal{B}_{2b}^{\alpha\beta}+\mathcal{B}_{2c}^{\alpha\beta}+\mathcal{O}(c^{-4})$, where we remind, $\mathcal{B}_{2a}^{\alpha\beta}=-\frac{\epsilon}{2} \frac{1}{1-\sigma} \sum_A m_A v_A^\alpha v^\beta_A \  \big[h^{00} \big(\Delta \delta(\textbf{y}-\textbf{r}_A)\big)\big]$, the precise expression for the latter,
\begin{equation*}
\begin{split}
M_{\mathcal{B}_{2}} \boldsymbol{R}\,=\, c^{-2}\int_{\mathcal{M}} d\textbf{x} \ \textbf{x} \ \mathcal{B}^{00}_{2}\,=&\,\frac{-\epsilon}{2(1-\sigma)}\Big\{\textbf{S}_1-\textbf{S}_2+\sum_Am_A \int_{\mathcal{M}}d\textbf{x} \ \Delta\big[h^{00}\textbf{x}\big]\delta(\textbf{x}-\textbf{r}_A)\Big\}\\
=\,&  \frac{4 \epsilon}{1-\sigma}\frac{G}{c^2}\sum_A\sum_{B\neq A} \frac{m_A\tilde{m}_B}{r_{AB}^3}\textbf{r}_{AB},
\end{split}
\end{equation*}
where $\textbf{S}_1=\oint_\mathcal{\partial M}dS^p[h^{00}\textbf{x}][\partial_p\delta(\textbf{x}-\textbf{r}_A)]$ and $\textbf{S}_2=\oint_\mathcal{\partial M}dS^p\partial_p[h^{00}\textbf{x}]\delta(\textbf{x}-\textbf{r}_A)$ are surface terms originating from multiple partial integration. In addition we have $\Delta [h^{00}\textbf{x}]=G\sum_B\tilde{m}_B\big\{[\Delta\textbf{x}] \frac{1}{|\textbf{x}-\textbf{r}_B|}+2[\partial_p \textbf{x}] \ \big[\partial^p \frac{1}{|\textbf{x}-\textbf{r}_B|}\big]+\textbf{x} \ \big[\Delta\frac{1}{|\textbf{x}-\textbf{r}_B|}\big]\big\}$, where $\Delta\textbf{x}=\textbf{0}$, $\partial_p \textbf{x}\partial^p \frac{1}{|\textbf{x}-\textbf{r}_B|}=\frac{\textbf{r}_B-\textbf{x}}{|\textbf{x}-\textbf{r}_A|^3}$ and $\Delta \frac{1}{|\textbf{x}-\textbf{r}_B|}=-4\pi \delta(\textbf{x}-\textbf{r}_A)$ \cite{PoissonWill,Jackson, LandauLifshitz}. Surface terms can be discarded in the near-zone as they are proportional to $\delta(\mathcal{R}-|\textbf{r}_A|)$ or its derivative, so that we have, $\textbf{S}_2=\oint_\mathcal{\partial M}dS^p\partial_p[h^{00}\textbf{x}]\delta(\textbf{x}-\textbf{r}_A)\propto \delta(\mathcal{R}-|\textbf{r}_A|)=0$. The last term in $\Delta[h^{00}\textbf{x}]$ gives rise to an integral which we already encountered in \cite{Alain1}, $\int_\mathcal{M}d\textbf{x}[\Delta h^{00}]\textbf{x}=-4\pi \textbf{r}_A\delta(\textbf{r}_B-\textbf{r}_A)=0$. The additional two contributions $\mathcal{B}^{00}_{2b}$ and $\mathcal{B}^{00}_{2c}$ do not contribute because they lead, after multiple partial integration, to terms of the following kind, $\sum_A\sum_{B\neq A} m_Am_B \ \int_{\mathcal{M}} d\textbf{x} \  \delta(\textbf{x}-\textbf{r}_A) \ \nabla^m \delta(\textbf{x}-\textbf{r}_B)\,=\,0, \quad \forall m\in \mathbb{N}$. To illustrate this we will have a closer look at the integral,
\begin{equation*}
\begin{split}
c^{-2}\int_\mathcal{M}d\textbf{x}\mathcal{B}^{00}_{2b}\textbf{x}\,=&\,\frac{-\epsilon\kappa}{2(1-\sigma)^2}\sum_Am_A\int_\mathcal{M}d\textbf{x}\ \textbf{x}\Big\{[\Delta h^{00}]\ [\Delta \delta(\textbf{x}-\textbf{r}_A)]+2[\partial_ph^{00}]\ [\Delta\partial^p\delta(\textbf{x}-\textbf{r}_A)]+h^{00}\ [\Delta^2\delta(\textbf{x}-\textbf{r}_A)]\Big\},
\end{split}
\end{equation*}
where $\Delta^2=\partial_p\partial^p\partial_m\partial^m$. The first term in this relation is proportional to $\int_\mathcal{M}d\textbf{x}\ \textbf{x}\ \delta(\textbf{x}-\textbf{r}_A)\ \delta(\textbf{x}-\textbf{r}_B)=0$. The second term is less straightforward, but nicely outlines the different kind of terms that have to be examined.
For clarity reasons we will omit the constant prefactor,
\begin{equation*}
\begin{split}
\int_\mathcal{M}d\textbf{x}\ \textbf{x} \ [\partial_p h^{00}]\ [\partial^p\partial_m\partial^m \delta(\textbf{x}-\textbf{r}_A)]=\textbf{S}_3-\textbf{S}_4+\textbf{S}_5-\int_\mathcal{M}d\textbf{x}\ \partial^m\partial_m\big[\partial_p[\textbf{x}(\partial^ph^{00})]\big]\ \delta(\textbf{x}-\textbf{r}_A),
\end{split}
\end{equation*}
where $\textbf{S}_3=\oint_{\partial\mathcal{M}}dS^p[\partial_ph^{00}]\ [\Delta\delta(\textbf{x}-\textbf{r}_A)] \ \textbf{x}$, $\textbf{S}_4=\oint_{\partial\mathcal{M}}dS^m\ \big[\partial_p[\textbf{x}(\partial^p h^{00})]\big]\ [\partial_m \delta(\textbf{x}-\textbf{r}_A)]$ and $\textbf{S}_5=\oint_{\partial\mathcal{M}}dS^m\ \partial_m\big[\partial_p[\textbf{x}(\partial^p h^{00})]\big]\ \delta(\textbf{x}-\textbf{r}_A)$. Surface terms vanish in the near-zone, so that we only need to consider the last integral containing the term,
\begin{equation*}
\begin{split}
\partial^m\partial_m\big[\partial_p[\textbf{x}(\partial^p h^{00})]\big]\,=\,&[\partial^m \partial_m(\partial_p\textbf{x})]\ (\partial^ph^{00})+2[\partial^m(\partial_p\textbf{x})]\ [\partial_m \partial^p h^{00}]+[\partial_p\textbf{x}]\ [\partial^m\partial_m\partial^p h^{00}]+\\
&[\partial^m\partial_m\textbf{x}]\ [\partial_p\partial^ph^{00}]+2[\partial^m\textbf{x}]\ \big[\partial_m[\partial_p\partial^p h^{00}]\big]+\textbf{x}\ [\partial^m\partial_m\partial_p\partial^p h^{00}].
\end{split}
\end{equation*}
As we have that $\textbf{x}=x^k\textbf{e}_k$, $\partial_p\textbf{x}=\textbf{e}_p$ and $\partial_m\partial_p\textbf{x}=\textbf{0}$, this last expression above gives only rise to terms proportional to $\nabla\delta(\textbf{x}-\textbf{r}_B)$ and $\nabla^2\delta(\textbf{x}-\textbf{r}_B)$. This result brings us back to the claim made earlier where we stated that the integral $c^{-2}\int_\mathcal{M}d\textbf{x}\mathcal{B}^{00}_{2b}\textbf{x}$ will be proportional to terms that have the following shape $\sum_A\sum_{B\neq A} m_Am_B \ \int_{\mathcal{M}} d\textbf{x} \  \delta(\textbf{x}-\textbf{r}_A) \ \nabla^m \delta(\textbf{x}-\textbf{r}_B)\,=\,0, \quad \forall m\in \mathbb{N}$. A very similar reasoning applies to the integral containing the $h^{00} \ [\Delta^2\delta(\textbf{x}-\textbf{r}_A)]$ piece as well as to the integral containing the $\mathcal{B}_{2c}^{00}$ contribution. This finally allows us to write, $M_{\mathcal{B}_{2}} \boldsymbol{R}\,=\,  \frac{4 \epsilon}{1-\sigma}\frac{G}{c^2}\sum_A\sum_{B\neq A} \frac{m_A\tilde{m}_B}{r_{AB}^3}\textbf{r}_{AB}+\mathcal{O}(c^{-4})$ where it should be noticed that this term eventually vanishes as well because it is proportional to $\sum_A\sum_{B\neq A}\frac{\textbf{r}_{AB}}{r^3_{AB}}=\textbf{0}$ ($\textbf{r}_{AB}=-\textbf{r}_{BA}$, $r_{AB}=|\textbf{r}_{AB}|=r_{BA}$), so that we finally find $ M_{\mathcal{B}_{2}} \boldsymbol{R}=\boldsymbol{0}$ at the 1.5 post-Newtonian order of accuracy. The next contribution can be computed in a very similar way,
\begin{equation*}
M_{D}\boldsymbol{R}\,=\,c^{-2} \int_{\mathcal{M}} d\textbf{x} \ D^{00} \ \textbf{x}\,=\, \int_{\mathcal{M}} d\textbf{x} \ \mathcal{S}(\sigma,\kappa)\big[\sum_A m_A\nabla^{2p+2n-m} \delta(\textbf{x}-\textbf{r}_A) \big] \ \big[\nabla^mh^{00}\big]+\mathcal{O}(c^{-4}),
\end{equation*}
where $\mathcal{S}(\sigma,\kappa)=\sum_{n=1}^{\infty} \frac{\kappa^n}{n!} \sum_{m=1}^{2n} \dbinom{2n}{m} \sum_{s=0}^{+\infty} \sigma^s  \sum_{p=0}^{+\infty} \frac{(s\kappa)^p}{p!}$ and $\dbinom{2n}{m}=\frac{(2n)!}{(2n-m)!m!}$ is the binomial coefficient. Only the lowest derivative terms contribute to the final result. Higher order derivative terms will lead, after partial integration(s), to contributions that are proportional either to one of the following two relations or to both at the same time, $\sum_A\sum_{B\neq A}m_A \tilde{m}_B \ \nabla^q\delta(\textbf{r}_A-\textbf{r}_B)=0,\ \forall q\geq 0$, $\nabla^k \textbf{x}=0, \ \forall k\geq 2$. The lowest derivative term acting on the Dirac distribution is obtained for $n=1, \ m=1, \ p=0$,
\begin{gather*}
M_D\boldsymbol{R}\,=\,\frac{\kappa}{1-\sigma}\sum_Am_A\int_\mathcal{M}d\textbf{x}\ \Big\{\dbinom{2}{1}[\partial_p\delta(\textbf{x}-\textbf{r}_A)]\ [\partial^ph^{00}]\ \textbf{x}+\dbinom{2}{2}\delta(\textbf{x}-\textbf{r}_A)\ [\Delta h^{00}]\ \textbf{x}\Big\}+\mathcal{O}(c^{-4})\\
=\, \frac{\kappa}{1-\sigma}\sum_Am_A\Big\{2\textbf{S}_5+\frac{8G}{c^2}\sum_{B\neq A}\frac{m_A\tilde{m}_B}{r^3_{AB}}\textbf{r}_{AB}\Big\}+\mathcal{O}(c^{-4}),
\end{gather*}
where $\textbf{S}_5$ results from partial integration and has already been outlined above. Here again $\Delta h^{00}\propto \delta(\textbf{x}-\textbf{r}_B)$, $[\partial^ph^{00}] \ [\partial_p \textbf{x}]=\frac{\textbf{r}_B-\textbf{x}}{|\textbf{r}_A-\textbf{x}|}$ and the usual regularization prescription $\frac{\delta(\textbf{x}-\textbf{r}_A)}{|\textbf{x}-\textbf{r}_A|}$ were used.
We have seen previously that surface terms, coming from partial integration, can be freely discarded in the near-zone domain $\mathcal{M}:|\textbf{x}|<\mathcal{R}$ as they are proportional to $\delta(\mathcal{R}-|\textbf{r}_A|)=0$. In analogy to the previous contribution the result is proportional to $\textbf{r}_{AB}$ and therefore we finally obtain $M_{D}\boldsymbol{R}=\boldsymbol{0}$ at the 1.5 post-Newtonian order of accuracy.
\subsection{The field contribution:}
This appendix-subsection is devoted to the determination of the near-zone field contribution to the centre-of-mass, $M_{N_{LL}}\boldsymbol{R}=c^{-2}\int_{\mathcal{M}}d\textbf{x} \ N^{00}_{LL} \ \textbf{x}$, where we recall the effective Landau-Lifshitz pseudotensor \cite{Alain1}, $N^{00}_{LL}=(1-\sigma)\tau^{00}_{LL}-\epsilon \Delta \tau^{00}_{LL}-\sigma \sum_{m=2}^{+\infty}\frac{\kappa^m}{m!}\Delta^m\tau^{00}_{LL}$, together with $\tau^{00}_{LL}=-\frac{7}{8\pi G}\partial_pV\partial^pV$. In accordance to the main text of this article we will review the three different contributions of $M_{N_{LL}}\boldsymbol{R}$ one after the other. The first integral gives rise to a non-vanishing contribution,
\begin{equation*}
\begin{split}
c^{-2}\int_{\mathcal{M}} d\textbf{x} \ \tau^{00}_{LL} \ \textbf{x}\,=&\,-\frac{7}{8c^2\pi G}\int_{\mathcal{M}} d\textbf{x} \  \Big[\partial_p(V\partial^pV)-V\Delta V\Big] \ \textbf{x} \\
=&\,-\frac{7}{8c^2\pi G}\int_{\mathcal{M}} d\textbf{x} \  \Big[\partial_p(V\partial^pV)+4\pi G \sum_A\tilde{m}_A\ \delta(\textbf{x}-\textbf{r}_A)V\Big] \ \textbf{x}\\
=&\,f_d(\mathcal{R})-\frac{7G}{16c^2} \sum_A\sum_{B\neq A}  \frac{\tilde{m}_A \tilde{m}_B}{r_{AB}}\textbf{r}_{AB}-\frac{7G}{2c^2} \sum_A\sum_{B\neq A}  \frac{\tilde{m}_A \tilde{m}_B}{r_{AB}}\textbf{r}_A,
\end{split}
\end{equation*}
where $r_{AB}=|\textbf{r}_A-\textbf{r}_B|$ is the relative separation between body $A$ and body $B$ and $\Delta |\textbf{x}-\textbf{r}_A|^{-1}=-4\pi \delta(\textbf{x}-\textbf{r}_A)$ \cite{PoissonWill,Jackson,LandauLifshitz}. To arrive at this result the following relation was used $[\partial_p(V\partial^pV)]\ \textbf{x}=\partial_p[(V\partial^pV)\ \textbf{x}]-V\partial^kV \textbf{e}_k$, where $\textbf{x}=x^k\textbf{e}_k$ is a position vector. The first term on the right-hand-side of this equation gives rise to an $\mathcal{R}$-depending quantity only,
\begin{equation*}
\begin{split}
-\frac{8c^2\pi G}{7}f_d(\mathcal{R})\,=\,\int_\mathcal{M}d\textbf{x}\ \partial_p\big[(V\partial^pV)\ \textbf{x}\big]\,=&\,\oint_{\partial\mathcal{M}}dS^p\ (V\partial_pV)\ \textbf{x}\\
=&\,G^2\sum_{A,B}\tilde{m}_A\tilde{m}_B\oint_{\partial\mathcal{M}}dS^p\ \frac{ \textbf{x}}{|\textbf{x}-\textbf{r}_A|}\frac{(r_B-x)_p}{|\textbf{x}-\textbf{r}_B|^3}\\
=&\,G^2\sum_{A,B}\tilde{m}_A\tilde{m}_B\int d\Omega\Big[N^pN^kN_p+N^pN_p \frac{r^k_B}{\mathcal{R}}\Big]\textbf{e}_k+\mathcal{O}(r_{AB}/\mathcal{R})\propto\mathcal{R}^{-1},
\end{split}
\end{equation*}
where the Gauss-Ostrogradsky-theorem was used as well as the angular-integrals $\int d\Omega\ N^p N^k N_p=0$ and $\int d\Omega\ N^p N_p=4\pi$ over the surface of a sphere \cite{PoissonWill}. Moreover the substitution $\textbf{y}=\textbf{x}-\textbf{r}_B$ was introduced, so that $\textbf{x}-\textbf{r}_A=\textbf{y}-\textbf{r}_{AB}$ and where $\textbf{r}_{AB}=\textbf{r}_A-\textbf{r}_B$ is the relative separation vector between body $A$ and body $B$. Taking into account that the modulus of the bodies' position vectors inside the near-zone domain is much smaller than the near-zone radius ($r_A\ll \mathcal{R}$) we have that $|\textbf{y}-\textbf{r}_A|<\mathcal{R}\leadsto y/\mathcal{R}<1+\mathcal{O}(r_A/\mathcal{R})$, which implies for the near-zone boundary $\partial\mathcal{M}\approx\partial\mathcal{M}_y$, where $\mathcal{M}_y:\ |\textbf{y}|<\mathcal{R}$ \cite{PoissonWill,WillWiseman,PatiWill1}. $\textbf{N}=\textbf{y}/y$ is a unit vector normal to the surface of the boundary defined by $y\approx\mathcal{R}$, the surface element $dS^p=\mathcal{R}^2 N^pd\Omega$ and $\frac{1}{|\textbf{y}-\textbf{r}_{AB}|}\big|_{|\textbf{y}|=\mathcal{R}}=\mathcal{R}^{-1}+\mathcal{O}(r_{AB}/\mathcal{R})$. The second term of $[\partial_p(V\partial^pV)]\textbf{x}$ is more demanding and therefore requires a more detailed investigation,
\begin{equation*}
\begin{split}
\int_\mathcal{M}d\textbf{x} \ V\partial^k V\textbf{e}_k\,=\,G^2\sum_{A,B}\tilde{m}_A\tilde{m}_B\ \Big\{\int_{\mathcal{M}_y}d\textbf{y}\ \frac{1}{|\textbf{y}-\textbf{r}_{AB}|}\frac{\textbf{y}}{y^3}-\oint_{\partial\mathcal{M}_y}\frac{1}{|\textbf{y}-\textbf{r}_{AB}|}\frac{\textbf{y}}{y^3} \ \textbf{r}\cdot d\textbf{S}\Big\},
\end{split}
\end{equation*} 
where we remind that $-(V\partial^kV)\textbf{e}_k=G^2\sum_{A,B}\frac{\tilde{m}_A\tilde{m}_B}{|\textbf{x}-\textbf{r}_A|}\frac{\textbf{x}-\textbf{r}_B}{|\textbf{x}-\textbf{r}_B|^3}$. We essentially follow the integration techniques presented in \cite{PoissonWill}, $\int_{\mathcal{M}}d\textbf{y} \ f(\textbf{y})=\int_{\mathcal{M}_y}d\textbf{y} \ f(\textbf{y})-\oint_{\partial\mathcal{M}_y}\textbf{r}\cdot d\textbf{S} \ f(\textbf{y})+\cdots$, where the substitution $\textbf{y}:=\textbf{x}-\textbf{r}_B$ and a translation of the domain of integration were performed $\mathcal{M}\leadsto \mathcal{M}_y+\partial\mathcal{M}_y$. We also recall that the near-zone domain $\mathcal{M}$ is defined by $|\textbf{x}|<\mathcal{R}$, that $\mathcal{M}_y$ is defined by $|\textbf{y}|<\mathcal{R}$ and $\partial\mathcal{M}_y$ is its boundary at $y=\mathcal{R}$. It is clear that the surface integral is smaller than the volume integral by a factor $r/\mathcal{R}$ and that the neglected terms are even smaller. For the first of the two integrals outlined above we obtain,
\begin{equation*}
\sum_{A,B\neq A}\ \int_{\mathcal{M}_y}d\textbf{y}\ \frac{\tilde{m}_A\tilde{m}_B}{|\textbf{y}-\textbf{r}_{AB}|}\frac{\textbf{y}}{y^3}\,=\,\frac{1}{2}\sum_{A,B\neq A}\frac{\tilde{m}_A\tilde{m}_B}{r_{AB}}\textbf{r}_{AB},
\end{equation*}
where we used the addition theorem for spherical harmonics $\frac{1}{|\textbf{y}-\textbf{r}_{AB}|}=\sum_{l=0}^{+\infty}\sum_{m=-l}^l\frac{4\pi}{2l+1}\frac{r^l_<}{r_>^{l+1}}Y^*_{lm}(\textbf{n}_{AB})Y^{lm}(\textbf{N})$, in which $r_<=\text{min}(y,r_{AB})$, $r_>=\text{max}(y,r_{AB})$ and $\textbf{n}_{AB}=\textbf{r}_{AB}/r_{AB}$ is a unit vector pointing from body $B$ towards body $A$ \cite{PoissonWill, Jackson,LandauLifshitz}. In addition the following two important identities $\sum_{m=-l}^lY^*_{lm}(\textbf{n}_{AB})\int Y_{lm}(\textbf{N})N^{\langle L'\rangle}d\Omega=\delta_{ll'}n^{\langle L\rangle}_{AB}$ and $\int_0^\mathcal{R} dy\ y^n\frac{r^l_<}{r^{l+1}_>}=\frac{2l+1}{(l-n)(l+n+1)}|\textbf{r}_{AB}|^n$ were employed \cite{PoissonWill} in the derivation of the integral outlined above. It should however be noticed that terms proportional to $\sum_A\sum_{B\neq A}  \frac{\tilde{m}_A \tilde{m}_B}{r_{AB}}\textbf{r}_{AB}=0$ vanish for a N-body system because we have that $\textbf{r}_{AB}=-\textbf{r}_{BA}$. Employing the techniques displayed in the beginning of this subsection, we find that the surface integral is proportional to the near-zone cut-off radius $\oint_{\partial\mathcal{M}_y}\frac{1}{|\textbf{y}-\textbf{r}_{AB}|}\frac{\textbf{y}}{y^3} \ \textbf{r}\cdot d\textbf{S}\,\propto \, \mathcal{R}^{-1}$. Discarding $\mathcal{R}$-depending contributions (they will eventually cancel out with terms coming from the wave-zone \cite{WillWiseman, PatiWill1,PatiWill2}), the integral finally reduces to $c^{-2}\int_\mathcal{M}d\textbf{x}\ \tau^{00}_{LL}\ \textbf{x}=-\frac{7G}{2c^2}\sum_{A,B\neq A}\frac{\tilde{m}_A\tilde{m}_B}{|\textbf{r}_A-\textbf{r}_B|}\textbf{r}_A$. Here again the standard regularization prescription \cite{PoissonWill,Blanchet1,Blanchet2}, introduced in the previous subsection, was used in order to avoid singularities. This allows us to come to the position-integral over the second term in the effective Landau-Lifshitz pseudotensor,
\begin{equation*}
 c^{-2}\int_{\mathcal{M}}d\textbf{x} \ \textbf{x} \  \Delta \tau^{00}_{LL}\,=\,-\frac{7}{4\pi c^2 G}\int_\mathcal{M}d\textbf{x}\ \textbf{x} \ \big\{ (\partial_p\partial_m\partial^mV)\ \partial_pV+(\partial_m\partial_pV)\ (\partial^m\partial^pV)\big\},
\end{equation*}
where $\Delta=\partial_m\partial^m$ is the Laplace-operator. We will see that the first of these two terms vanishes after integration over the near-zone domain,
\begin{equation*}
c^{-2} \int_{\mathcal{M}}d\textbf{x} \ \textbf{x} \ (\partial_p\Delta V) \ \partial^p V\,=\,-\pi G \sum_A\tilde{m}_A\Big\{\textbf{S}_6-\int_{\mathcal{M}}d\textbf{x}\ \partial_p[\textbf{x} \partial^ph^{00}] \ \delta(\textbf{x}-\textbf{r}_A)\Big\}\,\propto\,\sum_A\sum_{B\neq A}  \frac{\tilde{m}_A\tilde{m}_B}{r^3_{AB}}\textbf{r}_{AB}\,=\,\textbf{0},
\end{equation*}
where $\textbf{S}_6=\oint_{\mathcal{M}}dS^p\ \big[\textbf{x}\ \partial_ph^{00}\big]\ \delta(\textbf{x}-\textbf{r}_A)$ is a surface term proportional to $\delta(\mathcal{R}-r_B)$ coming from partial integration. The latter disappears in the near-zone and we additionally used $(\partial_p\textbf{x})\ (r_B-x)^p\ |\textbf{x}-\textbf{r}_B|^{-3}=(\textbf{r}_B-\textbf{x})\ |\textbf{x}-\textbf{r}_B|^3$ as well as $\Delta |\textbf{x}-\textbf{r}_A|^{-1}=-4\pi\ \delta(\textbf{x}-\textbf{r}_A)$ to work out the integral extending over the three-dimensional near-zone domain $\mathcal{M}:|\textbf{x}|<\mathcal{R}$. The second contribution of the integral outlined above splits into two different terms which will be reviewed separately in the remaining part of this appendix-section,
\begin{gather*}
\int_{\mathcal{M}}d\textbf{x}  \ (\partial_m\partial_pV) \  (\partial^m\partial^pV)\ \textbf{x}\,=\\
G^2\sum_{A,B\neq A} \tilde{m}_A \tilde{m}_B \ \int_{\mathcal{M}}d\textbf{x} \ \Big\{ \frac{-3}{|\textbf{x}-\textbf{r}_A|^3\ |\textbf{x}-\textbf{r}_B|^3}+ \frac{9\ (\textbf{x}-\textbf{r}_A)_m(\textbf{x}-\textbf{r}_A)_p\ (\textbf{x}-\textbf{r}_B)^m(\textbf{x}-\textbf{r}_B)^p}{|\textbf{x}-\textbf{r}_A|^5\ |\textbf{x}-\textbf{r}_B|^5}\Big\} \ \textbf{x}.
\end{gather*}
This result was obtained by employing the following relation, $\partial_m\partial_p\  |\textbf{x}-\textbf{r}_A|^{-1}=-\delta_{mp}\ |\textbf{x}-\textbf{r}_A|^{-3}+3\ (x-r_A)_p\ (x-r_A)_m\ |\textbf{x}-\textbf{r}_A|^{-5}$. In order to evaluate these integrals we will rely on the same integration techniques that were already introduced previously in this appendix-section. We perform the substitution $\textbf{y}:=\textbf{x}-\textbf{r}_B$ followed by a translation of the domain of integration  $\mathcal{M}\leadsto \mathcal{M}_y+\partial\mathcal{M}_y$. This allows us to reformulate the initial near-zone integral $\int_{\mathcal{M}}d\textbf{y} \ f(\textbf{y})=\int_{\mathcal{M}_y}d\textbf{y} \ f(\textbf{y})-\oint_{\partial\mathcal{M}_y}\textbf{r}\cdot d\textbf{S} \ f(\textbf{y})+\cdots$ in terms of two integrals, one extending over the domain $\mathcal{M}_y:|\textbf{y}|<\mathcal{R}$ and the other over its boundary $\partial\mathcal{M}_y:|\textbf{y}|=\mathcal{R}$. This brings us back to the initial integral above,
\begin{equation*}
\int_{\mathcal{M}}d\textbf{x} \  \frac{1}{|\textbf{x}-\textbf{r}_A|^3} \ \ \frac{1}{|\textbf{x}-\textbf{r}_B|^3} \ \textbf{x}=\,\int_{\mathcal{M}_y}d\textbf{y} \ \frac{1}{|\textbf{y}-\textbf{r}_{AB}|^3} \frac{\textbf{y}+\textbf{r}_B}{y^3}-\oint_{\partial\mathcal{M}_y}\textbf{r}\cdot d\textbf{S} \ \frac{1}{|\textbf{y}-\textbf{r}_{AB}|^3}\frac{\textbf{y}+\textbf{r}_B}{y^3}+\cdots.
\end{equation*}
The volume integral can be evaluated by making use of the addition theorem for spherical harmonics $\frac{1}{|\textbf{y}-\textbf{r}_{AB}|^m}=\frac{1}{r^{m-1}_>|\textbf{y}-\textbf{r}_{AB}|}+\mathcal{O}(\frac{r_<}{r_>})=\sum_{l=0}^{+\infty}\sum_{m=-l}^l\frac{4\pi}{2l+1}\frac{r^l_<}{r_>^{l+m}}Y^*_{lm}(\textbf{n}_{AB})Y^{lm}(\textbf{N})+\mathcal{O}(\frac{r_<}{r_>})$, by the relation, $\sum_{m=-l}^lY^*_{lm}(\textbf{n}_{AB})\int Y_{lm}(\textbf{N})N^{\langle L'\rangle}d\Omega=\delta_{ll'}n^{\langle L\rangle}_{AB}$ s well as by the integral,
\begin{equation*}
\int_0^\mathcal{R}dy\ y^n\frac{r_<^l}{r_>^{l+m}}\,=\,\int_0^{r_{AB}}dy\ \frac{y^{n+l}}{r_{AB}^{l+m}}+r_{AB}^l\int_{r_{AB}}^\mathcal{R}dy\ y^{n-l-m}\,=\,\frac{-2l-m}{(n+l+1)\ (n-l-m+1)}\ r_{AB}^{n-m+1}+\mathcal{O}\big(r_{AB}/\mathcal{R}\big),
\end{equation*}
where we remind that $r_{AB}\ll \mathcal{R}$. We also recall that $r_<=\text{min}(y,r_{AB})$, $r_>=\text{max}(y,r_{AB})$ and $\textbf{n}_{AB}=\textbf{r}_{AB}/r_{AB}$ is a unit vector pointing from body $B$ towards body $A$. The surface integral, which gives rise to an $\mathcal{R}$-dependent term only, was worked out by using $|\textbf{y}-\textbf{r}_{AB}|^{-3}\big|_{y=\mathcal{R}}=\mathcal{R}^{-3}+\mathcal{O}(r_{AB}/\mathcal{R})$, the angular integrals $(4\pi)^{-1}\int d\Omega\ N^p=0$ and $(4\pi)^{-1}\int d\Omega \ N^s N^p=\delta^{sp}/3$, where we remind that $\textbf{N}=\textbf{y}/y$ is a unit vector normal to the near-zone boundary surface and $d\Omega=\sin\theta d\theta d\phi$ is an element of solid angle specified by the angles $\theta$ and $\phi$ \cite{PoissonWill}. After a rather long but straightforward computation, using only the techniques displayed above, we obtain the following two results,
\begin{equation*}
\int_{\mathcal{M}_y}d\textbf{y} \ \frac{1}{|\textbf{y}-\textbf{r}_{AB}|^3} \frac{\textbf{y}+\textbf{r}_B}{y^3}\,=\,\frac{10\pi}{9}\frac{\textbf{r}_{AB}}{r_{AB}^3}+\frac{16\pi}{3}\frac{\textbf{r}_B}{r^3_{AB}} ,\quad \int_{\partial\mathcal{M}_y}\textbf{r}\cdot d\textbf{S} \ \frac{1}{|\textbf{y}-\textbf{r}_{AB}|^3}\frac{\textbf{y}+\textbf{r}_B}{y^3}\,\propto\,\mathcal{R}^{-3}.
\end{equation*}
The first term in the volume integral will eventually vanish because of the double sum running over $A$ and $A\neq B$ and $\textbf{r}_{AB}=-\textbf{r}_{BA}$ and the surface integral is proportional to the near-zone cut-off scale and can therefore be discarded too \cite{PoissonWill,WillWiseman,PatiWill1,PatiWill2}. The second position integral is more complicated in the sense that it contains even more terms that have to be worked out separately. However the integration techniques that are required to evaluate this integral do not change and were already extensively described in this appendix-subsection. We therefore content ourselves here to only provide the most important computational steps as well as all intermediate results. In analogy to the previous calculation we need to perform first the variable substitution $\textbf{y}=\textbf{x}-\textbf{r}_B$ followed by a shift of the domain of integration $\mathcal{M}\leadsto \mathcal{M}_y+\partial\mathcal{M}_y$. The integral essentially splits-up \cite{PoissonWill} in a volume and surface term,
\begin{gather*}
\,\int_{\mathcal{M}}d\textbf{x} \  \frac{(\textbf{x}-\textbf{r}_A)_m(\textbf{x}-\textbf{r}_A)_p}{|\textbf{x}-\textbf{r}_A|^5} \ \ \frac{(\textbf{x}-\textbf{r}_B)^m(\textbf{x}-\textbf{r}_B)^p}{|\textbf{x}-\textbf{r}_B|^5} \ \textbf{x}\,=\\ 
\int_{\mathcal{M}_y}d\textbf{y} \ \frac{(y-r_{AB})_p \ (y-r_{AB})_m }{|\textbf{y}-\textbf{r}_{AB}|^5}\frac{y^p \ y^m}{y^5} \ (\textbf{y}+\textbf{r}_B)-\oint_{\partial\mathcal{M}_y}\textbf{r}\cdot d\textbf{S} \ \frac{(y-r_{AB})_p \ (y-r_{AB})_m }{|\textbf{y}-\textbf{r}_{AB}|^5}\frac{y^p \ y^m}{y^5} \ (\textbf{y}+\textbf{r}_B)+\cdots.
\end{gather*}
It turns out to be convenient to divide these two terms into four volume contributions,
\begin{equation*}
\begin{split}
\int_{\mathcal{M}_y}d\textbf{y}\  \frac{y^4}{|\textbf{y}-\textbf{r}_{AB}|^5} \ \ \frac{(\textbf{y}+\textbf{r}_B)}{y^5}\,=&\, \frac{7\pi}{9}\frac{\textbf{r}_{AB}}{r_{AB}^3}+\frac{10\pi}{3} \frac{\textbf{r}_B}{r^3_{AB}},\\
 \int_{\mathcal{M}_y}d\textbf{y}\  \frac{r_{ABp}y^py^2}{|\textbf{y}-\textbf{r}_{AB}|^5} \ \ \frac{(\textbf{y}+\textbf{r}_B)}{y^5}\,=&\,\frac{304\pi}{225} \frac{\textbf{r}_{AB}}{r^3_{AB}}+\frac{14\pi}{15} \frac{\textbf{r}_B}{r^3},\\
  \int_{\mathcal{M}_y}d\textbf{y}\  \frac{r_{ABm}y^my^2}{|\textbf{y}-\textbf{r}_{AB}|^5} \ \ \frac{(\textbf{y}+\textbf{r}_B)}{y^5}\,=&\, \frac{304\pi}{225} \frac{\textbf{r}_{AB}}{r^3_{AB}}+\frac{14\pi}{15} \frac{\textbf{r}_B}{r^3},\\
\int_{\mathcal{M}_y}d\textbf{y} \ \frac{r_{ABm}y^m \ r_{ABp} y^p}{|\textbf{y}-\textbf{r}_{AB}|^5} \ \frac{(\textbf{y}-\textbf{r}_B)}{y^5}\,=&\,\frac{796\pi}{1225} \frac{\textbf{r}_{AB}}{r^3_{AB}}+\frac{584\pi}{315} \frac{\textbf{r}_B}{r^3_{AB}},
\end{split}
\end{equation*}
and four different surface terms with the surface element $dS^p=\mathcal{R}^2 N^pd\Omega$ on the boundary of the domain $\mathcal{M}_y$,
\begin{equation*}
\begin{split}
\oint_{\partial\mathcal{M}_y}\textbf{r}\cdot d \textbf{S} \ \frac{y^4}{|\textbf{y}-\textbf{r}_{AB}|^5} \ \ \frac{(\textbf{y}+\textbf{r}_B)}{y^5}\,\propto&\,\mathcal{R}^{-3}, \\
 \oint_{\partial\mathcal{M}_y}\textbf{r}\cdot d\textbf{S} \ \frac{r_{ABp}y^py^2}{|\textbf{y}-\textbf{r}_{AB}|^5} \ \ \frac{(\textbf{y}+\textbf{r}_B)}{y^5}\,\propto&\, \mathcal{R}^{-5},\\
  \oint_{\partial\mathcal{M}_y}\textbf{r}\cdot d\textbf{S} \ \frac{r_{ABm}y^my^2}{|\textbf{y}-\textbf{r}_{AB}|^5} \ \ \frac{(\textbf{y}+\textbf{r}_B)}{y^5}\,\propto&\,\mathcal{R}^{-5}, \\
\oint_{\partial\mathcal{M}_y}\textbf{r}\cdot d\textbf{S} \ \frac{r_{ABm}y^m \ r_{ABp} y^p}{|\textbf{y}-\textbf{r}_{AB}|^5} \ \frac{(\textbf{y}-\textbf{r}_B)}{y^5}\,\propto&\,\mathcal{R}^{-5}.
\end{split}
\end{equation*}
We observe that all four volume integrals give rise to a contribution proportional to $\textbf{r}_{AB}$ as well as to a term proportional to $\textbf{r}_B$.  While the latter will contribute to the effective barycentre of the $n$-body-system at the 1.5 post-Newtonian order of accuracy, the former will vanish after summation over $A$ and $B\neq A$. The four surface terms are all proportional to the near-zone scale $\mathcal{R}\lesssim \lambda_c$ and can therefore be discarded as they will eventually cancel out with terms coming from the wave-zone domain \cite{PoissonWill, WillWiseman, PatiWill1,PatiWill2}. Most of the contributions in the remaining piece of the effective Landau-Lifshitz pseudotensor, containing infinitely many derivative terms $\sum_{m=2}^{+\infty}\frac{\kappa^m}{m!}\Delta^m \tau^{00}_{LL}$, will not contribute to the effective barycentre at the 1.5 PN order of accuracy. After multiple partial integration they will be proportional to $\sum_A\sum_{B\neq A}\tilde{m}_A \tilde{m}_B \ \nabla^q\delta(\textbf{r}_A-\textbf{r}_B)\,=\,0,\ \forall q\geq 0, \ \text{or}\ \nabla^q \textbf{x}\,=\,0, \ \forall q\geq 2$ or to both terms at the same time. Surface terms, coming from (multiple) partial integration, are proportional to $\delta(\mathcal{R}-r_A)$ and will eventually vanish in the near-zone defined by $\mathcal{M}:|\textbf{x}|<\mathcal{R}$. However each derivative order $m\geq 2$ will produce a term proportional to $(\partial_{p_1}\cdots \partial_{p_m} V) \ (\partial^{p_1}\cdots\partial^{p_m}V)$. We explicitly illustrate this fact by analysing the derivative-term coming from $m=2$ in the sum displayed above, $\frac{\kappa^2}{2!}\Delta^2\tau^{00}_{LL}\,=\,-\frac{7\kappa^2}{16\pi G}\big[2(\partial_p\Delta^2V)\ (\partial^pV)+8(\partial^m\partial_p\Delta V)\ (\partial_m\partial^p V)+2(\partial_p\Delta V)\ (\partial^p \Delta V)+4(\partial_q\partial_m\partial_pV)\ (\partial^q\partial^m\partial^p V)\Big]$. While the first terms are essentially proportional to $(\partial_p\Delta V)(\partial^pV)\propto (\partial_p\delta(\textbf{x}-\textbf{r}_A))(\partial^pV)$ and therefore vanish after partial integration, only the last term is of the form mentioned above and will eventually contribute. The latter is proportional to,
\begin{equation*}
\partial_q\partial_m\partial_p V\,=G\sum_A\tilde{m}_A \ \Big\{ 3\frac{(\delta_{mp}(x-r_A)_q+\delta_{qp}(x-r_A)_m+\delta_{qm}(x-r_A)_p}{|\textbf{x}-\textbf{r}_A|^5}-5\frac{(x-r_A)_q\ (x-r_A)_m\ (x-r_A)_p}{|\textbf{x}-\textbf{r}_A|^7}\Big\},
\end{equation*}
we can deduce that we will eventually have $\frac{\kappa^2}{2!}\Delta^2\tau^{00}_{LL}\propto \big(\frac{\kappa}{r^2_{AB}}\big)^2$. Assuming that the bodies $A$ and $B$ are separated by astrophysical distances such that $\kappa\ll r^2_{AB}$ we see that this new term is smaller than the previous one by a factor $\frac{\kappa}{r^2_{AB}}\ll 1$. The next term, originating from $\frac{\kappa^3}{3!}\Delta^3\tau^{00}_{LL}$, will even be smaller than the leading term by a factor $\big(\frac{\kappa}{r^2_{AB}}\big)^2\ll 1$ this time. For the determination of the nonlocally modified 1.5 PN position vectors of a two body-system we have to recall the that to Newtonian order their respective position and velocity vectors are $\textbf{r}_1=(m_2/m)\textbf{r}$, $\textbf{v}_1=(m_2/m)\textbf{v}$ and $\textbf{r}_2=-(m_1/m)\textbf{r}$, $\textbf{v}_2=-(m_1/m)\textbf{v}$, where $\textbf{r}=\textbf{r}_1-\textbf{r}_2$ is the separation vector of the two bodies with masses $m_1$ and $m_2$. Substituting these vectors into the first post-Newtonian part of the effective two-body-barycentre, displayed in the main text, we obtain $\textbf{0}=m_1\textbf{r}_1+m_2\textbf{r}_2-\frac{\eta \gamma m}{2c^2}\big[v^2-(G(\sigma,\kappa,r)m)/r \big]\textbf{r}+\mathcal{O}(c^{-4},\kappa^2)$, where $G(\sigma,\kappa,r)=G\big[1-\frac{\sigma}{(1-\sigma)^2} \big(8-\sigma+\frac{831}{5}\frac{\kappa}{r^2}\big)\big]$ is the effective Newtonian coupling at the 1.5 PN order of accuracy. We attached the origin of the coordinate system to the barycentre $\boldsymbol{R}=\textbf{0}$. We also remind that $\eta=(m_1 m_2)/(m_1+m_2)^2$, $\gamma=(m_1-m_2)/(m_1+m_2)$ are dimensionless quantities and $m=m_1+m_2$ is the Newtonian mass of the binary-system. From this last equation together with the relative separation vector of the two bodies it is uncomplicated to deduce the effective two-body position vectors outlined in the main text.

\end{document}